\documentclass[twoside,12pt,openright]{book}

\usepackage{amsmath, amsthm, amssymb}
\usepackage{graphicx}

\newcommand{\prl}{Phys. Rev. Lett.~}
\newcommand{\prb}{Phys. Rev. B~}
\newcommand{\rmp}{Rev. Mod. Phys.~}

\makeatletter   
\def\cleardoublepage{\clearpage\if@twoside \ifodd\c@page\else%
\hbox{}%
\thispagestyle{empty}  
\newpage%
\if@twocolumn\hbox{}\newpage\fi\fi\fi}   
\makeatother 
 
\usepackage[bf]{caption}
\begin{document}

~

~

\begin{center}

{\Huge {\bf Influence of Spin and Interactions on} \\ {\bf Quantum Dots and Nano-Wires} \\ ~ \\ }

\Large{~ \\ {\bf Yuval Weiss} \\ ~
\\ ~ \\ ~ \\ Department of Physics \\
Bar-Ilan University, Israel \\ ~ \\ 
%
%
~ \\ ~ \\ {\it Ph.D. Thesis} \\ ~
\\ ~ \\ ~ \\ ~ \\ Submitted to the Senate of Bar-Ilan University \\
~ \\ Ramat-Gan, Israel ~~~~~~~~~~~~~~~~~~~~~~~~~~~~~~~~~ October 2007}

\end{center}


\baselineskip=18pt plus1pt




\setcounter{secnumdepth}{2}
\setcounter{tocdepth}{2}

\newpage \pagestyle{empty}

\cleardoublepage

~

~

~
\begin{center}

\Large{This work was carried out under the supervision of \\ ~ \\
Prof. Richard Berkovits\\ ~ \\
Department of Physics, Bar-Ilan University.}

\newpage \pagestyle{empty}
\cleardoublepage

~

~

~

\Large{
This research was generously supported by \\ ~ \\
Bar-Ilan University's President and Dean Scholarship \\ ~ \\
for outstanding Ph.D. students.\\ ~ \\
For that I am grateful.}

\end{center}


\newpage \pagestyle{empty}
\cleardoublepage
{\Huge \bf Acknowledgements}

~ \\ ~ \\

I am grateful to my supervisor, Professor Richard Berkovits, for 
his pleasant guidance along the recent years. The unique combination of
great physical knowledge and intuition on the one hand, 
together with enormous kindness and support on the other, 
makes him an ideal supervisor.

~

I'm indebted to my colleague, Moshe Goldstein, for the important contributions
he has made to many parts of this work, for the fruitful deliberations and
collaboration, and for the pleasant atmosphere in our room.

~

Other colleagues have assisted me, in various ways, to accomplish this work,
and it is my pleasure to acknowledge them as well:
Dr. Miri Sade, for the collaboration in my first DMRG code;
Hagai Vilchik and Dr. Avi Cohen, for initiating and participating in many theoretical discussions;
Liora Bitton and Noa Kurzweil, for bridging between the theoretical discussions to the real life
in the lab.

~

I'd like to thank my dear family for their continuous assistance and support:
My wife Nurit, for her love and care, and for letting and
encouraging me to do the things I like. 
My daughters Yehudit and No'omi, for being always proud of their Dad. 
My extended family, father, sister, brother and  
parents in law, and their families, for the constant help and interest.

~

At last I'd like to contribute a special acknowledgement to my dear mother, 
who has educated me, and trained me to attempt doing my tasks as best as I can. 
Despite her tragic death more than two years before 
the beginning of my PhD studies, she is the real driving force behind this work.

~

This thesis is dedicated to her memory. 

\newpage \pagestyle{empty}
\cleardoublepage

~ \\

~ \\

~ \\

\begin{center}
{\Large To my Mom}
\end{center}



\cleardoublepage
\newpage 

\pagenumbering{roman}
\pagestyle{headings}

\tableofcontents

\newcommand{\newchap}[1]    
	{                   
	\chapter*{#1}
	  \markboth{\hfill \MakeUppercase{{{#1}}}} {}
	  \markright{\MakeUppercase{{{#1}}} \hfill }
	}


\cleardoublepage	
\newpage 
\addcontentsline{toc}{chapter}{Abstract}
\newchap{Abstract}

The interest in the physical properties of mesoscopic systems experiences
a gradual increase in recent years. Significant results, which were obtained by both 
theoretical and experimental studies, together with the enormous promise of  
nano-technology applications, contribute to this interest.

In many mesoscopic systems, the length scale governing the electrons motion is
small enough to cause quantization of the energy levels. In this work we examine 
two types of such quantum systems: A {\it quantum dot} (QD) and a {\it nano-wire}. 
In a QD, the system is small enough in all dimensions, so that its energy spectrum is entirely 
discrete, while in a nano-wire the energy spectrum is quantized in two of the space coordinates, 
and continuous in the third direction.
During the last decade several experimental techniques have been developed for manufacturing
both kinds of devices, which are currently
an important tool for understanding low dimensions physics.

As a result, experiments on mesoscopic samples are conducted nowadays in many 
laboratories, and many interesting results are obtained. 
For example, some transport properties were recently shown
to exhibit interesting phenomena in the presence of both interactions
and disorder. An analytical treatment of these problems is unfortunately
difficult, since both the disorder and the interactions cannot be
considered as a small perturbation. A traditional numerical treatment of such systems 
(exact diagonalization) is
also problematic, since the dimension of the many-particle Hilbert space grows exponentially 
with the system size. Thus this numerical method is usually restricted to small systems.
Nevertheless, using analytical methods and sophisticated numerical calculations, the interplay 
between disorder and interactions is investigated in the current research for several types of systems.


In the first half of this work we study a system composed of a QD
with a single level which is
coupled to a one-dimensional (1D) interacting wire with spinless electrons.
We start by focusing on the filling of the dot level, the total occupation 
and the free energy of the system.
Using Green functions technique, we calculate these observables 
in the non-interacting limit. 
We then investigate numerically, using the density-matrix renormalization-group (DMRG) method, 
two phases of the interacting lead: 
The Tomonaga-Luttinger liquid (TLL) and the charge density wave (CDW) phases.
We explore the influence of interactions in the lead, as well as dot-lead interactions,
on the width of the dot filling as a function of the chemical potential, and on the position 
of the dot level. In the TLL phase the results are explained within the random phase approximation.
In the case of a CDW, we show that a semi-infinite lead 
coupled to the dot undergoes a first order quantum phase transition when the
dot's level crosses the wire's chemical potential. 

The Friedel oscillations in the wire, resulting from the
dot located at one of its edges, are then studied.
For the non-interacting case, we develop an exact formula of the oscillations in the 1D tight-binding model. 
When interactions in the wire are considered, the difference between the two phases of the lead
are explored. In the TLL phase the oscillations of a clean 
interacting sample decay as a power law, and once a disorder is introduced (Anderson insulator (AI) phase), 
the power law decay is multiplied by an exponential decay term due to the disorder.
The resulting decay length is shown to increase as a function of the interaction strength.
On the other hand, when the wire is in a CDW phase and the disorder 
is weak enough, the wire may still be described by a Mott insulator phase, and
the effect of interactions is the opposite.

We prove that the length scale governing the exponential decay, in the AI phase, may be
associated with the Anderson localization length and thus be used as a convenient
way to determine the dependence of the localization length on disorder and
interactions. Our results show a decrease of the localization length as a function
of the interaction strength, in accordance with previous predictions.

In the second half of the research
we study two cases of an isolated two-dimensional QD. 
We begin by exploring some properties of a 
disordered QD consisting of interacting spinless electrons.
Since the size of the relevant Hilbert space is huge, such a
problem cannot be solved by an exact diagonalization (except for very small systems).
We thus use the sophisticated particle-hole DMRG (PH-DMRG) method, showing that its approximation for
the ground-state energy is much more accurate than that of the Hartree-Fock (HF) method.
We also suggest an improvement of the PH-DMRG truncation algorithm, which reduces the error rate of the
traditional method by almost $30$ percents.

As an application of the improved PH-DMRG method we calculate the addition spectrum of the QD. 
We present the improvement of the PH-DMRG results comparing to the HF approximation in three 
aspects: the error rate, the average and the fluctuations of the addition spectrum.

Finally
we study the magnetization of a QD with spin $1/2$ electrons, 
in the presence of spin-orbit (SO) coupling. 
We calculate the g-factor and the expectation values of the spin operators in the ground state.
We find that when the dot is occupied by an even number of electrons, there is
a level crossing between the two lowest many-body eigenfunctions as a function of 
the SO scattering rate, resulting in a finite magnetization of the ground state.
This is a clear signature of the interplay between SO scattering and interactions,
and may have a significant influence on g-factor measurements.


~

~

~

~

~

~

~

~

~

~

~





\cleardoublepage
\newpage 
\addcontentsline{toc}{chapter}{List of Publications}
\newchap{List of Publications}

\begin{enumerate}
\item 
{\it Level coupled to a one-dimensional interacting reservoir: A density matrix renormalization group study}

{\small Miri Sade, {\bf Yuval Weiss}, Moshe Goldstein and Richard Berkovits

Phys. Rev. B {\bf 71}, 153301 (2005). }

\item 
{\it A DMRG study of a level coupled to a 1D interacting lead}

{\small {\bf Yuval Weiss}, Miri Sade, Moshe Goldstein and Richard Berkovits

Phys. Stat. Sol. (b) {\bf 243}, 399 (2006).}

\item 
{\it Driving a first order quantum phase transition by coupling a quantum dot
to a 1D charge density wave}

{\small {\bf Yuval Weiss}, Moshe Goldstein and Richard Berkovits

J. Phys.: Condens. Matter {\bf 19}, 086215 (2007).}

\item 
{\it Friedel oscillations in disordered quantum wires: 
Influence of electron-electron interactions on the localization length}

{\small {\bf Yuval Weiss}, Moshe Goldstein and Richard Berkovits

Phys. Rev. B {\bf 75}, 064209 (2007). }

\item 
{\it Disorder effect on the Friedel oscillations in a one-dimensional Mott insulator}

{\small {\bf Yuval Weiss}, Moshe Goldstein and Richard Berkovits

Phys. Rev. B {\bf 76}, 024204 (2007). }

\newpage

\item 
{\it A generation-based particle-hole density-matrix renormalization group study of interacting quantum dots}

{\small {\bf Yuval Weiss} and Richard Berkovits

Solid State Commun. {\bf 145}, 585 (2008).}

\item 
{\it Significant g-factor values of a two-electron ground state in quantum dots with spin-orbit coupling}

{\small {\bf Yuval Weiss}, Moshe Goldstein and Richard Berkovits

arXiv:0710.2772 (Submitted to Phys. Rev. B).}

\end{enumerate}

\cleardoublepage
\addcontentsline{toc}{chapter}{List of Figures}
\listoffigures

\cleardoublepage
\newpage 
\addcontentsline{toc}{chapter}{List of Abbreviations}
\newchap{List of Abbreviations}

\begin{enumerate}

\item [{$1D$}~~] One-Dimensional

\item [{$2D$}~~] Two-Dimensional

\item [{AI}~~] Anderson Insulator

\item [{CI}~~] Constant Interaction
\item [{CDW}~~] Charge Density Wave

\item [{DMRG}~~] Density Matrix Renormalization Group

\item [{FM}~~] Ferromagnetic
\item [{FO}~~] Friedel Oscillations

\item [{GOE}~~] Gaussian Orthogonal Ensemble
\item [{GUE}~~] Gaussian Unitary Ensemble
\item [{GSE}~~] Gaussian Symplectic Ensemble

\item [{HF}~~] Hartree Fock

\item [{LC}~~] Level Crossing

\item [{MG}~~] Mott Glass
\item [{MI}~~] Mott Insulator

\item [{NN}~~] Nearest Neighbor
\item [{NNN}~~] Next Nearest Neighbor
\item [{NRG}~~] Numerical Renormalization Group

\item [{PH-DMRG}~~] Particle-Hole DMRG

\item [{QD}~~] Quantum Dot
\item [{QPC}~~] Quantum Point Contact
\item [{QPT}~~] Qunatum Phase Transition
\item [{RMT}~~] Random Matrix Theory

\item [{SO}~~] Spin-Orbit

\item [{TLL}~~] Tomonaga-Luttinger Liquid

\end{enumerate}
\onecolumn

\pagenumbering{arabic}
\pagestyle{headings}

\newpage 

\cleardoublepage
\chapter{Introduction} \label{cpt:intro}

In recent years there is a gradually increasing interest in the physical 
properties of mesoscopic systems. Significant results, which were obtained by both 
theoretical and experimental studies,
together with the enormous promise of nano-technology applications, contribute to
this interest.

Mesoscopic physics deals with systems whose sizes are between the microscopic
and the macroscopic regimes \cite{Altsh94,Imry96}. 
Such an intermediate regime has a length-scale which is
small enough so that quantum physics phenomena are relevant, and still large enough
to enable laboratory experiments. This unique combination, 
gives a challenging topic for research, and attracts many recent investigations. 

The classification of a physical system as a mesoscopic one is usually done by
an examination of the electron motion, checking whether it is coherent, i.e., 
if the electron's phase is conserved \cite{alhassid00}. 
The loss of phase conservation can be caused by inelastic collisions with phonons
or with other electrons, thus one can say that for mesoscopic systems, the mean 
distance which the electron passes between such subsequent inelastic collisions, is 
significantly larger than the system size. Since the probability for inelastic
collisions decreases when the temperature is lowered, the recent advances in
low-temperature physics make experimental development of mesoscopic systems
feasible.

It is important to notice, however, that there is another important type of 
collisions, which does not cause a phase decoherence. For example, when an electron 
experiences an elastic collision with an impurity, its phase is conserved, so that 
the motion is coherent. Such collisions, therefore, can be used to further classify 
the system. Usually one defines the electron's {\it mean free path} as the mean 
distance which the electron passes between subsequent elastic collisions with 
impurities. Denoting the mean free path by $l$, the ratio between $l$ and either the 
system size $L$ or the inverse Fermi momentum $k_F^{-1}$ can characterize the 
motion. For $l>L$ the electron is hardly
influenced by the impurities, so that its motion is ballistic. In the other limit,
when there are lots of impurities, i.e., the disorder is strong, so that $l<k_F^{-1}$
(where $k_F$ is the Fermi momentum),
the electron's motion is limited in space, and the electronic wave functions are 
localized. In the intermediate regime, for $k_F^{-1}<l<L$, the motion is diffusive \cite{alhassid00}. 

In many mesoscopic systems, the length of one (or more) of the dimensions is
small enough to cause quantization of the energy levels.
In our research we will examine a {\it quantum dot}, in which the system is
small enough in all dimensions, so that its energy spectrum is entirely discrete.
In some parts of this work, we will be interested also in a one-dimensional lead 
which is connected to the dot. Such a device, composed of a dot and one or two 
leads, can be used for various transport and charge sensing measurements, and can
be an important tool for understanding low dimensions physics.

In this chapter we shortly describe the two main parts of the 
physical model. We first briefly introduce the quantum dot, after which we will
take a short tour in the one-dimensional world. We conclude this chapter
with the motivation for the research, and the outline of the rest of the thesis.

\section{Quantum Dots}

A {\it quantum dot} (QD) is a mesoscopic physical device,
in which the motion of the electrons is restricted in space, so that the 
energy levels are quantized in all three dimensions. As a result, the QD 
has a completely discrete energy spectrum \cite{Beenakker91,meirav95,Kelly95,kouwenhoven98}.

Experimentally, there are several physical realizations, such as metallic or semiconducting 
nano-particles, in which the electrons are restricted to a certain region in space.
Most experimental QDs are based, however, on semiconductor heterostructures.
A two-dimensional (2D) plane, into which the electrons are confined, is created, and
by further applying a voltage above or below specific points in
the plane, one is able to restrict the electrons' motion inside a small region in
this 2D plane. An example for such a construction is shown in Fig.~\ref{fig:qd_pics}.
Therefore, a common QD situation is that of electrons moving in a confined 2D area, 
which is separated from other regions by potential barriers.

\begin{figure}[htbp]
\centering
\caption[An example of a quantum dot]
{\label{fig:qd_pics} A typical example of a 2D QD (drawn schematically on the 
left, while the actual electron micrograph is shown on the right). The electrons are confined
vertically to the 2D interface between GaAs and AlGaAs, and negative voltages applied to 
the metallic gates confine them also laterally. The transport of electrons through the dot 
is indicated by red arrows. The mean free path and the coherence length are of the order of 
10 microns, so that the electronic motion is coherent and ballistic 
(picture taken from Ref.~\cite{kouwenhoven98}).}
\end{figure}

\subsection{Hamiltonian for a QD}
\label{sec:intro_Hqd}
As many QDs are defined on a 2D electron gas, one of
the convenient ways to model them numerically is to use the {\it Anderson model},
which is derived from a tight-binding description of a 2D dot. 
Assuming that an electron is restricted to lattice sites, with the 
ability to hop from one site to another (usually limited to be one of its 
neighbors), one can therefore replace the continuous space with an effective 2D 
lattice model, for example of $A$ rows and $B$ columns, which sum up to $A  B$ 
sites.

For each one of the $AB$ sites there is a well defined {\it on-site energy}, 
$\epsilon$, which is the potential that the electron feels when it is located 
on that site. In order to model impurities in the QD, a common method is to 
take a different on-site energy for each site. These energies are usually 
determined randomly, using a uniform distribution between $[-W/2,W/2]$, where
$W$ describes the strength of the disorder. Therefore, by using the fermionic 
operators $\hat a^\dagger_i$ and $\hat a_i$ for a creation and annihilation, 
respectively, of an electron at site $i$, the on-site energy term in the 
Hamiltonian can be written as $\sum_i \epsilon_i \hat a^\dagger_i \hat a_i$.

The hopping element will be written, in a similar notation, as
$-t \sum_{\langle i,j \rangle} (\hat a^\dagger_i \hat a_j + H.c.)$, where $\langle i,j \rangle$ denotes
nearest neighbor (NN) sites $i$ and $j$, and the overlap integral between 
NNs is represented by $t$ and is taken to be constant.
Finally, we get the following Hamiltonian:
\begin{eqnarray} \label{eqn:Intro_Hdot}
\hat H^0_{\rm QD} = 
\sum_i \epsilon_i \hat a^\dagger_i \hat a_i
-t \sum_{\langle i,j \rangle} (\hat a^\dagger_i \hat a_j + H.c.).
\end{eqnarray}

\subsubsection{Spin-Orbit Coupling}
For spin $1/2$ particles the Hamiltonian Eq.~(\ref{eqn:Intro_Hdot}) should be slightly
modified in order to describe the physics of the system. It is convenient to separate
the site index from the spin index, and to use two indices:
$i$ which describes the site index, and $\sigma=~\uparrow,\downarrow$ 
which represents the spin projection.
For a system of $N$ sites with spin $1/2$ electrons, one should thus diagonalize a matrix of size
$2N \times 2N$.

In some cases there is no {\it spin-orbit coupling}, i.e., the orbital and the spin degrees 
of freedom are completely decoupled. In these cases,
the system, and the resulted Hamiltonian matrix, has an orthogonal symmetry. Effectively,
the Hamiltonian remains identical to $\hat H^0_{\rm QD}$ described above, and with the new 
indices notation it can be written as
\begin{eqnarray} \label{eqn:Intro_Hdot2}
\hat H^0_{\rm QD} = 
\sum_{i\sigma} \epsilon_i \hat a^\dagger_{i\sigma} \hat a_{i\sigma}
-t \sum_{\langle i,j \rangle,\sigma} (\hat a^\dagger_{i\sigma} \hat a_{j\sigma} + H.c.).
\end{eqnarray}

However, when spin-orbit coupling is taken into account, the system's symmetry 
is no longer orthogonal, but rather symplectic \cite{dyson62,porter65}. 
Such a coupling can result, e.g. from the magnetic moments of the impurities, allowing
an electron to change its spin projection during hopping between sites.
The Hamiltonian for such a system can be thought of as an $N \times N$ matrix, in which every element is 
a quaternion with $4$ degrees of freedom. Diagonalization of such a Hamiltonian results in 
pairs of degenerate eigenvalues, which is known as the Kramers degeneracy \cite{Merzbacher}.

In practice, in common QD devices (see e.g. Fig.~\ref{fig:qd_pics}), the electrons
are confined to a 2D interface separating between regions having different electric potentials. 
As a result, a perpendicular electric field is affecting the electrons, leading to the 
spin-orbit {\it Rashba} term \cite{rashba84} in the Hamiltonian, 
$H_{Rashba} = \alpha E_z (\vec \sigma \times \vec p)_z$.

A useful method for considering such a term for a lattice,
which was suggested by Ando \cite{ando89}, is by replacing the overlap integral $t$ 
of Eq.~(\ref{eqn:Intro_Hdot2})
by two $2 \times 2$ matrices $V_x$ and $V_y$ which encompass the probabilities to change 
or preserve the spin projection in transitions in the $\hat x$ and $\hat y$ directions.
If we now denote the 2D-lattice site with row and column indices $m,n$, respectively,
the resulting Hamiltonian reads
\begin{eqnarray} \label{eqn:Intro_Hdot_so}
\hat H^0_{\rm QD} &=& 
\sum_{m,n,\sigma} \epsilon_{m,n} \hat a^\dagger_{m,n,\sigma} \hat a_{m,n,\sigma} \\ \nonumber
&-& \sum_{m,n,\sigma,\sigma^\prime} ( V_{x} \hat a^\dagger_{m,n,\sigma} \hat a_{m,n+1,\sigma^\prime} +
 V_{y} \hat a^\dagger_{m,n,\sigma} \hat a_{m+1,n,\sigma^\prime} + H.c.).
\end{eqnarray}

The $V$ matrices written by Ando are composed of the identity matrix and Pauli
matrices $\sigma_x,\sigma_y$ with the prefactors $V_1$, for a motion without changing the spin,
and $V_2$, for spin-flips. Note that the Rashba term leads to an influence of  
$\sigma_x$ on the motion in the $y$ direction, and of $\sigma_y$ in the $x$ direction. 
The matrices are thus 
\begin{equation} \label{eqn:Intro_Hso_V}
V_x = \left( \begin{array}{cc}
	V_1 & V_2 \\
	-V_2 & V_1 \\
	\end{array} \right)~~;~~~
	V_y = \left( \begin{array}{cc}
	V_1 & -iV_2 \\
	-iV_2 & V_1 \\
	\end{array} \right)~.
\end{equation}

A correct building of the $V$ matrices is essential in order to conserve the 
system symplectic symmetry.
The main manifestation of such a system symmetry is a sign flip under a $2\pi$ rotation, 
preserving a regular symmetry under rotations of $4\pi$.

\subsubsection{Magnetic Field}
In order to insert an in-plane magnetic field dependence into the Hamiltonian, one adds the 
term $H_B =\vec \sigma \cdot (\mu_B \vec H)$ to the Hamiltonian, where $\sigma$ represents a vector of Pauli spin matrices,
and $\mu_B = \frac{e\hbar}{2mc}$ is the Bohr magneton. In the presence of
spin-orbit coupling, the in-plane magnetic field 
yields a non-diagonal term. For example, for a magnetic field in the
$\hat x$ direction, i.e., $\vec H = H \hat x$, one has $H_B = \mu_B H \sigma_x$. As a result, for the $i$-th 
energy level, the magnetic field couples the elements $|i,\uparrow>$ and $|i,\downarrow>$ through
\begin{equation*} \label{eqn:Intro_H_B}
H_B  \left( \begin{array}{c} 
		i,\uparrow \\
		i,\downarrow \\
	\end{array} \right)
= \mu_B H \left( \begin{array}{cc}
		0 & 1 \\
		1 & 0 \\
		\end{array} \right)
		\left( \begin{array}{c} 
		i,\uparrow \\
		i,\downarrow \\
	\end{array} \right)
= \mu_B H \left( \begin{array}{cc}
		i,\downarrow \\
		i,\uparrow \\
	\end{array} \right)~.	
\end{equation*}

A Perpendicular magnetic field gives a simpler term, since $\sigma_z$ is diagonal in the spin
indices $\uparrow$ and $\downarrow$. 
However, in that case one should consider also orbital effects and, as a result, the hopping term we take
includes the diamagnetic coupling through the Peierls substitution \cite{Ferry97}.
Taking a Landau gauge for the vector potential $A = H y \hat x$ one gets a revised hopping 
element, since in the $\hat x$ direction \cite{berkovits94} the hopping element $t$, or the matrix $V_x$, if 
spin-orbit coupling is considered \cite{usaj04,tsitsi04}, should be multiplied by 
$\exp \left( i \frac {2 \pi m H s^2}{\phi_0} \right)$, where $m$ is the row number, 
$s$ is the lattice constant, and $\phi_0 = hc/e$ is the magnetic flux unit.


\subsubsection{Interactions}
When the interactions between electrons cannot be neglected, an appropriate term
should be added to the Hamiltonian. Since in metallic materials there is a
substantial screening effect, one usually restricts the interaction terms to have a short
range. 
For spinless electrons it is thus sufficient in many cases to deal with NN interactions, 
so that the interaction term can be written as
\begin{eqnarray} \label{eqn:Intro_Hdot_int2}
\hat H^{(nn)}_{\rm int} = U_{\rm nn} \displaystyle \sum_{\langle i,j \rangle}
{\hat a}^{\dagger}_{i}{\hat a}^\dagger_{j}{\hat a}_{j}{\hat a}_{i},
\end{eqnarray}
while in other cases one may consider the full long-ranged Coulombic term, i.e.,
\begin{eqnarray} \label{eqn:Intro_Hdot_int1}
\hat H^{(Coulomb)}_{\rm int} = U_c \displaystyle \sum_{i < j} \frac{1}{r_{ij}} 
{\hat a}^{\dagger}_{i}{\hat a}^\dagger_{j}{\hat a}_{j}{\hat a}_{i},
\end{eqnarray}
where $r_{ij}$, the distance between the sites $i$ and $j$, is measured in units of
the lattice constant.

When the spin of the electron is also considered, the most important interaction
term is the on-site interaction between spin-up and spin-down electrons. This
is the Hubbard term which is written as
\begin{eqnarray} \label{eqn:Intro_Hdot_Hubbard}
\hat H^{(Hubbard)}_{\rm int} = U_H \displaystyle \sum_{i}
{\hat a}^{\dagger}_{i,\uparrow}{\hat a}_{i,\uparrow}{\hat a}^\dagger_{i,\downarrow}{\hat a}_{i,\downarrow}.
\end{eqnarray}

The innocent look of the interaction terms hides the huge impact they have on the 
calculation of the system's physical properties. This results from the fact that the 
system states are multi-particle states, and thus the dimension of the Hilbert space
grows exponentially with the system size. This issue is described in detail
in chapter~\ref{cpt:numerics}.

\subsection{Coulomb Blockade}
\label{sec:intro_CB}
Once the QD is connected to two metallic leads, applying a bias voltage on the 
leads can cause an electric current through the dot. An interesting phenomena
occurs when the QD is weakly coupled to the leads, so that the transport of
electrons to and from the dot is by tunneling. Usually, tunneling
of an electron into the dot is blocked by the Coulomb repulsion of the electrons
which are already inside the dot. However, by changing the gate voltage one can 
compensate for that energy difference, and for an appropriate value of $V_g$ the 
number of electrons in the dot can increase by one, and a peak in the conductance 
through the dot appears.

This phenomenon, named the {\it Coulomb blockade}, was first observed \cite{giaever68}
as early as in 1968. The gradual increase of the gate voltage was found to cause
jumps in the current through the quantum dot in specific values of the gate potential. 
These jumps are easily seen in $I$-$V$ (or $\frac{dI}{dV}$ vs. $V$) curves.
The energy values in which the jumps occur were shown to be sample-dependent, 
but subsequent measurements of the I-V characteristics for the same sample gives the 
same values each time. 

For zero temperature, the conductance is possible only through one of the empty 
discrete energy levels in the dot, $\epsilon_N$.
The transition is allowed only if the total energies when there are $N-1$ or $N$
electrons in the dot are identical. The electrostatic energy in the dot can be written 
in the {\it constant interaction (CI) model} \cite{alhassid00} as 
$U(N) = -eNV_g + e^2 N (N-1)/2C$, where $C$ is the total capacitance between the dot
and the leads. Let's suppose that there are currently $N-1$ electrons in a dot. 
The condition for the transition of the $N$-th electron into the dot is 
\begin{eqnarray} \label{eqn:Intro_N_minus_1}
\sum_{i=1}^{N-1} \epsilon_i + U(N-1) = \sum_{i=1}^{N} \epsilon_i + U(N). 
\end{eqnarray}
Substituting the formula for $U(N)$ given above,
leads to the relation which determines $\mu_N = e V_g$, the chemical potential of the 
dot in which the conductance peak occurs, as 
\begin{eqnarray} \label{eqn:Intro_mu_N}
\mu_N = e V_g = \epsilon_N + (N-1)e^2/C.
\end{eqnarray}

After the $N$-th electron was added to the dot, a further increase of $V_g$ blocks again 
the transport channel, until the next electron will be in the appropriate position to enter the dot. 
As a result, peaks will appear in the curve of $\frac{dI}{dV}$ as a function of the 
gate voltage, for specific values of $V_g$. In practice, one can use this method
in order to obtain a control on the number of electrons in the dot. Starting from
an initial state in which the dot is not charged, one can gradually increase
the gate voltage and count the current peaks, and thus know exactly how many 
electrons have transferred into the dot.

\subsection{Addition Spectrum}
\label{sec:intro_delta2}
Using Eq.~(\ref{eqn:Intro_mu_N}) it is easy to define the addition spectrum as the 
change in chemical potential required to add the $N$-th electron to the dot, i.e., 
the distance between the subsequent peaks $N-1$ and $N$:
\begin{eqnarray} \label{eqn:Intro_Delta2}
\Delta_2^{(N)} = \mu_{N}-\mu_{N-1} = \epsilon_{N} - \epsilon_{N-1} + e^2/C.
\end{eqnarray}

Thus, based on the assumptions of the CI model, one finds that the addition spectrum 
consists of the level spacing with an extra charging energy.
The reason that the distance between peaks seems sometimes constant, is that 
in most cases the charging energy is very large comparing to the level spacing fluctuations.
However, by subtracting this constant charging energy from the measured distances, 
one can reconstruct the probability distribution of the level
spacings. For disordered quantum dots in the diffusive regime, the level spacings
have a well known probability distribution, the Wigner (or Wigner-Dyson) distribution,
which was first predicted in the 1950's in calculations of energy levels in the nucleus
\cite{wigner53,dyson62}. The distribution has a different form for different symmetry classes 
(orthogonal, unitary and symplectic), which are sometimes denoted by GOE, GUE and GSE 
(for Gaussian orthogonal ensemble, and similarly for the other two).
The Wigner-Dyson distribution has become an important part of the random matrix theory 
\cite{porter65,Mehta91}.

Unfortunately, several experiments on QDs which were performed in the last decade are not consistent 
with the results of the random matrix theory \cite{sivan96,simmel97,patel98}. 
For example, while it was found that the mean level spacing can be described by the CI model, 
the $\Delta_2$ distribution does not fit the predicted Wigner-Dyson distribution.
It is thus clear that for obtaining an accurate description of some measurements,
the effect of interactions should be considered beyond the CI model.

In order to take into account the interactions in the system, one needs to define
the addition spectrum in an alternative way, without any assumption on the interactions.
In general, the gate energy in which the $N$'th conductance peak occurs must be equal to the 
difference in the ground-state energies for $N-1$ and $N$ electrons. 
Therefore, $\mu_N = E_{gs}(N) - E_{gs}(N-1)$, which leads to the exact definition 
of the addition spectrum as
\begin{eqnarray} \label{eqn:Intro_Delta2_exact}
\Delta_2^{(N)} = \mu_{N}-\mu_{N-1} = E_{gs}(N) - 2E_{gs}(N-1) + E_{gs}(N-2).
\end{eqnarray}
Such an expression requires the knowledge of three different ground-state
energies, with consecutive electronic occupation.

\section{One-Dimensional Lead}
\label{sec:int_1d}
The previous section mentioned some of the physical results of connecting the QD to  
external leads: Current can now be transferred through the QD; The coupling of the dot 
levels with the leads causes a broadening of the energy levels width; If this
coupling is weak enough the discrete levels of the QD can still be seen experimentally.
Of course, more complicated properties of the QD can be measured as well. However, it is
interesting enough to explore these simple phenomena for different kinds of the
external leads. What will be the influence of the lead on the QD when the lead is not 
described by a simple Fermi liquid theory? Does it have a similar influence, or, perhaps,
new physics may be explored?

For answering these questions, we would like to connect a special kind of external lead to 
the QD: a one-dimensional (1D) lead. A device is defined as 1D if its energy spectrum is 
quantized in two of the space coordinates, while it is continuous in the third direction. 
During the last decade several experimental techniques have been developed for manufacturing 
various 1D devices, such as carbon nanotubes \cite{swnt1,swnt2,swnt3,avouris99,mwnt}, 
polymer nanofibers \cite{polyfiber} and semiconducting nanowires \cite{pescini99,mose}. 
For example, Fig.~\ref{fig:nanowire1} shows a silicon quantum wire, which is suspended in a 
highly doped silicon film in a silicon-on-insulator substrate.

\begin{figure}[htbp]
\centering
\caption[Example of a one-dimensional device]
{\label{fig:nanowire1} SEM micrograph of a suspended silicon quantum wire in a highly n-doped 
silicon-on-insulator film. The wire width is $80 nm$ and the length is $1.5 \mu m$
(taken from Ref.~\cite{pescini99}).}
\end{figure}

It is thus clear that the creation of a 1D lead is nowadays experimentally possible.
As a result, a door to a new world of physical models is opened. In this section we give
a brief introduction to that world.

\subsection{Hamiltonian for a 1D Lead}
We start by presenting the Hamiltonian of a 1D lead. As in the QD case,
we restrict ourselves to the tight-binding description of the electronic orbitals,
and thus we start the discussion from a 1D lattice. The electrons'
motion is thus not continuous, as they jump from one site to another. The potential
energy of the electrons originates from the on-site energy of each lattice site. If 
every site has the same on-site energy, all energies can be rescaled with respect 
to that energy, and it can be excluded from the Hamiltonian. However, when there is a 
difference between lattice sites, e.g. when the unit cell consists of two or more atoms,
or when impurities are involved, this term must be included.

The kinetic energy of the electrons is written through the hopping element, which is in
charge for the electrons' motion along the lead. As in the QD case, we restrict this term
to the physical intuition of NN hopping only. This, of course, has a 
general justification since the hopping probability decays exponentially with distance.

As a result, for electrons with spin but without spin-orbit coupling\footnote[1]
{Spin orbit coupling in the lead is not discussed in this thesis.},
one gets the following Hamiltonian:
\begin{eqnarray} \label{eqn:Intro_Hlead1}
\hat H^0_{\rm lead} = 
\sum_{i,\sigma} \epsilon_i \hat c^\dagger_{i,\sigma} \hat c_{i,\sigma}
-t \sum_{i,\sigma} (\hat c^\dagger_{i,\sigma} \hat c_{i+1,\sigma} + H.c.),
\end{eqnarray}
where $c^\dagger_{i,\sigma}$ ($c_{i,\sigma}$) is the creation (annihilation) operator\footnote[2]
{For convenience, along this thesis the dot operators are denoted by $\hat a$
whereas the lead operators are denoted by $\hat c$.}
for an electron with spin $\sigma$ in lattice site number $i$, which has an on-site energy 
$\epsilon_i$.
For spinless fermions the Hamiltonian is even simpler, and has the form
\begin{eqnarray} \label{eqn:Intro_Hlead2}
\hat H^0_{\rm lead} = 
\sum_{i} \epsilon_i \hat c^\dagger_{i} \hat c_{i}
-t \sum_{i} (\hat c^\dagger_{i} \hat c_{i+1} + H.c.).
\end{eqnarray}

The addition of interactions is straightforward. As noted in the previous subsection,
usually one can consider only the most important term, which is the Hubbard interaction
in the spin $1/2$ case, and the NN interactions for spinless fermions.
From now on
we will focus on the spinless fermionic case, without the existence of a disorder. Therefore, 
if we assume a positive background in the lattice sites, one can write the interaction term as
\begin{eqnarray} \label{eqn:Intro_Hlead_int}
\hat H^{int}_{\rm lead} = I
\sum_{i} (\hat c^\dagger_{i} \hat c_{i} - 1/2) ( \hat c^\dagger_{i+1} \hat c_{i+1} -1/2),
\end{eqnarray}
and combining it with the previous term results in the form
\begin{eqnarray} \label{eqn:Intro_Hlead}
\hat H_{\rm lead} = 
-t \sum_{i} (\hat c^\dagger_{i} \hat c_{i+1} + H.c.)
+I \sum_{i} (\hat c^\dagger_{i} \hat c_{i} - \frac{1}{2}) 
( \hat c^\dagger_{i+1} \hat c_{i+1} -\frac{1}{2}).
\end{eqnarray}

Note that since multiplying the entire Hamiltonian by a constant factor
does not change the physics of the problem, practically this Hamiltonian has only one 
free parameter, which is the ratio $I/t$ (given that $t$ is taken as positive).

\subsection{1D Spin Chain}
Interestingly, the spinless fermionic interacting Hamiltonian which is described by 
Eq.~(\ref{eqn:Intro_Hlead}), can be shown to be equivalent to an entirely different type of 
1D system, that of a spin chain \cite{giamarchi03,mikeska04}. 
Let's assume we have a chain of spin $1/2$,
i.e., we have a chain in which there is a spin $S_i$ on every site. The spin $S_i$ is 
defined by the relation $S_i = \hbar \sigma_i/2$, where $\sigma_i$ are the Pauli matrices,
and the three components of the spin have the regular angular momentum commutation relations.

The Heisenberg Hamiltonian for such a chain considers only NN interactions between
spins, with a rotational symmetry in all directions. The Heisenberg Hamiltonian is written as
\begin{eqnarray} \label{eqn:Intro_Hspin0}
\hat H_{\rm Heisenberg} = 
\sum_{i} ~ J \vec S_i \vec S_{i+1} =
\sum_{i} ~ J (\hat S^x_{i} \hat S^x_{i+1} + \hat S^y_{i} \hat S^y_{i+1} 
+ \hat S^z_{i} \hat S^z_{i+1}).
\end{eqnarray}
When $J$ is positive the resulted ground state is anti-ferromagnetic, whereas for a negative
sign of $J$, a ferromagnetic state is favored.

When the rotational symmetry is broken, however, the Hamiltonian is modified to have different
coupling constants ($J$'s) in each direction. A famous case involves symmetry breaking in the 
$\hat z$ direction, whereas the system is still invariant for rotations in the $\hat x$ - $\hat y$ 
plane. The result, known as the XXZ Hamiltonian, is thus
\begin{eqnarray} \label{eqn:Intro_Hspin0_2}
\hat H_{\rm XXZ} &=& 
\sum_{i} ~ \left[ J_{xy} (\hat S^x_{i} \hat S^x_{i+1} + \hat S^y_{i} \hat S^y_{i+1} )
+ J_z \hat S^z_{i} \hat S^z_{i+1} \right],
\end{eqnarray}
with $\Delta=J_z/J_{xy}$ as a single free parameter .

We now wish to show the equivalence of Eqs.~(\ref{eqn:Intro_Hlead}) and (\ref{eqn:Intro_Hspin0_2})
\footnote[3]{In the next steps we follow the route of Ref.~\cite{giamarchi03}.}.
The first step is to relate the spin-down state to an empty site in the fermionic language, and
the spin-up state with an occupied one. The dimension of the Hilbert space is exactly the same
since each site contributes $2$ possibilities in both representations. Such a mapping can be easily
obtained by defining $S^+_i = c^\dagger_i$ and $S^z_i = c^\dagger_i c_i -1/2$, and it is easy 
to show that the commutation relations in each site are obeyed. However, the inter-site commutation
relations of the spin chain and the anti-commutation relations of the fermionic case are different.

In order to preserve the anti-commutation relations without destroying the local commutation relations,
Jordan and Wigner has proposed the mapping
\begin{eqnarray} \label{eqn:Intro_JW_map}
S^+_i &=& c^\dagger_i \exp(i \pi \sum_{j < i} c^\dagger_{j} \hat c_{j}), \\ \nonumber
S^z_i &=& c^\dagger_i c_i -1/2.
\end{eqnarray}

Using the Jordan-Wigner transformation, one can rewrite the terms of the XXZ Hamiltonian in
the fermionic language. For example, the $\hat x$ - $\hat y$ term in the Hamiltonian can be written as
\begin{eqnarray} \label{eqn:Intro_JW_spm1}
\hat S^x_{i} \hat S^x_{i+1} + \hat S^y_{i} \hat S^y_{i+1} &=&
\hat S^x_{i+1} \hat S^x_{i} + \hat S^y_{i+1} \hat S^y_{i} \\ \nonumber
&=& \frac{1}{2} (\hat S^+_{i+1} \hat S^-_{i} + \hat S^-_{i+1} \hat S^+_{i}),
\end{eqnarray}
for which it is easy to substitute the Jordan-Wigner mapping, and to get
\begin{eqnarray} \label{eqn:Intro_JW_spm2}
\hat S^+_{i+1} \hat S^-_{i} &=& c^\dagger_{i+1} \exp(i \pi \sum_{j < i+1} c^\dagger_{j} \hat c_{j}) ~
\exp(-i \pi \sum_{j < i} c^\dagger_{j} \hat c_{j}) c_{i}  \\ \nonumber
&=& c^\dagger_{i+1} \exp(i \pi c^\dagger_{i} \hat c_{i}) c_{i} \\ \nonumber
&=& c^\dagger_{i+1} ( 1 + i \pi c^\dagger_{i} \hat c_{i} + \ldots ) c_{i}  \\ \nonumber
&=& c^\dagger_{i+1}  c_{i},
\end{eqnarray}
since all the other elements end with $c_{i}c_{i}$ and thus vanish.

In a similar way one gets $\hat S^-_{i+1} \hat S^+_{i} = c^\dagger_{i}  c_{i+1}$,
and thus 
\begin{eqnarray} \label{eqn:Intro_Hspin1}
\hat H_{\rm XXZ} &=& 
\sum_{i} ~ \left[ \frac{J_{xy}}{2} ( c^\dagger_{i+1}  c_{i} + H.c. )
+ J_z (c^\dagger_{i+1} c_{i+1} -\frac{1}{2}) (c^\dagger_{i} c_i -\frac{1}{2}) \right].
\end{eqnarray}
 
The final step is to make a canonical transformation in which the momentum of the fermions 
is shifted by $\pi$, i.e., we multiply $c_j$ by a factor $\exp(i \pi j)=(-1)^j$. With such a
transformation the first term in Eq.~(\ref{eqn:Intro_Hspin1}) gets a minus sign,
so finally
\begin{eqnarray} \label{eqn:Intro_Hspin2}
\hat H_{\rm XXZ} &=& 
\sum_{i} ~ \left[ -\frac{J_{xy}}{2} ( c^\dagger_{i+1}  c_{i} + H.c. )
+ J_z (c^\dagger_{i+1} c_{i+1} -\frac{1}{2}) (c^\dagger_{i} c_i -\frac{1}{2}) \right].
\end{eqnarray}

Taking $t=\frac{J_{xy}}{2}$ and $I=J_z$ one gets to the original fermionic Hamiltonian 
Eq.~(\ref{eqn:Intro_Hlead}). 
We thus conclude that the XXZ spin chain and the spinless fermionic system have identical
physical properties. The significance of that conclusion will become clear shortly.

\subsection{Phase Diagram}
\label{sec:int_pd}
As noted in the case of the QD Hamiltonian, when interactions are included 
in the Hamiltonian, the system in general cannot be exactly solved. However,
for some special cases (many of them low-dimensional systems) specific methods 
were developed, so that their physical properties can be found exactly. 

One such method for exactly solving 1D systems is the Bethe ansatz
technique. It was first suggested \cite{bethe31} by Bethe, in 1931, for the XXZ spin chain problem with
periodic boundary conditions \cite{karbach98}, and 
since then it was extended and used in more complicated systems. As a result, the 
physical properties of the XXZ spin chain, as a function of $\Delta=J_z/J_{xy}$, 
are known exactly \cite{baxter73}, and we now review the main results which are important for our purpose.

We first point out that the transformation $J_{xy} \rightarrow -J_{xy}$ and 
$J_z \rightarrow J_z$ of the XXZ spin chain is
identical to replacing the operators $S_i^x, S_i^y$ by $(-1)^i S_i^x$ and $(-1)^i S_i^y$.
Therefore, when taking $\Delta \rightarrow -\Delta$, the physical system remains 
almost the same, except for a change of the spin-spin coupling between
anti-ferromagnetic and ferromagnetic couplings\footnote[4]{Note that 
the transformation $J_{xy} \rightarrow -J_{xy}$ is accompanied, in the fermionic language,
by an inversion of the energy band, because $t$ changes its sign.}.

When $\Delta$ is exactly $-1$ or $+1$ the system is isotropic, whether in 
a ferromagnetic or an anti-ferromagnetic state. Different phases of the system
evolve for the three regimes divided by these points: $\Delta<-1$, $-1<\Delta<1$
and $1<\Delta$.

For $\Delta<-1$ the XXZ spin chain is in a ferromagnetic phase. In the ground state of
such a system all the spins point to the same direction, usually defined along the $\hat z$
axis. Obviously, without a magnetic field, the ground state is doubly degenerate with
$S_z=\pm N/2$. The first excited states, which can be exactly found by the Bethe ansatz method,
are magnons with $S_z=\pm (N/2-1)$, and there is an energy gap which is linear in 
$|\Delta|-1$. In the limit $\Delta \rightarrow -1$, therefore, the system becomes gapless.

The $1<\Delta$ phase is similar to the $\Delta<-1$ one, but instead of a ferromagnetic state
the ground state is anti-ferromagnetic (N\'eel). 
The ground state is doubly degenerate regarding the two possible anti-ferromagnetic orders, 
$\uparrow \downarrow \uparrow \downarrow \cdots$ or $\downarrow \uparrow \downarrow \uparrow \cdots$,
and the excited states produce a gap which grows as $|\Delta|-1$. 
Again, in the limit $\Delta \rightarrow 1$ the system becomes gapless.

\begin{figure}[htbp]
\centering
\includegraphics[trim=0mm 0mm 0mm 0mm, clip, width=5in]{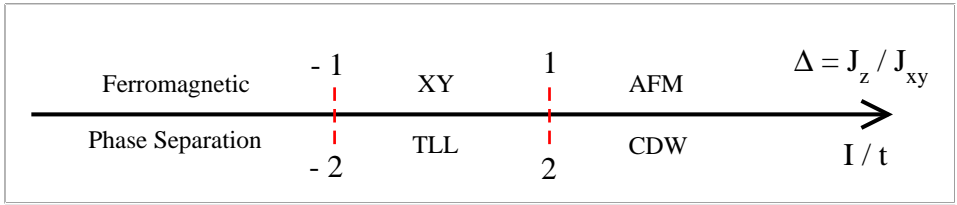}
\caption[Phase diagram of the spin chain and the spinless fermionic lattice]
{\label{fig:tll_pd} The phase diagram of the spin chain and the spinless fermionic lattice. The different
phases are shown as a function of the parameter $\Delta = J_z/J_{xy}$ in the spin
chain case, and $I / t$ in the fermionic case.}
\end{figure}

The interesting regime is the intermediate one, $-1<\Delta<1$, for which a full solution using
the Bethe ansatz technique shows that the system is gapless. This phase is called the XY
phase. We remark that this phase includes the non-interacting point $\Delta=0$.
The three different phases are schematically shown in Fig.~\ref{fig:tll_pd}.

We now want to understand these results in the fermionic case. We
use the equivalence shown in the previous section, in which the spin-up (down) 
case in the spin chain was shown to be identical to an occupation (absence) of a fermion.
The XXZ ferromagnetic phase tells us that the fermionic ground state for $I<-2t$
is either entirely occupied or entirely empty\footnote[5]{When a system is explicitly 
enforced to have a certain filling factor (e.g. to be
half filled) the electrons prefer to gather, and the result is called a "phase separation".}.
The energy gap points out that this phase is an insulator.

Similarly, the XXZ anti-ferromagnetic phase is also related to an insulating fermionic 
phase occurring for $I>2t$, which is the charge density wave. 
The degenerate XXZ ground state is related to the two possible 
charge density wave orders in which alternating sites are occupied,
either like $\bullet \circ \bullet \circ \cdots$ or $ \circ\bullet \circ \bullet \cdots$.

The middle phase, the gapless XY phase in the XXZ chain, is projected onto a metallic fermionic
phase, with $-2t<I<2t$. It can be shown that the XY phase can not be described by the
Fermi liquid theory; Rather it belongs to a different universality class, called {\it Tomonaga-Luttinger liquid}. 
This class will be briefly introduced in the next section.

\section{A Short Introduction to Tomonaga-Luttinger Liquids}
\label{sec:int_TLL}
The inclusion of electron-electron interactions in the Hamiltonian has usually
a very significant effect on the possibility to exactly calculate physical
quantities of the system. These difficulties are in general both theoretical and numerical.
However, a physical interacting system may sometimes be equivalent 
to a non-interacting system, and thus can be exactly analyzed.
A good example for that is the case of interacting electrons in two or three dimensions,
a system which is usually referred to as a Fermi liquid.
Landau, in his famous Fermi liquid theory \cite{landau57}, 
has proven that instead of calculating directly the Fermi liquid properties, one can use 
its equivalence to a Fermi gas of quasi-particles for which these properties can be easily 
found \cite{nozieres64}. These quasi-particles can be shown to share the same important physical properties, 
such as charge and spin, with the original electrons, so that instead of the difficult 
task of solving a system of interacting electrons, one can more easily solve a system of quasi-electrons gas.

Unfortunately, in one dimension it can be shown \cite{voit94,sefer_katom}
that an excitation which is composed of quasi-particles is not stable, so that quasi-particles
cannot describe the 1D system. The stable excitations, on the other hand,
do not have the physical properties of the electrons. In general, the Fermi liquid theory 
cannot be used in one dimension.

A significant progress in the 1D world was obtained by the works of Tomonaga \cite{tomonaga50}
and Luttinger \cite{luttinger63}, and we now briefly review their main steps. The first step is the linearization
of the dispersion relation of a free electron, which moves in a 1D system of length
$L$ with periodic boundary conditions. It is clear that the energy levels which contribute 
to the transport properties are those which are close to $k_F$,
so that one usually neglects the levels which are far below or far 
above $k_F$. Since in one dimension a Fermi surface does not exist, but there are only two Fermi 
points ($\pm k_F$) instead, the contributing electrons are split to "right-movers" ($k \approx k_F$) 
and "left-movers" ($k \approx -k_F$), and thus one can write up to first order in $k-|k_F|$
\begin{eqnarray} \label{eqn:Intro_TLL_linear}
\epsilon_+ &=& \epsilon_F + \frac {d\epsilon}{dk} |_{k=+k_F} (k-k_F) \\ \nonumber
\epsilon_- &=& \epsilon_F + \frac {d\epsilon}{dk} |_{k=-k_F} (k+k_F).
\end{eqnarray}

\begin{figure}[htbp]
\centering
\includegraphics[trim=0mm 0mm 0mm 0mm, clip, width=3.5in]{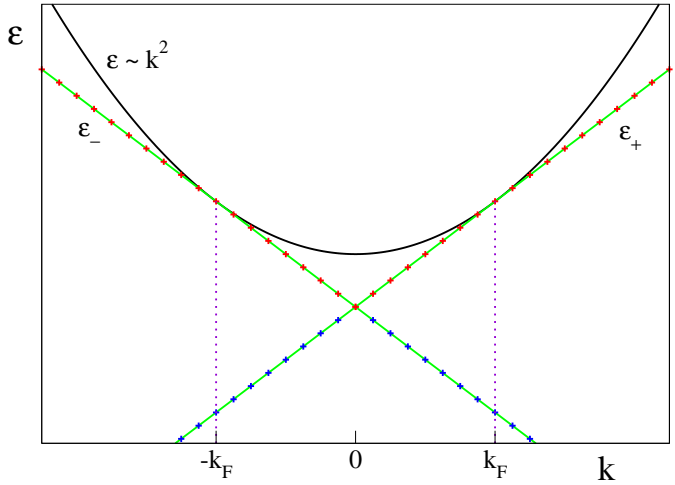}
\caption[Tomonaga and Luttinger models]
{\label{fig:tll_disp} 
The dispersion relation $\epsilon \sim k^2$ and the approximation by the right and left 
branches near $\pm k_F$. The states taken by Tomonaga's model are schematically signed by the red crosses,
while the states added by the Luttinger model are colored blue.}
\end{figure}

These two branches are the linear lines shown in Fig.~\ref{fig:tll_disp},
which can be used to approximate the quadratic dispersion relation in the vicinity 
of the Fermi points. Since $\frac {d\epsilon}{dk} |_{k= \pm k_F} = \pm v_F$, one gets
\begin{eqnarray} \label{eqn:Intro_TLL_linear2}
\epsilon_{\pm} = \epsilon_F + v_F (\pm k-k_F).
\end{eqnarray}

This energy dependence can be denoted by the following fermionic Hamiltonian,
which neglects the constant Fermi energy,
\begin{eqnarray} \label{eqn:Intro_TLL_H1}
\hat H = \sum_{k,r=\pm1} v_F (rk-k_F) \hat a^\dagger_{r,k} \hat a_{r,k},
\end{eqnarray}
where $r=\pm 1$ is used to discriminate between right and left movers, and 
$\hat a^\dagger_{r,k}$ ($\hat a_{r,k}$) denotes the creation (annihilation) of
an electron occupying the $k$-th energy level with a direction $r$.

In order to proceed from Eq.~(\ref{eqn:Intro_TLL_H1}) it is clear that the first question
is what are the values that $k$ can take. If we restrict $k$ to be in the vicinity of
$\pm k_F$, where the approximation is rigorously valid, the model cannot be exactly solved. Tomonaga
was the first to offer a creative idea: Since the states which are far below $k_F$ should
not change the basic physics of the system, one can include these states in the model, so that
it will become easier to solve. He offered to include, in the sum over $k$, all the states
which are either on the right-moving branch with $k>0$, or on the left-moving one 
with $k<0$. Such states are schematically drawn in red in Fig.~\ref{fig:tll_disp}.
While this slight change in the model is still not enough in order to exactly solve the
Hamiltonian, yet an approximate solution can be found, as was shown by Tomonaga himself.

The next significant step was suggested by Luttinger. His idea was to include the bottom of
the two branches entirely, i.e., take the right branch for $-\infty < k < k_F$, and the left branch
for $-k_F < k < \infty$ (i.e., add also the blue states in Fig.~\ref{fig:tll_disp}). 
With this change the model becomes exactly solvable, and nowadays
the model is mainly named after Luttinger.

We now present the main steps required to solve the Hamiltonian. First we 
define an operator (for each branch), which changes the electrons momentum by $k$: 
$\rho_k = \sum_q a^\dagger_{q+k} a_q$. It turns out that in the right branch
$\rho_k$ with $k>0$ creates a particle-hole pair, while a $k<0$ term annihilates it,
and the opposite happens in the left branch. Since $\rho_k$ is built out of two fermionic
operators, it is not surprising to find out that its commutation relations are bosonic,
up to a multiplicative factor. It is thus convenient to exactly define the bosonic
operators $B_k^\dagger$ and $B_k$ as a function of $\rho_{\pm k}$, so that
$[B_k,B^\dagger_{k^\prime}]=\delta_{k,k^\prime}$.

Defining two number operators, which count the particles on the right and left branches, as
$N_r = \rho_0 = \sum_{n} a_{r,n}^\dagger a_{r,n}$, and using the Kronig identity, one can express
the Hamiltonian Eq.~(\ref{eqn:Intro_TLL_H1}), as a function of the bosonic operators, as
\begin{eqnarray} \label{eqn:Intro_TLL_H2}
\hat H = \sum_{k \ne 0} v_F |k| B^\dagger_k B_k + \frac {\pi v_F}{L} (N_+^2 + N_-^2).
\end{eqnarray}

This Hamiltonian represents three types of excitations. The first term is simply
a momentum change of an electron in one of the branches. The second term results from
two other optional excitations:
the jump of an electron from one branch to the other (current excitations), and the 
addition of an even number $N$ of electrons to the system, $N/2$ to every branch
(charge excitations).

By transforming the density operators to real space, one can get a more 
convenient representation of the Hamiltonian. We write
\begin{eqnarray} \label{eqn:Intro_TLL_rho_x}
\rho^{\pm}(x) = \sum_{k} \exp(-ikx) \rho^{\pm}_k = N_{\pm} + \sum_{k \ne 0} \exp(-ikx) \rho^{\pm}_k,
\end{eqnarray}
and with these definitions and a bit of simple algebra one can show that
\begin{eqnarray} \label{eqn:Intro_TLL_H3}
\hat H = \frac {\pi v_F}{L^2} \int_0^L dx [\rho^+(x)\rho^+(x) + \rho^-(x)\rho^-(x)].
\end{eqnarray}

Looking at the form of Eq.~(\ref{eqn:Intro_TLL_H3}) it is obvious that the Hamiltonian
can be easily written as a function of the "charge density" $\rho^+(x) + \rho^-(x)$,
and the "current density" $\rho^+(x) - \rho^-(x)$. This can be obtained by defining\footnote[6]
{The two fields $\phi (x)$ and $\theta (x)$ are named and defined in 
different ways by different authors. We adopt the version of Ref.~\cite{giamarchi03}.}
the fields $\phi (x)$ and $\theta (x)$:
\begin{eqnarray} \label{eqn:Intro_TLL_phi_theta}
\phi (x) &=& - (N_+ + N_-) \frac {\pi x}{L} - \frac {i \pi}{L} \sum_{p \ne 0} \frac {\exp(-ipx)}{p} (\rho^+_p + \rho^-_p) \\ \nonumber
\theta (x) &=& (N_+ + N_-) \frac {\pi x}{L} + \frac {i \pi}{L} \sum_{p \ne 0} \frac {\exp(-ipx)}{p} (\rho^+_p - \rho^-_p).
\end{eqnarray}

With this definition one finds out that $\bigtriangledown \phi (x) = - \frac {\pi}{L} [\rho^+(x) + \rho^-(x)]$
and $\bigtriangledown \theta (x) = \frac {\pi}{L} [\rho^+(x) - \rho^-(x)]$, leading
to the nice form 
\begin{eqnarray} \label{eqn:Intro_TLL_H4}
\hat H = \frac {1}{2 \pi} \int_0^L dx v_F [(\bigtriangledown \phi (x))^2 + (\bigtriangledown \theta (x))^2].
\end{eqnarray}

The advantage of this Hamiltonian form, is that it is a free-particle bosonic Hamiltonian
so that its solution is exactly known. However, since
we haven't yet considered the electron-electron interactions, it may not be such a
surprise. But, as we now show, the great importance of the Luttinger model is that
up to some prefactors, interactions do not change this Hamiltonian form.

\begin{figure}[htbp]
\centering
\includegraphics[trim=0mm 0mm 0mm 0mm, clip, width=3.5in]{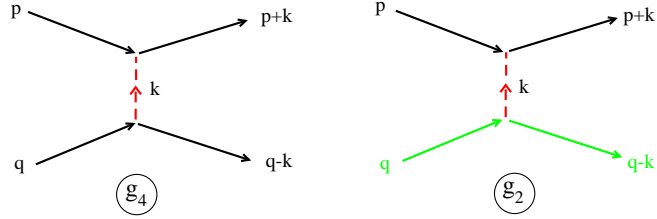}
\caption[Interaction processes considered in the Luttinger model]
{\label{fig:tll_g2_g4}
The interaction process $g_2$ and $g_4$ which are considered in the Luttinger model. 
The first describes a momentum transfer between 
electrons from different branches, while the second describes a transfer of momentum between 
electrons in the same branch (different branches are presented in different colors).}
\end{figure}

For spinless electrons there are two kinds of interactions, which are historically
denoted as $g_2$ and $g_4$. In Fig~\ref{fig:tll_g2_g4} these processes are shown\footnote[7]
{Historically the interaction processes were given the notations $g_1, g_2, g_3$ 
and $g_4$. However, $g_1$ has the same form, for spinless electrons, as $g_2$, and
$g_3$ denotes umklapp processes, which are not included in the Luttinger model.}.

The Hamiltonian terms of these interaction processes are easily written in the
same form of the non-interacting Hamiltonian:
\begin{eqnarray} \label{eqn:Intro_TLL_g42_1}
\hat H_{\rm g_4} &=& \frac {1}{L} \frac {g_4} {2} \sum_k \sum_{p,q} a^\dagger_{p+k} a_p a^\dagger_{q-k} a_q 
= \frac {1}{L} \frac {g_4} {2} \sum_k (\rho^+_k\rho^+_{-k} + \rho^-_k\rho^-_{-k}), \\ \nonumber
\hat H_{\rm g_2} &=& \frac {1}{L} g_2 \sum_k \sum_{p,q} a^\dagger_{p+k} a_p a^\dagger_{-(q-k)} a_{-q}
= \frac {1}{L} g_2 \sum_k \rho^+_k\rho^-_{-k}.
\end{eqnarray}

After the transformation to real space one gets
\begin{eqnarray} \label{eqn:Intro_TLL_g42_2}
\hat H_{\rm g_4} &=& \frac {g_4} {4 \pi^2} \int dx [(\bigtriangledown \phi (x))^2 + (\bigtriangledown \theta (x))^2], \\ \nonumber
\hat H_{\rm g_2} &=& \frac {g_2} {4 \pi^2} \int dx [(\bigtriangledown \phi (x))^2 - (\bigtriangledown \theta (x))^2].
\end{eqnarray}

Finally, incorporating the interaction terms into the free Hamiltonian Eq.~(\ref{eqn:Intro_TLL_H4})
one gets 
\begin{eqnarray} \label{eqn:Intro_TLL_H5}
\hat H = \frac {1}{2 \pi} \int_0^L dx [\frac{u}{\kappa} (\bigtriangledown \phi (x))^2 + u \kappa (\bigtriangledown \theta (x))^2],
\end{eqnarray}
where $u$ is defined through $u = v_F \sqrt{(1+g_4/2\pi v_F)^2-(g_2/2\pi v_F)^2}$
and has units of velocity, while $\kappa$ is a dimensionless parameter which is defined by
\begin{eqnarray} \label{eqn:Intro_TLL_K_ro}
\kappa = \sqrt{\frac{1+g_4/2\pi v_F-g_2/2\pi v_F} {1+g_4/2\pi v_F+g_2/2\pi v_F}}.
\end{eqnarray}

The value of $\kappa$ describes the interactions in the system. $\kappa>1$ for systems with
attractive interactions, while $\kappa<1$ for repulsive interactions. The value
$\kappa=1$ corresponds to a non-interacting system.

In conclusion, we see that a free-bosonic Hamiltonian can be used to describe
the physics of an interacting 1D system. Thus the physical properties
of such a system can be exactly calculated. The next significant milestone 
was by Haldane in 1981 \cite{haldane81}, when he has proven that Luttinger's theory 
defines a new universality class which can be used also to solve other systems. 
The problems which belong to this universality class, and for which the 
Tomonaga-Luttinger theory can be used, are called, after Haldane's suggestion, 
Luttinger liquids (or Tomonaga-Luttinger liquids).

\section{Motivation and Outline}

In the previous sections we have briefly described a few physical properties
of 1D lattices and of QDs.
In the following chapters of this thesis we will delve deeper into these two systems,
demonstrating some new findings for both, as well as discuss new features which
arise once we couple both systems to each other.

Recently, in the work of Kane and Fisher \cite{kane92}, it was shown that by 
coupling an impurity between two Luttinger-liquid leads the conductivity vanishes. 
Namely, the Luttinger liquid has a metallic behavior only in its pure form, and 
once disorder is introduced it immediately becomes an insulator.
Indeed, it was recently demonstrated that the conductance through a 1D disordered device 
produces infinitely sharp Coulomb blockade peaks at zero temperature \cite{nazarov03}.

So can we say that nothing happens when a QD is coupled to a lead which is described by the 
Luttinger liquid theory? Clearly, as we just said, the conductivity should vanish. 
Yet, one can ask what happens to other properties, such as the broadening of the dot's level, 
and the charge distribution in the lead.

Does the coupling leave a signature on the dot's energy levels? 
Can we expect to measure a broadening of the levels as in a regular weak-coupling regime, or 
maybe the discrete form of the levels should re-appear because of the Kane-Fisher result?
What happens to the charge distribution inside the lead when we connect the QD? Does it
show Friedel Oscillations? What happens when the lead is disordered? 
What happens to these phenomena when the lead's phase is changed between Tomonaga-Luttinger
liquid and, let's say, charge density wave?

~

In this work we try to answer these questions.
In chapters \ref{cpt:ch1} and \ref{cpt:ch2} we will examine a 1D lead of spinless fermions 
which is connected to a QD. We restrict the QD, for the sake of simplicity, to have only 
a single energy level. While in chapter \ref{cpt:ch1} we deal with a clean system, in chapter 
\ref{cpt:ch2} we investigate the influence of a weak disorder on such a system.
In chapters \ref{cpt:ch3} and \ref{cpt:ch4} we ignore the leads, and examine closely two
kinds of properties related to an isolated QD. In chapter \ref{cpt:ch3} we 
present the results of an innovative numerical method for the ground-state energy 
of a disordered QD. We also utilize this method in order to examine the dot's addition spectrum,
comparing our results to those obtained by the Hartree-Fock approximation.
Chapter \ref{cpt:ch4} is dedicated to the investigation of the g-factor of doubly-occupied QDs,
in the presence of spin-orbit coupling and electron-electron interactions.

Since much of the work utilizes numerical calculations, we choose to begin, 
in chapter \ref{cpt:numerics}, with a detailed review of the arsenal of numerical methods which are to be used.


\cleardoublepage
\chapter{Numerical Methods} \label{cpt:numerics}

\section{Introduction}

In this chapter we present the methods with which the numerical results
of the following chapters are obtained. 
The main physical quantity which we are generally looking for, 
is the energy spectrum 
of our systems, including wave functions and their energies.
Four methods are detailed here,
and while their basic goal is quite the same, they have very different 
assumptions, algorithms  and frameworks, and thus, they are utilized
in different situations.

In a few simulations the lowest
energy state - the ground state - will be sufficient, while in other cases
we might need to calculate the entire spectrum. In some of the systems we
will be able to find exactly the physical properties, while in others we
are forced to use approximations. And finally, some of 
the algorithms use the real-space as their framework, while others are built
upon the momentum-space.

In the following sections we describe in detail the four methods
utilized. We start with the good old "exact diagonalization" method, which 
serves very successfully for the last few decades. Following it we present
another very well-known method, the Hartree-Fock method, with which interactions
between the electrons can be approximately treated. We then present a newer
method, the density-matrix renormalization-group method, which was first
established by S. R. White in 1992 \cite{white92,white93}, and which since then has got
several hundreds of applications by a few dozens of groups around the world.
We finish by presenting a bit more sophisticated\footnote[1]{and much more 
complicated ...} version of this method, 
the particle-hole density-matrix renormalization-group, 
which was used only rarely, but seems a very promising method. 

\section{Exact Diagonalization}
The most simple method in which a system's energy spectrum can be calculated
is by an exact diagonalization of the system's Hamiltonian.
Numerically, the Hamiltonian is represented by a matrix, and the
{\it physical} Hamiltonian diagonalization is nothing else but a
{\it mathematical} matrix diagonalization. 
In order to write the physical Hamiltonian, and to represent 
it by a matrix, one can choose the basis in which physical states
are written, and accordingly the way in which the 
matrix indices will be related to these physical states.

Let's start with a simple example, that of a tight-binding Hamiltonian of 
a two-dimensional (2D) lattice with $A$ rows and $B$ columns, which is commonly used 
in order to represent a lateral quantum dot (QD) (see chapter \ref{cpt:intro}). We
denote the total number of sites by $N$ (i.e., $N=AB$). The motion of the electrons is 
by hopping from one lattice site to one of its nearest neighbors (NNs), which we denote
as usual by $\langle \cdots \rangle$. For simplicity we assume here that the QD is clean 
(without disorder), and we set the zero energy level at 
the constant energy of the electrons when they are located at the lattice sites. 
An intuitive choice of basis states leads to a notation in which 
state $|i \rangle$ refers to an electron in the $i$-th site of the lattice,
so that $1 \le i \le N$.
Denoting by $\hat a^\dagger_j$ and $\hat a_j$ the physical operators for creation 
and annihilation of an electron at site $j$, one gets the Hamiltonian
of the system (compare to Eq.~(\ref{eqn:Intro_Hdot})) as 
\begin{eqnarray} \label{eqn:Hlead}
\hat H = -t \sum_{\langle i,j \rangle} (\hat a^\dagger_i \hat a_j + H.c.),
\end{eqnarray}
where $t$ is the overlap integral between NN sites,
and which is assumed to be identical for all sites.

We now move on to writing down the matrix related to this Hamiltonian.
Using the same indices notation, we will use the vector indices $1,2, ..., N$
to represent the states $|1 \rangle, |2 \rangle, ..., |N \rangle$ and the matrix element $(i,j)$
will now represent the physical quantity $\langle i|\hat H|j \rangle$. The size of
the Hamiltonian matrix is thus $M = N$, and diagonalizing the Hamiltonian is 
done by a diagonalization of the $M \times M$ Hamiltonian matrix.
The results, the eigenvalues and eigenvectors we get, are the
energies and the wave functions of the system.

How far can one go with such a method? A computational limit may result 
from the fact that an $M \times M$ matrix diagonalization
is an $O(M^3)$ process, so that if, for example, the matrix size is doubled, 
e.g. for treatment of larger systems, or in order to consider the spin of the electron,
the computational time is multiplied by a factor of $8$. Nevertheless, with the current 
technology, the diagonalization task is still possible for quite large physical systems.

Problems begin when interactions between electrons are considered.
The process we have detailed so far, can help one find only the single-particle 
energies and wave functions. Let's assume that we'd like to put $2$ electrons in
a lattice of $A$ rows and $B$ columns. If there is no interaction between
the electrons, then the many-body wave function is a Slater determinant
of the single wave functions. In other words, when we put the first electron in
a single-level $|i \rangle$ and the second in $|j \rangle$, the resulted wave function
is $|i,j \rangle = \frac{1}{\sqrt{2}} \left( |i \rangle |j \rangle - |j \rangle |i \rangle \right)$ 
with an energy $E_{i,j} = E_i + E_j$. But when 
the electrons have some kind of an interaction between them, this is
completely wrong. In that case the many-body wave function is more complicated,
and in order to get the correct wave functions for two interacting electrons
one has to include all the states $|i,j \rangle$ in the matrix indices.
In general, for $n_e$ interacting electrons with spin, in a lattice of $A$ 
rows and $B$ columns, the matrix size will be $M = \binom{2AB}{n_e}$, 
with an exponential growth as a function of the system size, and it becomes too much
for our simple $O(M^3)$ diagonalization method, even for modest system sizes.

\subsection{Lanczos Diagonalization Method}

A partial solution for that problem is achieved by changing the diagonalization
algorithm we use. If we agree to pay some price, we can use a more efficient 
diagonalization method, which works in about $O(M^2)$ steps. 
The price we pay is that we can't get the entire spectrum with such a method, 
but only a few of the levels. The profit is a decrease in the calculation time, 
which results in an increase of the size limitation for our systems. 

There are two such diagonalization algorithms that are more 
efficient than the $O(M^3)$ one, the Lanczos method and the Davidson method.
Both methods retrieve only a few eigenvectors, 
the ones with either the highest or the lowest eigenvalues\footnote[2]{Actually they 
can retrieve the entire spectrum, but then they won't be efficient comparing to a 
regular exact diagonalization.}.
Both also assume that the matrices are sparse, and require
an external efficient matrix-vector multiplication procedure.

In the numerical work reported in this thesis, the Lanczos algorithm is used
\cite{lanczosbook}.
The idea is to replace the original matrix that we want to diagonalize by a new
tridiagonal matrix with the same eigenvalues and eigenvectors, 
so that the diagonalization task will be much easier. Furthermore, since 
one usually needs only the lowest or highest eigenstates, the size of the 
tridiagonal matrix $T$ can be much smaller than the original matrix size.
One uses a recursive relation which is known as "Lanczos recursion",
in order to produce the new set of vectors ("Lanczos vectors"), with whom the 
original matrix transforms to the tridiagonal form. The Lanczos vectors
are orthogonalized using a Gram-Schmidt process, and during their production, 
a few tridiagonal matrices $T$, of varying sizes, are kept. These $T$ matrices 
are then diagonalized iteratively, starting from the smallest, and the 
eigenvalues and eigenvectors obtained are mapped onto those of the original 
matrix, which is usually larger. If the mapping is not successful, a new 
iteration starts with the next $T$ matrix that was kept \cite{netlib}.

To conclude, by using the Lanczos method one can enlarge the system
size and try to include also interaction between electrons. However,
there is still a serious size limitation when using exact diagonalization
methods, and the largest interacting systems investigated so far with that 
method were of $6 \times 6$ with $4$ spinless electrons. Of course that
in special Hamiltonians with unique symmetry circumstances one might
be able to increase this limit, yet, in regular cases the value of the
upper size limit is very disappointing.

\section{The Self-Consistent Hartree-Fock Method}
The Hartree-Fock (HF) approximation is a well known method to deal with
interactions in a "mean-field" way. The idea is to replace
the exact potential felt by an electron due to all the others with
a more convenient term. Instead of solving the exact Hamiltonian and find
the real many-body wave functions, one uses the HF approximation in order to find 
a basis of "one-body" wave functions which incorporate implicitly the interactions.

Suppose for example that we want to calculate the spectrum of a QD, which is represented as
an interacting 2D system of spinless electrons without disorder. 
Such a system is governed by the Hamiltonian (see chapter \ref{cpt:intro})
\begin{eqnarray} \label{eqn:H_hf0}
{\hat H} = -t \displaystyle \sum_{\langle m,n \rangle}({\hat a}^{\dagger}_{m}{\hat a}_{n} + H.c.) +
\displaystyle \sum_{\langle m,n \rangle}V{\hat a}^{\dagger}_{m}{\hat a}^\dagger_{n}{\hat a}_{n}{\hat a}_{m},
\end{eqnarray}
where the first term, to be denoted by $\hat H_0$, represents the 
hopping matrix elements between sites $m$ and $n$ which are NNs,
and the second ($\hat H_{int}$) represents NN 
interactions\footnote[3]{The modification of the interactions to Coulombic form is 
straightforward.}.

Writing the interaction term in k-space by using the relation 
${\hat a}_{m} = \sum_\alpha{{\hat b}_\alpha \phi_\alpha(m)}$, 
where $\phi_\alpha$ is the $\alpha$-th wave function ($\alpha$ runs over the 
momentum-space levels), gives
\begin{eqnarray} \label{eqn:H_hf1}
{\hat H} = {\hat H_0} + 
\sum_{\alpha,\beta,\gamma,\delta} 
{\hat b}^{\dagger}_{\alpha}{\hat b}^{\dagger}_{\beta}{\hat b}^{}_{\gamma}{\hat b}^{}_{\delta}
~ V \sum_{\langle m,n \rangle} \phi^*_\alpha(m) \phi^{*}_\beta(n) \phi_\gamma(n) \phi^{}_\delta(m).
\end{eqnarray}

Instead of rewriting the entire Hamiltonian with $\binom{AB}{n_e}$ indices, 
Hartree has proposed that when we are interested on site $m$, we'll do an average,
in the vacuum state, over the indices of the other site, i.e., take 
$\langle 0|~b^\dagger_\beta \phi^{*}_\beta(n) b_\gamma \phi_\gamma(n)~|0 \rangle$.
The meaning of the vacuum state $|0 \rangle$ is a summation
over all the states below $k_F$, so that 
$\langle 0|~b^\dagger_\beta \phi^{*}_\beta(n) b_\gamma \phi_\gamma(n)~|0 \rangle=
\sum_{k<k_F} \phi^{*}_k(n) \phi_k(n)  $. 
We thus get
\begin{eqnarray} \label{eqn:H_int_h}
{\hat H_{int}^{Hartree}} &=& \sum_{\langle m,n \rangle} \sum_{\alpha,\delta} {\hat b}^{\dagger}_{\alpha}
\phi^*_\alpha(m) ~{\hat b}^{}_{\delta} \phi^{}_\delta(m)
~ V \sum_{k<k_F} \phi^{*}_k(n) \phi_k(n) \\ \nonumber
&=& \sum_{\langle m,n \rangle} \hat a^\dagger_m \hat a_m ~ V \sum_{k<k_F} \phi^{*}_k(n) \phi_k(n),
\end{eqnarray}
which tells that one should add $V \sum_{\langle m,n \rangle} \sum_{k<k_F} \phi^{*}_k(n) \phi_k(n)$
to the matrix element $H(m,m)$.

A similar argument, which was proposed by Fock, leads to an addition of the 
non-diagonal elements, but with a minus sign because of the commutation relations. 
The averaging is now done over $\langle 0|~b^\dagger_\beta \phi^{*}_\beta(n) b_\delta \phi^{}_\delta(m)~|0 \rangle$, thus
giving $\sum_{k<k_F} \phi^{*}_k(n) \phi_k(m)$, so that
\begin{eqnarray} \label{eqn:H_int_f}
{\hat H_{int}^{Fock}} &=& \sum_{\langle m,n \rangle} \sum_{\alpha,\gamma} {\hat b}^{\dagger}_{\alpha}
\phi^*_\alpha(m) ~{\hat b}^{}_{\gamma} \phi^{}_\gamma(n)
~ V \sum_{k<k_F} \phi^{*}_k(n) \phi_k(m) \\ \nonumber
&=& \sum_{\langle m,n \rangle} \hat a^\dagger_m \hat a_n ~ V \sum_{k<k_F} \phi^{*}_k(n) \phi_k(m),
\end{eqnarray}
and therefore we should subtract $V \sum_{k<k_F} \phi^{*}_k(n) \phi_k(m)$
from the matrix element $H(m,n)$, where $m$ and $n$ are nearest neighbors.

The Hamiltonian matrix is now written in the $A \cdot B$ real-space indices. 
We first copy all the elements of $H_0$ which are easily calculated in 
real-space coordinates. For the interaction we have these two contributions of
Hartree and Fock terms. 
In the diagonal elements we should add the Hartree term, given by Eq.~(\ref{eqn:H_int_h}),
and from the non-diagonal elements we should subtract the contribution of the
Fock term, Eq.~(\ref{eqn:H_int_f}).
But unfortunately the $\phi$'s, which are required for these two terms, 
are the unknown wave functions we seek for.

If we could have written this Hamiltonian, then its diagonalization would give us the
correct wave functions. So what we do is to employ a self-consistent method: we take some 
initial wave functions (it can be simply the one-body wave functions), 
substitute them in Eqs.~(\ref{eqn:H_int_h}) and (\ref{eqn:H_int_f}), 
and then diagonalize $\hat H$ and find a new set of wave functions. 
We then take this new set and calculate again $\hat H_{int}$
and $\hat H$, and diagonalize $\hat H$ again. These steps are repeated until
we get a stable solution, i.e., the output wave functions are identical (up to a
defined accuracy) to the input wave functions.

At the end, after the convergence of the self-consistent method, we have a new set of 
"one-body" wave functions which consider the interactions in a mean-field way.
With this new basis one can compute the physical properties in a simple way. For example,
since this is effectively a "one-body" basis, the ground-state energy of $n_e$ electrons
occupying the lattice is related to the sum of the lowest $n_e$ eigenenergies in the new basis.
However, it is not identical to this sum, since one should compensate for double
counting in the Hartree and Fock terms. It is easy to prove that the ground-state
energy is thus
\begin{eqnarray} \label{eqn:HF_Egs}
E_{gs} &=& \sum_{k<k_F} {\epsilon_{k}} - \\ \nonumber
&~& \frac{V}{2} \sum_{\langle m,n \rangle} \sum_{l,k<k_F} { \left[ \phi^{*}_k(n) \phi_k(n) \phi^{*}_l(m) \phi_l(m)
- \phi^{*}_k(n) \phi_k(m) \phi^{*}_l(m) \phi_l(n) \right] },
\end{eqnarray}
where the first sum runs over the single-particle eigenvalues, and the second
one counts for the Hartree and the Fock terms \cite{Merzbacher}.

\section{The Density-Matrix Renormalization Group Method }
\label{sec:num_dmrg}
In this section we describe a method which is used in order to get the lowest 
lying levels of a system without the usage of exact
diagonalization, the density-matrix renormalization-group (DMRG) method.
This method works for much larger systems than those one can treat by using the
exact methods, and gives much more accurate results than those one can get by 
using the HF method.

The DMRG method gives only an approximate result for the physical properties.
However, it has proven itself as very accurate in various systems. 
Nevertheless, when one starts an investigation of a new system,
they must assure that the results are accurate and represent the
real physics.
Another severe limitation of the DMRG method is its implicit restriction to 
one-dimensional (1D) and quasi 1D systems. 

The DMRG method is actually a development of some earlier
numerical renormalization group methods. The basic idea behind many 
of the numerical renormalization group (NRG) methods 
is very simple: We cannot diagonalize the full Hamiltonian
since it is too large. However, if we need only the lowest levels, and 
if we can increase the Hilbert space iteratively (for example by adding a site
after another) we can omit in each step 
the states which seem the most unimportant, and then continue
with just part of the Hilbert space. Afterwards we add another site, the Hilbert space
increases again, and we again throw the unimportant states. We continue
this process site after a site, until we finally get to the size of system we 
want. If we did a good job during the iteration process, i.e.,
all the states we've omitted had indeed a negligible contribution to the final ground
state, then we have a good approximation of the ground state.

So the million-dollar question is how to decide which of the states we'd like to keep, 
and which can be omitted. The old methods used a pretty logical criteria, in which the states
being kept are those with the lowest energies.
In fact, this method was first developed by Wilson \cite{wilson} for the problem
of an impurity coupled to a 1D non-interacting lead. It was 
shown that the system can be mapped into a new system with a tridiagonal form,
representing an impurity level which is coupled either to nearby sites, or to states 
with very low energies. In addition, it was proven that the hopping elements of the new
system decrease algebraically between consecutive sites. In such a problem, it was shown 
that the NRG method gives accurate results. However, attempts to use this method for cases in 
which the leads contain non-trivial physics (for example when interactions between the electrons
in the lead are not neglected) weren't so fruitful. 

One of the reasons for the NRG failure in those cases, is that there is
no guarantee that a state which wasn't important for a $10$-site system, is not 
important also for the $100$-site one. Furthermore, the boundary conditions are changed
during the process of sites addition. The states of the $N$-site-system iteration
do not have the same boundary conditions as those of the iteration with $N+1$ sites.

\subsection{DMRG: Quick Overview}

In 1992, S. R. White proposed a different approach regarding the decision
which states to keep \cite{white92,white93}, the DMRG method. 
Since then, it has become one of the most useful numerical methods for treatment of 
1D systems, with many applications and variations \cite{Peschel99,schollwock05}.

\begin{figure}[htbp]
\centering
\includegraphics[trim=0mm 0mm 0mm 0mm, clip, width=4in]{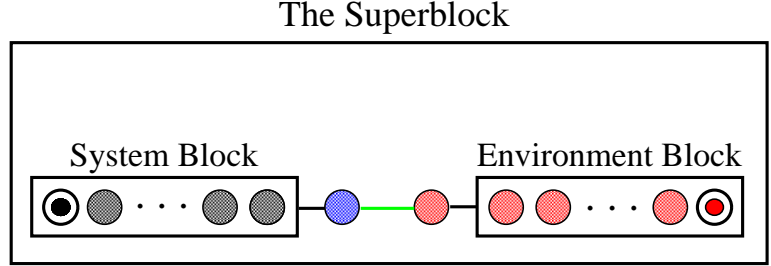}
\caption[The DMRG superblock composition]
{\label{fig:dmrg_block}
The DMRG superblock composition. The system block (in black), containing the dot 
(the most left site) coupled to a lead with $N-1$ sites, is coupled to a new site
(in blue). Then the entire system block is coupled (green line) to its mirror picture
(red sites) which represents the environment to form the superblock. 
For a block of $p$ states, one gets a superblock of $4p^2$ states.}
\end{figure}

The main idea of this method is to include in the Hamiltonian of the current iteration, 
not only the "system" sites that were already added, but also some "environment" sites
which are coupled to the current system. Not only does the inclusion of more sites 
assure the correct boundary conditions at each iteration, but it also forces the 
diagonalization process to take into account information from larger system sizes.
This leads to the diagonalization of a Hamiltonian representing a larger
physical system, thus getting a ground state of a larger basis. 
To return to the correct system size, one now defines a density matrix 
from which the environment degrees of freedom are
traced out. 
By a diagonalization of this density matrix one can identify which of the system states are 
the most important for the following steps, since they get the highest eigenvalues.

Before getting deeper into the technical details of the method, let's describe 
in general the iteration process. We start an iteration when our system 
already contains a block of $N$ sites, whose physical operators are
represented by matrices of size $p$. This is named the "System Block" schematically 
shown (in black) in Fig.~\ref{fig:dmrg_block}. 
We assume that this size, $p$, is the constant number of states we want to keep
during the iteration process\footnote[4]{Increasing the value of $p$ will increase 
the DMRG accuracy on one hand, but on the other hand it affects dramatically the 
required calculation time and memory resources.}.
Now we would like to add the next site (the blue one).
We thus enlarge our basis set by a factor of $2$ (assuming for the moment that
the electron is spinless), resulting in an enlargement of all
the operator matrices to the size $2p$. 
The idea of White is, as mentioned above, to diagonalize a larger system's 
Hamiltonian, called a "superblock", by an addition of some 
environment sites. This can be done simply by taking a mirror picture 
of the system we have as the environment (the red sites), connecting the system 
and environment by the regular Hamiltonian terms (represented by the green line).
This results, in principle, in a superblock size of $4p^2$, which is now
diagonalized (usually using Lanczos or Davidson methods), in order to 
find its $4p^2$-sized ground state. From this ground state one constructs a
density matrix of size $2p$, by summing over the environment indices, so that 
the remaining $2p$ indices represent our initial block with the addition of the new site.
By taking the upper half of the density-matrix eigenvectors (i.e., those with the larger
eigenvalues) as a projection operator, one can now transform all the $2p$-sized operators to the
new basis, of size $p$. This new block is the starting point of a new iteration.

\subsection{DMRG: a Technical Overview}

We now move to describe the technical details related to the Hamiltonian
that we use in different parts of the thesis. This Hamiltonian describes the
motion of spinless electrons on a 1D lattice ("lead") of size $L$, 
which is connected at one side to a QD. 
Such a Hamiltonian should consist of both $\hat H_{QD}$ 
and $\hat H_{lead}$ described in the previous chapter (Eqs.~(\ref{eqn:Intro_Hdot})
and (\ref{eqn:Intro_Hlead2})). 
We now restrict the QD to have only a single site, and
in addition, we add a hopping term between 
the dot and the first site in the lead, whose matrix element will be denoted by $V_0$. 
If no interaction is assumed\footnote[5]{The inclusion of NN interaction 
is straightforward, see the discussion in the next subsection.}, one gets
\begin{eqnarray} \label{eqn:H_dmrg}
{\hat H} = \epsilon_{0}{\hat a}^{\dagger}{\hat a} 
-V_0 ({\hat a}^{\dagger}{\hat c}_{1} + {\hat c}^{\dagger}_{1}{\hat a})
-t \displaystyle \sum_{j=1}^{L-1}({\hat c}^{\dagger}_{j}{\hat c}_{j+1} + H.c.),
\end{eqnarray}
where $a^{\dagger}$ and $a$ represent the creation and annihilation operators 
of an electron in the single state of the dot, of energy $\epsilon_{0}$.
We now use the DMRG method in order to get the ground state of the Hamiltonian Eq.~(\ref{eqn:H_dmrg}). 

\subsubsection{Step 1: Building the First Block}

In the very first step one treats a very 
short system, e.g. a dot level coupled to a lead with only one site. 
Since this system has in total $2$ sites, it has only $4$ basis states. 
The $4 \times 4$ Hamiltonian matrix of this small system is now written 
exactly:
\begin{eqnarray} \label{eqn:H_b}
{\hat H_B} = \epsilon_{0}{\hat a}^{\dagger}{\hat a}
-V_0 ({\hat a}^{\dagger}{\hat c}_{1} + {\hat c}^{\dagger}_{1}{\hat a}).
\end{eqnarray}

We now want add a new lead site to the system, thus increasing the 
matrix size by a factor of $2$. Actually we can assume that we are 
in the beginning of a new general iteration, since the next steps
do not depend on the
number of sites the system already contains, 
specifically this is of course true also for the first iteration.

\subsubsection{Step 2: Addition of a New Site}

Let us denote the current block matrix size, containing already
$n$ lattice sites, by $p$. This block is represented by the black sites in Fig.~\ref{fig:dmrg_block}.
We now add the next site, number $n+1$ (the blue site), and the new
matrix size after the addition is $2p$.
The new Hamiltonian, denoted as $H_{B\bullet}$, should now
contain the same elements as $H_B$, the diagonal elements of the new
site, and also the connection between the new site to the previous one:
\begin{eqnarray} \label{eqn:H_bo}
{\hat H_{B\bullet}} = {\hat H_B} -t ({\hat c}^{\dagger}_{n}{\hat c}_{n+1} + 
{\hat c}^{\dagger}_{n+1}{\hat c}_{n}).
\end{eqnarray}

Since the expression of the Hamiltonian is done using the operators
$c^\dagger_n$ and $c_n$, one can see that there is a need to store the operators  
of the previous step. Later we shall see that 
the operator $c^\dagger_n c_n$ may be also used during the calculation.
Since the DMRG process change the basis
states continuously, and the states kept are not a complete basis
of the Hilbert space, in general
\begin{eqnarray}
\langle i|c^\dagger_n~c_n~|j \rangle \ne 
\sum_m \langle i|c^\dagger_n |m \rangle \langle m| c_n |j \rangle,
\end{eqnarray}
therefore this operator, as well as every other operator that will be
required later, should be specifically kept.  

Aside from increasing the size of $\hat H$, we perform a similar procedure for
any operator we shall need. These operators are rewritten in the new $2p$-sized basis,
which includes the previous basis and the new added site. In the next steps
we will, in general, decrease the Hilbert-space size back to the original size 
$p$.\footnote[6]{At the first few iterations of the process, the size of the new basis
might be still smaller than $p$, the matrix size we want to keep. In this case
we continue to add sites, each time doubling our matrices size, until the matrix size
is greater than $p$. Then we continue to the next two steps, which are needed for
the truncation of the Hilbert space.}

\subsubsection{Step 3: The Superblock}

We  now have a system of $n+1$ sites. Recalling that the DMRG idea 
is to place this system inside some sort of an environment, one can see that the 
simplest way to do it is to take a mirror picture of the system as the 
environment, and couple these two subsystems\footnote[7]{Alternatively one can
grow a clean lead without a dot at an initial DMRG stage, and then couple it
to the $B \bullet$ block at its right edge.}. We thus reflect the $B \bullet$ system 
in order to get an environment block $\bullet B^R$ ($B^R$ denotes the reflected block), and
couple the two new sites in the middle by the regular hopping term (see Fig~\ref{fig:dmrg_block}).

By that we get the superblock Hamiltonian
\begin{eqnarray} \label{eqn:H_boob}
{\hat H_{B\bullet - \bullet B^R}} = {\hat H_{B\bullet}} -t ({\hat c}^{\dagger}_{n+1}{\hat c}_{n+2} +
{\hat c}^{\dagger}_{n+2}{\hat c}_{n+1}) + {\hat H_{\bullet B^R}},
\end{eqnarray}
of size $(2p)^2 = 4p^2$.

\subsubsection{Step 4: The Density Matrix and Hilbert-Space Truncation}

The next step is to find the ground state $\Psi_0$ of the superblock, usually
by using the Lanczos algorithm\footnote[8]{Usually the superblock is not diagonalized as a $4p^2$-size matrix,
but in sectors containing different numbers of electrons. This can be done since the
Hamiltonian does not couple states with different number of electrons. Using such a 
method, one should diagonalize each sector, and finally choose the eigenvector with
the lowest eigenvalue among all sectors.}. Dividing the indices of the
ground state to $\Psi_0 = \sum_{i_S,j_E} C_{i_S,j_E} |i_S \rangle |j_E \rangle$, where
S (E) denoted a system (environment) index, we use 
$\Psi_0$ to build the density matrix $\rho$ by tracing out the 
environment degrees of freedom:
\begin{eqnarray} \label{eqn:rho_DMRG}
(\rho)_{i_S,i^\prime_S} = \sum_{j_E} {\Psi_{i_S,j_E} \Psi^\dagger_{i^\prime_S,j_E}}.
\end{eqnarray}

The density matrix, whose size is $2p$, is now diagonalized and the $p$ eigenvectors
which have the highest eigenvalues are used as a projection operator 
$\hat O$ of size $2p \times p$. 
If we denote $\rho |u_\alpha \rangle = \omega_\alpha |u_\alpha \rangle$, 
then we choose the eigenvectors $|u_\alpha \rangle$ with the highest $\omega_\alpha$'s 
to form an $2p \times p$ matrix $O$. 
For every stored operator $Q$ we now use 
$\hat O^\dagger Q \hat O \rightarrow Q$ in order to 
decrease its size to the initial size, as it was at the beginning of the iteration.

Now that the system block has $n+1$ sites, and all of the operators are back in 
their initial size, we can return to step $(2)$, and add the next site.

As steps $(2) - (4)$ are repeated iteratively, the physical size of the system
increases, while the Hilbert-space size remains the same. The iteration process
is stopped when the physical observable we want to measure, e.g. the occupation
of the dot level, converges.

\subsection{More Complicated Models}
\label{sec:num_nnn}
The generalization of the previous subsection to the case of interacting electrons
depends crucially on the details of the interaction term, specifically
on its range. 
The inclusion of NN interactions is straightforward, since
in each iteration we just have to add the term 
$I {\hat c}^{\dagger}_{n}{\hat c}_{n} {\hat c}^{\dagger}_{n+1}{\hat c}_{n+1}$.
The operators ${\hat c}^{\dagger}_{n}$ and ${\hat c}_{n}$ are of the
previous iteration, and since they are needed for the hopping term, they are kept anyway.
The operators ${\hat c}^{\dagger}_{n+1}$ and ${\hat c}_{n+1}$ are of the
new added site, so they can be written explicitly.

The same argument holds for the superblock creation, in which interaction should 
just be added between the two new sites (in the green line of Fig.~\ref{fig:dmrg_block}).

\begin{figure}[htbp]
\centering
\includegraphics[trim=0mm 0mm 0mm 0mm, clip, width=4in]{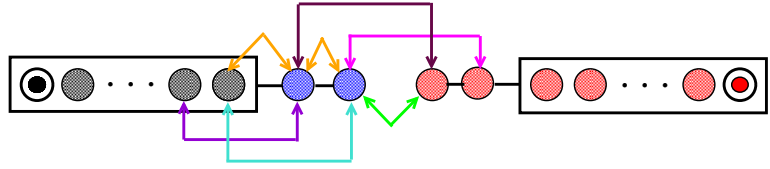}
\caption[DMRG superblock for next-nearest-neighbor interaction model]
{\label{fig:dmrg_nnn}
The DMRG superblock composition for next-nearest-neighbor interaction model. 
The terms added in step $2$ are drawn in orange ($t,I$), blue and violet ($I_2$),
while the terms of step $3$ are colored green ($t,I$), magenta and maroon ($I_2$).}
\end{figure}

If the interactions include terms of longer range, the case is different.
Let's look at the model of next-nearest-neighbor interactions, in which there are
two interaction terms, one for sites which are NNs (denoted by $I$) and
the second ($I_2$) for sites in distance of $2$ lattice sites.

The simplest solution to that case is that in every iteration we add two sites, instead
of a single site as explained in the previous section. In addition, we will store in memory 
the operators of the two sites added in the previous iteration. 
Let's denote the new added sites by $N_1$ and $N_2$, and the sites added in the previous
iteration (and whose operators are kept) by $P_1$ and $P_2$. 
A schematic picture of the superblock is shown in Fig.~\ref{fig:dmrg_nnn}.
In step $(2)$ the addition of the new sites will be simple, where we will make the following
connections: $N_1$ will be connected to $P_1$ with $I_2$ and to $P_2$ with $I$, 
$N_2$ will be connected to $P_2$ with $I_2$, 
and $N_1$ and $N_2$ are also connected with $I$.

In step $(3)$ the superblock formation is also as simple as that, since there are $4$ new
states in the middle, and the interactions do not connect any other sites.
So actually the algorithm is identical to the one detailed above.
The main difference is the sizes: the addition of $2$ new sites increases the operators
basis by a factor of $4$, so that the superblock Hamiltonian size, which is
to be diagonalized, is $16 p^2$. This decreases, in principle, the upper limit of $p$, 
so that the accuracy might decrease.

Such a solution works, of course, only for a limited range of interactions. If we want
to include more interaction terms, it quickly becomes impossible to increase the number of
sites we add in each iteration. Another solution might be to add in each iteration only 
a single site, but to keep the operators of all previous added sites in memory, 
and use them in the connection of the new site, and in the formation of the superblock.
However, in this solution steps $(2)$ and $(3)$ require much heavier calculations, e.g.
multiplications of large matrices, and they also include mixing of operators from the system 
and the environment sides. In general, this leads to a much slower calculation.

As a last remark we note that the simple solution we described for the next-nearest-neighbor 
interactions model, i.e., by adding $2$ sites at a time, is practically similar to a solution for the case of
spin $1/2$ electrons with NN interactions, in which instead of adding two sites, 
we add in each iteration a new single site with $2$ spin states.

\subsection{The Finite-Size DMRG}
While the accuracy of the DMRG method described above
was pretty good for several systems, there are other problems for which
the DMRG results are not good enough. In 1993, only a short time after his
first publication of the DMRG method, White has proposed
an improved version of the algorithm, the finite-size DMRG \cite{white93}.

The idea is to use the DMRG process on a fixed-sized system, instead of
an infinite one. The application is based on the DMRG iteration
process, but requires more computer resources, mainly memory. In this method
the operators of every iteration, i.e., of each and every size of the 
physical system (e.g. the lead), are stored for future use. Let's assume
we'd like a system of a single-level dot coupled to a lead of size $L-1$, so that
the total length of the system is simply $L$. 
During the iterations, one can combine two saved operators by connecting operators 
of the left hand side of the system of size $x$ to those of the right hand side
of the system, with size $L-x$. Enumeration over $x$ from $x=1$ to $x=L-1$ will give 
us a complete path through the lead.

\begin{figure}[htbp]
\centering
\includegraphics[trim=0mm 0mm 0mm 0mm, clip, width=5in]{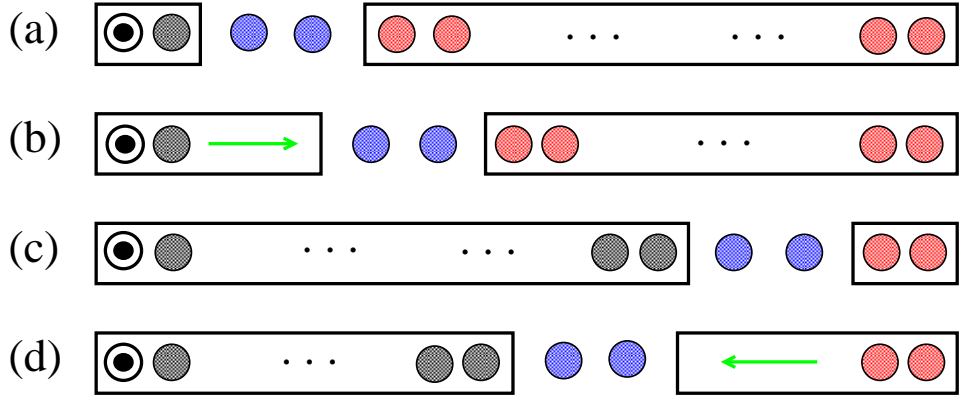}
\caption[A sweep of the finite-size DMRG algorithm]
{A sweep of the finite-size DMRG algorithm.
(a) Starting from the minimal system size, (b) the system grows and the environment
size decreases, (c) the minimal size of the environment was reached, (d) the
environment growth. The entire process of (a)-(b)-(c)-(d)-(a) is a sweep.}
\label{fig:finite_dmrg} 
\end{figure}

Let's now explain exactly how it works. At first, we build a lead 
of size $L$ by the regular DMRG method. The only difference is that
in the process we keep a copy of all the operators representing the physical systems of sizes 
$1, 2, ... , L$. We denote these operators as the environment operators, and they represent
the right hand side of our desired system - see Fig.~\ref{fig:finite_dmrg} - a clean lead. The
operators which represent the left hand side of the system, 
containing the dot and the left part of the lead are called the system operators.

We now start again with a new system from the left hand side, the part of the dot, 
by taking a new first block composed of a dot and a single lead site. This
is the first block, whose physical length is $2$.
To this block we add a new lead site, and now we'd like to get a superblock. 
In the regular method we would take this $3$-sites system and couple it
to its mirror picture, getting a superblock of length $6$.
Instead, we couple it to a block which will make the total size exactly $L$.
Such a block is composed of a previously-stored environment-block of size $L-4$, and an extra new site.
The total physical length of the system which is diagonalized - the superblock - is thus $L$.
A schematic picture of this process is given in Fig.~\ref{fig:finite_dmrg}(a).

The superblock is now diagonalized and the density matrix is formed. The
projection process over half the space gives the system operators for a 
$3$-sites system. We now continue by adding site $\#4$, and coupling 
the new system to an environment of size $L-5$, in a similar way,
and finally getting the system operators for the next iteration. 
We can continue this process iteratively (see Fig.~\ref{fig:finite_dmrg}(b)) 
until we reach the other end - when our system has $L-4$ sites on its own,
and the size of the environment is the minimal size - $2$ sites 
(Fig.~\ref{fig:finite_dmrg}(c)).

During the process we keep copies of all the system operators, since
we are just about to use them: after reaching the right end, we want
to come back from right to left (see Fig.~\ref{fig:finite_dmrg}(d)), 
connecting the new growing environment to the stored system operators, 
while the system side goes back from size $L-4$ to $2$. 
During these steps we update the environment operators, to be used in the next time,
when we'll move from left to right again.

We can continue this process in a zigzag manner, going from the dot
side to the right, and then back all the way, where each back and forth iteration - 
(a)-(b)-(c)-(d)-(a) in Fig.~\ref{fig:finite_dmrg} - is called a sweep. 
When we are moving to the right we update the system operators, and when 
we are on the way back we update the environment ones.
 
Usually after a few sweeps the physical observables converge,
the results are sufficiently accurate, and then we stop the iteration.

\subsection{Finite or Infinite}
The last subsection about the DMRG method will be devoted to the number of particles 
($n_e$) in the "finite-sized" system. Most DMRG applications use $n_e$ as a quantum
number which is fixed during the renormalization process. 
In practice, one can order the block states in a block-diagonal form by their electrons number, 
and conserve this form when the superblock Hamiltonian 
is created. Furthermore, one might restrict the superblock state to have a certain 
number of electrons, by coupling only states which give together the required
number. In such a way, and since the Hamiltonian does not couple states with a
different number of particles, $n_e$ will remain a good quantum number even
after the truncation.
This of course has some impact on the superblock composition, which now has a
block-diagonal form where each block contains states with a certain number of particles. 
In general, it makes the numerical method work much better, though it is mainly relevant 
to cases in which there is an external constraint on $n_e$. 

In our DMRG application, however, we intentionally do not keep $n_e$ as a constant.
Instead, we diagonalize the term $\hat H - \mu \hat N_e$, where $\hat N_e$ is the
number operator, and $\mu$ is the chemical potential. We keep $\mu$ as a 
constant and let the renormalization physics determine what $n_e$ will be. In
practice, we form superblock states of different particle numbers, and we do
not keep the order of states in the blocks, and this has an important impact
on the renormalization. Even though in the superblock ground state of the first 
iteration we get an integral number of particles, it does not need to remain
integral after the truncation process takes place, so that in the next iterations
the number is not a good quantum number any more.

A direct advantage of this method is the option to get non-integral values 
for $n_e$. As will be shown in the following chapters of this thesis, this
property can be essential for obtaining, from a finite system, results which
are accurate for an infinite system. As a simple example, let's think of a 
physical system which has a particle-hole symmetry, so that with taking $\mu=0$
should be exactly half filled. Without interactions, such a system can be exactly
diagonalized (even if it is a large one), and the half filling property can be
exactly checked. However, if the size of the system $L$ is odd, even for $L \gg 1$, 
an exact diagonalization cannot give the "correct" answer (which is obviously correct
just for the infinite case). With our implementation such an "infinite-system result" 
can be reproduced even by using systems of only a few hundred sites.
From the experimental point of view, the results we obtain with this method, for example
on 1D systems, can effectively describe a situation of a 
finite section of a 1D wire which is coupled to an external electron reservoir.

\section{The Particle-Hole DMRG Method}
\label{sec:num_phd}
The main disadvantage of the DMRG method is the need to have a well-defined 
order in which the system is enlarged. Regularly this limits the use of
DMRG to 1D, or at least quasi 1D, systems. 

In fact, a short time after the first appearance of the DMRG, a different 
scheme of the DMRG process usage was proposed \cite{xiang96}. If one adopts 
the momentum-space point of view, instead of that of the real-space, 
a new meaning replaces the traditional 1D approach: the order of the energy 
states. The order in which states are added can now be defined physically by their
energies, leading to the opportunity to start from the lowest lying states, and add
the next ones iteratively. In each iteration one adds a new energy state, thus
increasing the size of the Hilbert space, and then decreases this size to the 
original size, in a similar fashion to that is done in the regular DMRG process.

Since the first idea to use the momentum-space was proposed in Ref.~\cite{xiang96}, 
some implementations of this method were written in various fields \cite{nishimoto02,white_martin99}. 
In particular interest for us is the implementation named particle-hole DMRG (PH-DMRG),
since we utilize in this thesis a similar idea for QDs.
Several implementations of this sophisticated method have been reported so far, 
focusing on ultrasmall superconducting grains \cite{dukelsky99,dukelsky00,gobert04-1,gobert04-2}
and nuclear shell-model calculations \cite{dukelsky01,dukelsky02,dimitrova02}.
Some of these results, accompanied by a brief introduction to the algorithm, can be found 
in \cite{schollwock05-2,dukelsky04}.

\subsection{PH-DMRG: Quick Overview}
The name particle-hole DMRG is based on a special implementation of the momentum space,
where the renormalization starts from the Fermi energy and sweeps to both sides.
Lets assume that we want to find the ground state of a system composed of
a 2D lattice of $A$ rows and $B$ columns, in which there are
$n_e$ interacting spinless electrons\footnote[9]{The choice of this example here is 
based on the usage of this method in this thesis.}. 
An exact diagonalization
of such a system requires, as noted above, the diagonalization of an $\binom{AB}{n_e}$ 
matrix, which is often too large.

So now let us describe how the PH-DMRG proceeds. First, a single electron basis should
be chosen, and the most intuitive choice is the single electron eigenvalues, arranged 
according to their energies. However, the single-particle eigenvectors do not 
consider the interaction at all, and it can have a huge impact. A better way is to 
use the Hartree-Fock eigenvectors as the starting point to the PH-DMRG process.

\begin{figure}[htbp]
\centering
\includegraphics[trim=0mm 0mm 0mm 0mm, clip, width=5.5in]{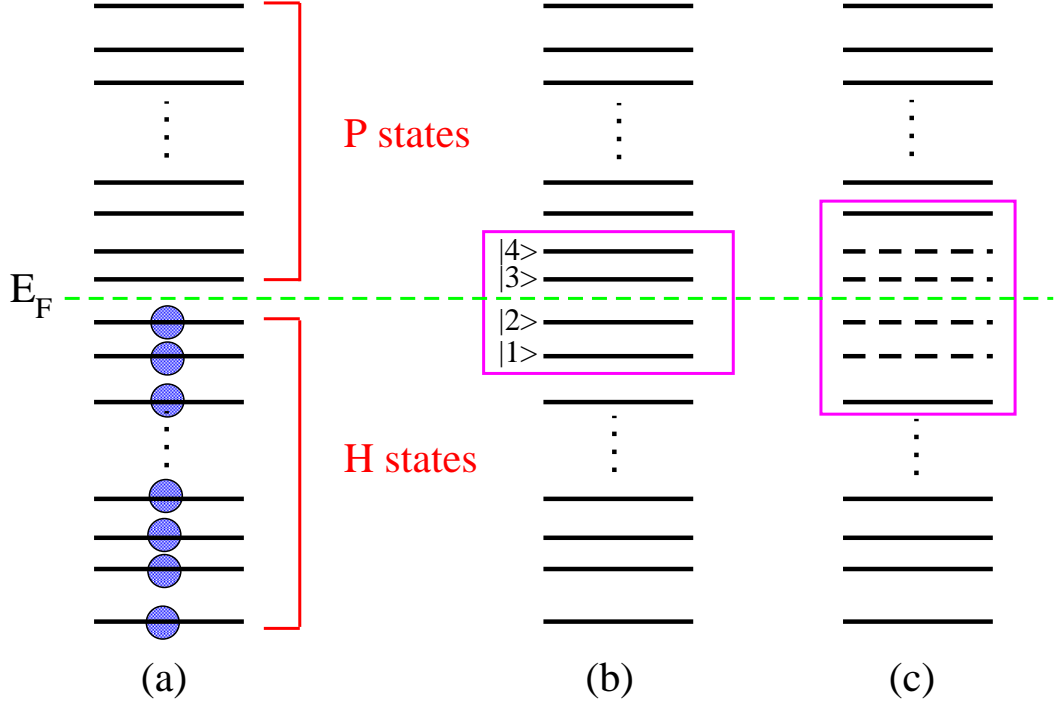}
\caption[PH-DMRG superblock composition for the first iterations]
{The PH-DMRG superblock composition for the first iterations.
(a) The energy states after employing the Hartree-Fock algorithm, showing the
occupied states (below $E_F$) and the unoccupied ones, dividing the states to
P- and H-blocks. (b) The first superblock composition, composed of the $4$ 
levels around $E_F$. (c) The superblock composition of the next iteration,
after basis change of the $4$ levels from the previous iteration (dashed lines)
and the addition of $2$ new levels.}
\label{fig:phdmrg_blocks}
\end{figure}

So now we have a set of wave functions sorted by their energies. In order to 
take into account exactly $n_e$ particles, the simplest way is to fill
the lowest $n_e$ states with electrons, leaving all the other states empty. 
This is depicted schematically in Fig.~\ref{fig:phdmrg_blocks}(a).
The idea of the PH-DMRG process is to change this basis of states, in order
to decrease the energy. This is done by considering some higher states also, 
while controlling the number of particles. Namely, putting a particle in a 
new energy state requires taking out an electron from a filled level, or, in other 
words, introducing a hole in one of the low states.

Therefore we keep two lists of energy states. The first one contains the energy
states above the Fermi energy, which are initially unoccupied.
These states will be called the "Particle" states (or simply "P-states"), 
since one can add a particle to the system by filling a state from this list. 
The second list is of the "Hole" states ("H-states"), which are the (initially) 
occupied energy states. In the situation we've just described, there will be total
of $n_h=n_e$ hole states, and $n_p=AB-n_e$ particle states. We maintain the "Particle" and the 
"Hole" lists separately during the process, creating two kinds of blocks, which are
combined once in a iteration to form the superblock.

Now let's explain in general the iteration process. At first we start with a small number
of energy states in the vicinity of the Fermi energy, for example we can take 
one state from each list to be in the relevant block, i.e., the particle block (P-block)
will contain the first state above $E_F$ and the hole block (H-block) the state just below it. 
This is of course true only in the first iteration, and we now want to present 
a general iteration process, so let's assume that we start a new iteration in 
which the number of energy states which are already contained in the blocks 
(whether it is the P-block or the H-block) are $n-1$.

We increase the size of the two blocks by adding a particle state to the particle block 
and a hole state to the hole block, thus increasing the number of states to $n$ in each block. 
From these two blocks we build the superblock Hamiltonian by taking an equal number 
of states, $f$, from the two blocks, where $f$ represents the number of particle-hole 
pairs. For example, in the first iteration (see Fig.~\ref{fig:phdmrg_blocks}(b))
there are $4$ single-particle states in the superblock, 
let's denote them by $|i \rangle$, with $i=1,2,3,4$.
There will be a single superblock state in which $f=0$, i.e., without any 
electron-hole pair, so that the lowest two states are filled and the higher are empty, 
and this state can be denoted by $|1,2 \rangle$. There will be four superblock states with 
one p-h pair: these are the states $|1,3 \rangle$, $|1,4 \rangle$, $|2,3 \rangle$ or $|2,4 \rangle$. And finally 
it includes the state $|3,4 \rangle$ in which two p-h pairs exist. So in the first superblock,
when $4$ single-level states are treated, the Hamiltonian is of size $6$, being 
$\binom{4}{2}$.

The diagonalization of the superblock can lead us towards a new basis, by using similar
considerations as in the regular DMRG process. We project the superblock ground state 
over the space of the particle states by summing over the hole indices, and 
create a P density matrix. We do the same for the Hole states, by summing over the particle 
indices and get an H density matrix as well, and we use these two matrices in order to get a new
P basis and a new H basis, which replace the ones that we've started the iteration
with. In each state of the new basis we keep a copy of the number operator of particles 
(or holes) involved in this state, in order to be able to couple it later 
(in the next iteration) in the same manner. 

Then we are in a good position to move on to the next energy states. We increase 
the number of states involved in the game by taking the next states - a P-state 
will join the P-block, and an H-state to the H-block (Fig.~\ref{fig:phdmrg_blocks}(c)). 
As the Hamiltonian is built out of operators, we should keep a copy of all the operators we need.
In each step the number of energy states increases, and so is the number of required operators.
Since every operator is an $p \times p$ matrix, the memory requirements can be a serious limit 
of the PH-DMRG method.

\subsection{PH-DMRG: a Technical Overview}
We now present the technical details in relation with the Hamiltonian we use in this 
thesis\footnote[10]{The PH-DMRG method is used in chapter $5$.}.
The Hamiltonian describes the motion of interacting spinless electrons in a 2D 
disordered lattice (a QD). We will write the Hamiltonian of the system as
\begin{eqnarray} \label{eqn:H_phdmrg}
{\hat H} = {\hat H}_{0} + {\hat H}_{int},
\end{eqnarray}
where ${\hat H}_0$ includes the disorder and the hopping elements, and
${\hat H}_{int}$ is the interaction. In real-space we can write (see Eq.~(\ref{eqn:Intro_Hdot}))
\begin{eqnarray} \label{eqn:H_0_phdmrg}
{\hat H}_0 = \sum_{m} \epsilon_{m}{\hat a}^{\dagger}_{m}{\hat a}_{m} 
-t \displaystyle \sum_{\langle m,n \rangle}({\hat a}^{\dagger}_{m}{\hat a}_{n} + H.c.),
\end{eqnarray}
where the first term contains the on-site energies, which are diagonal in the lattice sites 
($m$ and $n$ denote both rows and columns for simplicity), and the second presents
the hopping elements between sites $m$ and $n$, where the notation $\langle ... \rangle$ denotes NNs. 

The interaction term for NN interactions\footnote[11]{The modification of the 
interactions to Coulombic form is straightforward, and will be discussed in 
chapter \ref{cpt:ch3}.} is (Eq.~(\ref{eqn:Intro_Hdot_int2}))
\begin{eqnarray} \label{eqn:H_int_nn}
{\hat H}_{int} = V
\displaystyle \sum_{\langle m,n \rangle}{\hat a}^{\dagger}_{m}{\hat a}^\dagger_{n}{\hat a}_{n}{\hat a}_{m}.
\end{eqnarray}



In order to work in k-space, one needs to write all these terms in their k-space
equivalence. Using the relation ${\hat a}_{m} = \sum_\alpha{{\hat b}_\alpha \phi_\alpha(m)}$, 
one gets\footnote[12]{Along this section the operators ${\hat a}$ and the indices
$m,n$ are used for real-space, whereas the operators ${\hat b}$ and the Greek indices
are used for k-space. The indices $i,j,k,l$ will be used for the matrix indices. }
\begin{eqnarray} \label{eqn:H_0_1_k}
\sum_{m} \epsilon_{m}{\hat a}^{\dagger}_{m}{\hat a}_{m} =
\sum_{\alpha,\beta} {\hat b}^{\dagger}_{\alpha}{\hat b}^{}_{\beta} [\sum_{m} \epsilon_{m} \phi^*_\alpha(m) \phi^{}_\beta(m) ].
\end{eqnarray}

The hopping term is less simple, but without too much effort one gets
\begin{eqnarray} \label{eqn:H_0_2_k}
\sum_{\langle m,n \rangle} ({\hat a}^{\dagger}_{m}{\hat a}_{n} + {\hat a}^{\dagger}_{n}{\hat a}_{m}) =
2 \sum_{\alpha,\beta} {\hat b}^{\dagger}_{\alpha}{\hat b}^{}_{\beta} \sum_{\langle m,n \rangle} \phi^*_\alpha(m) \phi^{}_\beta(n),
\end{eqnarray}
so that we can write 
\begin{eqnarray} \label{eqn:H_0_k}
{\hat H}_0 =
\sum_{\alpha,\beta} {\hat b}^{\dagger}_{\alpha}{\hat b}^{}_{\beta} 
[\sum_{m} \epsilon_{m} \phi^*_\alpha(m) \phi^{}_\beta(m)-2t\sum_{\langle m,n \rangle} \phi^*_\alpha(m) \phi^{}_\beta(n) ].
\end{eqnarray}

Since shortly we'll begin multiplying the matrices which represent the operators
(${\hat b}^\dagger_\alpha$ and ${\hat b}^{}_\beta$ are $p \times p$ matrices) in 
order to get the Hamiltonian matrix, 
it is worth noticing that the term in the parentheses does not depend on the matrix
indices, so it is calculated only once for each $\alpha$ and $\beta$.

In the same way it is easy to write the interaction element in k-space, 
as
\begin{eqnarray} \label{eqn:H_int_k_1}
{\hat H}_{int} =
\sum_{\alpha,\beta,\gamma,\delta} 
{\hat b}^{\dagger}_{\alpha}{\hat b}^{\dagger}_{\beta}{\hat b}^{}_{\gamma}{\hat b}^{}_{\delta}
~ V \sum_{\langle m,n\rangle} \phi^*_\alpha(m) \phi^{*}_\beta(n) \phi_\gamma(n) \phi^{}_\delta(m),
\end{eqnarray}
where again, the second summation is done once for each set of $\alpha,\beta,\gamma,\delta$.

In order to simplify future calculations we define the anti-symmetric interaction term
\begin{eqnarray} \label{eqn:V_abcd}
V_{\alpha \beta \gamma \delta} &=& V \sum_{\langle m,n \rangle} 
\phi^*_\alpha(m) \phi^{*}_\beta(n) ~[\phi_\gamma(n) \phi^{}_\delta(m) - \phi_\delta(n) \phi^{}_\gamma(m)] \\ \nonumber
&~& ~~~~~~~ - \phi^*_\alpha(n) \phi^{*}_\beta(m) ~[\phi_\gamma(n) \phi^{}_\delta(m) - \phi_\delta(n) \phi^{}_\gamma(m)] \\ \nonumber
&=&  V \sum_{\langle m,n \rangle}  \left[ \phi^*_\alpha(m) \phi^{*}_\beta(n)  - \phi^*_\alpha(n) \phi^{*}_\beta(m)\right] 
~ \left[ \phi_\gamma(n) \phi^{}_\delta(m) - \phi_\delta(n) \phi^{}_\gamma(m) \right].
\end{eqnarray}

We then get 
\begin{eqnarray} \label{eqn:H_int_k}
{\hat H}_{int} = 
\sum_{\alpha,\beta,\gamma,\delta} 
\frac{V_{\alpha \beta \gamma \delta}}{4} ~
{\hat b}^{\dagger}_{\alpha}{\hat b}^{\dagger}_{\beta}{\hat b}^{}_{\gamma}{\hat b}^{}_{\delta}.
\end{eqnarray}

After having the Hamiltonian written in our favorite notation, we can start 
describing the iteration process. The algorithm main steps are roughly similar
to those of the regular DMRG method.
Nevertheless, the details are very different, so we now explain in detail the iteration steps.
Recalling that the first iteration is done after an application of the Hartree-Fock
method, we assume the existence of a well-defined ordered set of wave functions 
(sorted by their energies).

\subsubsection{Step 1: Building the First P- and H-Blocks}

The first blocks contain only one state each. Actually we create here the matrices for
all the operators we'll later need. The Hamiltonian of the superblock contains only terms with
an even number of operators, i.e., terms like $\hat b^\dagger \hat b$ and 
$\hat b^\dagger \hat b^\dagger \hat b \hat b$, but since we prepare the P- and H-operators
separately while the Hamiltonian is built out of combinations of them, 
we must keep all the possible combinations, including 
those with an odd number of operators. We thus keep the following combinations:
$\hat b$, $\hat b^\dagger \hat b$, $\hat b \hat b$, $\hat b^\dagger \hat b^\dagger \hat b$
and $\hat b^\dagger \hat b^\dagger \hat b \hat b$.
The operators $\hat b^\dagger$, $\hat b^\dagger \hat b^\dagger$ and 
$\hat b^\dagger \hat b \hat b$ are omitted
since they can be calculated using Hermitian adjoint relations.
Every $\hat b$ in the above terms, in addition to representing a $p \times p$ matrix, should have a subscript 
Greek index, denoting the energy state to which it is related. 
Nevertheless, in the case of the four-state-operators one does not need to store every operator
on its own, since only their summation 
$\hat D_4 = \sum_{\alpha,\beta,\gamma,\delta} 
\frac{V_{\alpha \beta \gamma \delta}}{4} ~
{\hat b}^{\dagger}_{\alpha}{\hat b}^{\dagger}_{\beta}{\hat b}^{}_{\gamma}{\hat b}^{}_{\delta}$
is required for the superblock. Therefore, only this sum, $\hat D_4$, is stored. In the very first iteration, 
there is only one P-state ($\alpha_0$) and only one H-state ($\beta_0$), so that the list of 
operators we create is doubled: there are $5$ P-block operators and $5$ H-block operators.

In practice, after building explicitly $\hat b$ and $\hat b^\dagger$, we can define the other ones by a 
multiplication. Note that this simple method works only in the first time, because after
the first truncation the set of wave functions are not a complete basis any more, 
exactly as was the situation in the regular DMRG algorithm.
Thus for example
\begin{eqnarray} 
\langle i|~\hat b_1~\hat b_2~|j \rangle \ne \sum_m \langle i| \hat b_1 |m \rangle \langle m| \hat b_2 |j \rangle,
\end{eqnarray}
so we must keep all the operators explicitly in the memory.

\subsubsection{Step 2: Addition of a New State to the Block}

We now move to the next part: adding a new state. Let's think about adding a state
to one of the blocks assuming there are already $n-1$ states inside, and we are adding
the $\alpha_n$ state. What we need to do is to add to the list of operators all the combinations
with $\hat b_{\alpha_n}$ or $\hat b^\dagger_{\alpha_n}$. We also increase in this step the
size of every operator by a factor of $2$ in order to include the new state, so that the
operator matrices are now of size $2p$.

For simplicity, let's denote the operators of the previous basis (all the $\hat b$'s and
$\hat b^\dagger$'s of the first $n-1$ states) simply by $\hat b$, while $\hat b_\alpha$ 
will now represent the operator of the new state (instead of $\hat b_{\alpha_n}$).
We do the same for the indices, so that the new state vector will be denoted by $|i,i_\alpha \rangle$, 
where $i$ encodes all the "old" indices, and $i_\alpha=0,1$ is the index of the new energy state. 
We now wish to calculate the operator matrix elements in the new basis, 
and in general we will look at the element $\langle i,i_\alpha| \hat O | j,j_\alpha \rangle$, 
where $\hat O$ can be any one of our operators. Since $|i,i_\alpha \rangle = |i \rangle |i_\alpha \rangle$, 
we can write it as $\langle i_\alpha| \langle i|$~$\hat O$~$|j \rangle |j_\alpha \rangle$.

Let's start from the one-state operator $\hat b$ which is the easy 
one\footnote[13]{As noted above, one does not need to keep $\hat b^\dagger$, since it can
be calculated using $\hat b$.}.
If the operator $\hat O$ is
of the old basis, denoted in general as $\hat b$, one has obviously
\begin{eqnarray} \label{eqn:add_state_1op_c}
\langle i_\alpha| \langle i|~\hat b~|j \rangle |j_\alpha \rangle = \langle i| \hat b |j \rangle \delta_{i_\alpha,j_\alpha},
\end{eqnarray}
since $\hat b$ does not operate in the subspace of $i_\alpha$ and $j_\alpha$.
Note that $\langle i| \hat b |j \rangle$ is already known (and saved) from previous iterations.
On the other hand, if $\hat O = \hat b_\alpha$ we need at first to replace the $\hat b_\alpha$ and $j$, 
and since these are fermionic operators it gives a factor of $(-1)^{n_j}$, where
$n_j$ is the number of particles in state $j$. Thus we get
\begin{eqnarray} \label{eqn:add_state_1op_b}
\langle i_\alpha| \langle i|~\hat b_\alpha~|j \rangle |j_\alpha \rangle  = 
(-1)^{n_j} \delta_{i,j} \langle i_\alpha| \hat b_\alpha |j_\alpha \rangle ,
\end{eqnarray}
where $\langle i_\alpha| \hat b_\alpha |j_\alpha \rangle$ operates only in the subspace of the new 
energy state, so it can be written explicitly .

It's about time to get to the two-state operators, $\hat b \hat b$ and $\hat b^\dagger \hat b$.
Since the process is similar for both of them, we will discuss only the $\hat b \hat b$ case,
and it is divided to four possibilities: $\hat b_\alpha \hat b_\alpha$, $\hat b_\alpha \hat b$, 
$\hat b \hat b_\alpha$ or $\hat b \hat b$. 

The idea is the same as in the one-state operators, so to make a long story short,
we'll write here only the final expressions. Whenever there are two operators (or more) of the same basis 
we put an extra subscript, in order to differentiate between them. However, the order
of operators is always conserved.
\begin{eqnarray} \label{eqn:add_state_2op}
\langle i_\alpha| \langle i|~\hat b_1~\hat b_2~|j \rangle  |j_\alpha \rangle &=& 
\langle i|~\hat b_1~\hat b_2~|j \rangle  \delta_{i_\alpha,j_\alpha}\\ \nonumber
\langle i_\alpha| \langle i|~\hat b~\hat b_\alpha~|j \rangle  |j_\alpha \rangle &=& 
(-1)^{n_j} \langle i| \hat b |j \rangle  \langle i_\alpha| \hat b_\alpha |j_\alpha \rangle  \\ \nonumber
\langle i_\alpha| \langle i|~\hat b_\alpha~\hat b~|j \rangle  |j_\alpha \rangle  &=& 
- (-1)^{n_j} \langle i| \hat b |j \rangle  \langle i_\alpha| \hat b_\alpha |j_\alpha \rangle \\ \nonumber
\langle i_\alpha| \langle i|~\hat b_{\alpha,1}~\hat b_{\alpha,2}~|j \rangle |j_\alpha \rangle  &=& \delta_{i,j} 
\langle i_\alpha|~\hat b_{\alpha,1}~\hat b_{\alpha,2}~|j_\alpha \rangle.
\end{eqnarray}

Note that the minus sign (in the third line) results from the fermionic commutation relation 
between $\hat b_\alpha$ and $\hat b$.

Using the same method we can write down the $8$ possibilities of the three-state operator
$\hat b_\mu^\dagger \hat b_\nu^\dagger \hat b_\lambda$ in the enlarged basis. However, using the anti-symmetric
fermionic relation $\hat b_\mu^\dagger \hat b_\nu^\dagger \hat b_\lambda = -\hat b_\nu^\dagger \hat b_\mu^\dagger \hat b_\lambda$
for $\mu \ne \nu$ (with $\hat b_\mu^\dagger \hat b_\mu^\dagger \hat b_\lambda = 0$),
one can keep only those operators with $\mu < \nu$. Therefore, the only relevant possibilities
are 
\begin{eqnarray} \label{eqn:add_state_3op}
\langle i_\alpha| \langle i|~\hat b^\dagger_1~\hat b^\dagger_2~\hat b_3~|j \rangle  |j_\alpha \rangle  &=& 
\langle i|~\hat b^\dagger_1~\hat b^\dagger_2~\hat b_3~|j \rangle  \delta_{i_\alpha,j_\alpha}\\ \nonumber
\langle i_\alpha| \langle i|~\hat b^\dagger_1~\hat b^\dagger_\alpha~\hat b_2~|j \rangle |j_\alpha \rangle  &=& 
- (-1)^{n_j} \langle i|~\hat b^\dagger_1~\hat b_2~|j \rangle  \langle i_\alpha|\hat b^\dagger_\alpha|j_\alpha \rangle   \\ \nonumber
\langle i_\alpha| \langle i|~\hat b^\dagger_1~\hat b^\dagger_2~\hat b_\alpha~|j \rangle  |j_\alpha \rangle  &=& 
(-1)^{n_j} \langle i|~\hat b^\dagger_1~\hat b^\dagger_2~|j \rangle  \langle i_\alpha|\hat b_\alpha|j_\alpha \rangle   \\ \nonumber
\langle i_\alpha| \langle i|~\hat b^\dagger~\hat b^\dagger_{\alpha,1}~\hat b~_{\alpha,2}|j \rangle  |j_\alpha \rangle  &=& 
\langle i|~\hat b^\dagger~|j \rangle  \langle i_\alpha|~\hat b^\dagger_{\alpha,1}~\hat b_{\alpha,2}~|j_\alpha \rangle .
\end{eqnarray}

Similarly, from the $16$ four-state operators $\hat b_\mu^\dagger \hat b_\nu^\dagger \hat b_\lambda \hat b_\kappa$
only those with $\mu < \nu$ and $\lambda < \kappa$ are necessary, which again leads to
only four possibilities. However, since we store only the sum of these four-state operators 
multiplied by $V_{\mu \nu \lambda \kappa}$, which we have already denoted as $\hat D_4$, 
we first enlarge the basis for the case in which all of the operators are from the old basis, using
\begin{eqnarray} \label{eqn:add_state_4op_a}
\langle i_\alpha| \langle i|~\hat D_4~|j \rangle  |j_\alpha \rangle  &=& 
\langle i|~\hat D_4~|j \rangle  \delta_{i_\alpha,j_\alpha}.
\end{eqnarray}
Next we add to $\hat D_4$ all the new cases, in which $\nu$ and / or $\kappa$ represent the
new energy level. Namely, we \underline{add} to $\langle i_\alpha| \langle i|~\hat D_4~|j \rangle  |j_\alpha \rangle$
\begin{eqnarray} \label{eqn:add_state_4op_b}
(\nu&=&\alpha)~~~~~~~~~
(-1)^{n_j} \langle i| ~\hat b^\dagger_1~\hat b_2~\hat b_3~  |j \rangle   \langle i_\alpha| \hat b^\dagger_\alpha   |j_\alpha \rangle 
V_{1,\alpha,2,3}           \\ \nonumber
(\kappa&=&\alpha)~~~~~~~~~
  (-1)^{n_j} \langle i| ~\hat b^\dagger_1~\hat b^\dagger_2~\hat b_3~  |j \rangle   \langle i_\alpha|  \hat b_\alpha  |j_\alpha \rangle 
V_{1,2,3,\alpha}          \\ \nonumber
(\nu=\kappa&=&\alpha)~~~~~~~~~
- \langle i|  ~\hat b^\dagger_1~\hat b_2~ |j \rangle   \langle i_\alpha| ~\hat b^\dagger_{\alpha,1}~\hat b_{\alpha,2}~   |j_\alpha \rangle 
V_{1,\alpha_1,2,\alpha_2},
\end{eqnarray}
and finally we store only $\hat D_4$.

At last, we remind that every operator $\hat b$ and $\hat b^\dagger$ listed 
in Eqs.~(\ref{eqn:add_state_1op_c}) - 
(\ref{eqn:add_state_4op_b})
should have an index which runs over all the states that have already entered to the game 
in the previous iterations.
One should also note that such a block enlargement is done 
separately for both the particle and the hole blocks.

\subsubsection{Step 3: The Superblock}
After the enlargement of the blocks we can combine the P- and H-blocks to form the superblock.
In the superblock it's the time we really build the Hamiltonian of the system.
This is done by matching for every P-state which contains $f$ particles, an H-state
with the same number $f$ of holes. 
Since the operators of each block are matrices of size $2p$, and
they operate on different spaces, the maximal size of the superblock Hamiltonian matrix will be 
$4p^2$. The matrix indices are a direct product of a P-index and an H-index.

So if we denote the P-indices by $i$ for the row and $k$ for the column, and
the H-indices by $j$ and $l$, accordingly, we now want to calculate the
matrix element $\langle j| \langle i| \hat H |k \rangle |l \rangle$. The Hamiltonian is given by 
$\hat H = \hat H_0 + \hat H_{int}$, where $\hat H_0$ and $\hat H_{int}$ 
are written in k-space and given by
Eqs.~(\ref{eqn:H_0_k}) and (\ref{eqn:H_int_k}), in which the operators 
are either the P-operators or the H- ones.

For the calculation of $\hat H_0$ we need to enumerate over the two state
indices, $\alpha$ and $\beta$, and divide the cases to four possibilities, depending on
whether $\alpha$ and $\beta$ are states from P- or H-blocks. By denoting 
the P (H) operators by an appropriate subscript we get
\begin{eqnarray} \label{eqn:super_H0}
\langle j| \langle i|~\hat b^\dagger_{\alpha,p}~\hat b_{\beta,p}~|k \rangle  |l \rangle  &=& 
\langle i|~\hat b^\dagger_{\alpha,p}~\hat b_{\beta,p}~|k \rangle  \delta_{j,l} \\ \nonumber
\langle j| \langle i|~\hat b^\dagger_{\alpha,p}~\hat b_{\beta,h}~|k \rangle  |l \rangle  &=& 
(-1)^{n_k} \langle i|~\hat b^\dagger_{\alpha,p}~|k \rangle  \langle j|\hat b_{\beta,h}~|l \rangle  \\ \nonumber
\langle j| \langle i|~\hat b^\dagger_{\alpha,h}~\hat b_{\beta,p}~|k \rangle  |l  \rangle &=& 
- (-1)^{n_k} \langle i|~\hat b_{\beta,p}~|k \rangle  \langle j|\hat b^\dagger_{\alpha,h}~|l \rangle  \\ \nonumber
\langle j| \langle i|~\hat b^\dagger_{\alpha,h}~\hat b_{\beta,h}~| k \rangle  | l \rangle  &=& 
\delta_{i,k} \langle j|~\hat b^\dagger_{\alpha,h}~\hat b_{\beta,h}~|l \rangle .
\end{eqnarray}

The calculation of $\hat H_{int}$ is just a bit more lengthy, because here we
have $4$ state operators ($\alpha,\beta,\gamma,\delta$), which result, in 
principle, in $16$ possibilities, depending on the configuration.
However, because of the enumeration over these indices, and because 
of our definition of the anti-symmetric interaction term 
$V_{\alpha \beta \gamma \delta}$, some of the possibilities are equivalent. 
For example, let's look at the terms $ppph$ (three P-operators and then an H-operator)
and $pphp$. Using the fermionic relations and the antisymmetry of $V$, we can write
\begin{eqnarray} \label{eqn:super_Hint_equiv}
V_{\alpha \beta \gamma \delta} \hat b^\dagger_{\alpha,p}~\hat b^\dagger_{\beta,p}~\hat b_{\gamma,h}~\hat b_{\delta,p}  = 
(- V_{\alpha \beta \delta \gamma}) 
(- \hat b^\dagger_{\alpha,p}~\hat b^\dagger_{\beta,p}~\hat b_{\delta,p}~\hat b_{\gamma,h}),
\end{eqnarray}
and thus these two terms give the same contribution. In the same manner we get
$phpp \sim hppp$ ($\sim$ denotes an equivalence), as well as $hhhp \sim hhph$, and $phhh \sim hphh$.
The term $phph$ is equivalent to three other terms: $hphp$, $phhp$ and $hpph$.

There are therefore only the following $9$ possibilities, which are (excluding the
factor $\frac {V_{\alpha \beta \gamma \delta}}{4}$ which is the same for all of them):
\begin{eqnarray} \label{eqn:super_Hint}
 (1)~ pppp ~&\Rightarrow&~ 
\langle i|~\hat b^\dagger_{\alpha,p}~\hat b^\dagger_{\beta,p}~\hat b_{\gamma,p}~\hat b_{\delta,p}~|k \rangle  \delta_{j,l} \\ \nonumber
 (2)~ ppph ~&\Rightarrow&~ 
2 (-1)^{n_k} \langle i| ~\hat b^\dagger_{\alpha,p}~\hat b^\dagger_{\beta,p}~\hat b_{\gamma,p}~   |k \rangle  \langle j| \hat b_{\delta,h}  |l \rangle    \\ \nonumber
 (3)~ hppp ~&\Rightarrow&~ 
-2 (-1)^{n_k} \langle i| ~\hat b^\dagger_{\beta,p}~\hat b_{\gamma,p}~\hat b_{\delta,p}~   |k \rangle  \langle j| \hat b^\dagger_{\alpha,h}  |l \rangle    \\ \nonumber
 (4)~ pphh ~&\Rightarrow&~ 
\langle i| ~\hat b^\dagger_{\alpha,p}~\hat b^\dagger_{\beta,p}~   |k \rangle  \langle j| ~\hat b_{\gamma,h}~\hat b_{\delta,h}~  |l \rangle    \\ \nonumber
 (5)~ phph ~&\Rightarrow&~ 
-4 \langle i| ~\hat b^\dagger_{\alpha,p}~\hat b_{\gamma,p}~   |k \rangle  \langle j|~ \hat b^\dagger_{\beta,h}~\hat b_{\delta,h}~  |l \rangle    \\ \nonumber
 (6)~ hhpp ~&\Rightarrow&~ 
\langle i| ~\hat b_{\gamma,p}~\hat b_{\delta,p}~   |k \rangle  \langle j| ~\hat b^\dagger_{\alpha,h}~\hat b^\dagger_{\beta,h}~  |l \rangle    \\ \nonumber
 (7)~ phhh ~&\Rightarrow&~ 
2 (-1)^{n_k} \langle i| \hat b^\dagger_{\alpha,p}   |k \rangle  \langle j| ~\hat b^\dagger_{\beta,h}~\hat b_{\gamma,h}~\hat b_{\delta,h}~  |l \rangle    \\ \nonumber
 (8)~ hhhp ~&\Rightarrow&~ 
-2 (-1)^{n_k} \langle i| \hat b_{\delta,p}   |k \rangle  \langle j| ~\hat b^\dagger_{\alpha,h}~\hat b^\dagger_{\beta,h}~\hat b_{\gamma,h}~  |l \rangle    \\ \nonumber
 (9)~ hhhh ~&\Rightarrow&~ 
 \delta_{i,k} \langle j|~\hat b^\dagger_{\alpha,h}~\hat b^\dagger_{\beta,h}~\hat b_{\gamma,h}~\hat b_{\delta,h}~|l \rangle .
\end{eqnarray}

We again remark that the enumeration over $\alpha,\beta,\gamma,\delta$ is done, in practice, only for
cases $(2) - (8)$, since for the cases $(1)$ and $(9)$ we use the operator $\hat D_4$ which was
already calculated as the required sum (two such operators exist, for both particle- and hole-states).

Finally we arrive at the superblock Hamiltonian matrix. Its size can vary
since we take only super-states with an equal number of particles and holes, and the number of such
combinations depends on the states kept in each block, which can be changed in every iteration.
The upper limit, of course, is $4p^2$, where $p$ is the block size of both P and H.
Anyway, we diagonalize the Hamiltonian, i.e., we find its ground state $\Psi_0$ for which
$\hat H |\Psi_0 \rangle = E_0 |\Psi_0 \rangle $, and we move on to the next step.

\subsubsection{Step 4: The Density Matrix and Hilbert-Space Truncation}

From the ground state $|\Psi_0 \rangle $ we build two density matrices, one for the P-block
and the other for the H-block. Since actually $|\Psi_0 \rangle $ encodes the indices for
both blocks, i.e.,
\begin{eqnarray} \label{eqn:psi0_ph}
|\Psi_0 \rangle  = \sum_{i_p,j_h} C_{ij} |i_p \rangle  |j_h \rangle ,
\end{eqnarray}
we can trace out half of these indices in order to build each density matrix. For the
P density matrix $\rho^{(p)}$ we define (omitting the subscript $0$ from $\Psi_0$)
\begin{eqnarray} \label{eqn:rho_p}
(\rho^{(p)})_{i_p,i^\prime_p} = \sum_{j_h} \Psi_{i_p,j_h} \Psi^\dagger_{i^\prime_p,j_h},
\end{eqnarray}
while $\rho^{(h)}$ is defined as
\begin{eqnarray} \label{eqn:rho_h}
(\rho^{(h)})_{j_h,j^\prime_h} = \sum_{i_p} \Psi_{i_p,j_h} \Psi^\dagger_{i_p,j^\prime_h}.
\end{eqnarray}

The next step is to diagonalize these matrices and get, for each,
the eigenvalues with the highest eigenvalues as a truncation operator.
If we denote $\rho |u_\alpha \rangle  = \omega_\alpha |u_\alpha \rangle $ (for each one of
the blocks), then we choose the eigenvectors $|u_\alpha \rangle $ with the highest $\omega_\alpha$'s 
to form an $2p \times p$ matrix $O$. We then use the $O$ matrix in order
to transform all the enlarged operators (from step $2$) to their original 
small size $p$, as it was at the beginning of the current iteration.

Now, of course, we are able to add another couple of states (one P and one H)
to the game, and thus we return to step $2$.

As steps $(2) - (4)$ are repeated iteratively, the number of states involved in the
physical game increases, and the states we are considering are further
away from the Fermi energy. One may thus expect that as this distance increases, the 
influence of the additional levels will decrease. 

A few minor remarks: in the first few iterations, when the block size is still
smaller than $p$, the truncation is not done, the operators continue to the
next iteration as they are, and the block size grows. When we are not
at half filling occupation, the number of P-states is different from that of the H-states.
For example when $n_p > n_h$, after we finish adding all the H-states we continue
in the process but add in each iteration only the next P-state. In the iterations that 
follow, the H-blocks will remain the same.


\cleardoublepage

\chapter[A QD coupled to a 1D interacting reservoir]
{A level coupled to a $1D$ interacting reservoir}
\label{cpt:ch1}

In this chapter, we explore the ground state properties of a one-dimensional 
system, consisting of a gate-controlled dot coupled to an interacting reservoir, using 
the numerical density-matrix renormalization-group method. First, we calculate a few
physical observables of the system in the thermodynamic limit and zero temperature
in the non-interacting case. We then move to investigate the Tomonaga-Luttinger 
liquid phase of the interacting lead. We concentrate on the
influence of interactions in the lead, as well as dot-lead interactions, 
on the width of the dot filling as a function of the chemical potential, 
and on the position of the dot level. 

We also study other phases of the lead, i.e., the charge density wave and the ferromagnetic phases. 
With particle-hole symmetry, it is known that for different values of the interaction 
strength the lead is described by different phases (see section \ref{sec:int_pd}).
We show that a semi-infinite charge density wave coupled to the dot undergoes a quantum phase transition when the 
dot's level crosses the wire's chemical potential. On the other hand,
in the ferromagnetic phase there is a simple level crossing at the same point.

\section{Introduction}
The properties of one-dimensional (1D) interacting systems
have attracted much interest going back half a century \cite{tomonaga50,luttinger63,voit94}. 
Much recent effort has concentrated on understanding the conductivity
and I-V characteristics of a Tomonaga-Luttinger liquid (TLL) coupled to an impurity \cite{kane92}. 
These properties are probed experimentally by measurements of the temperature dependent
conduction through 1D systems \cite{cond}, and tunneling spectroscopy into 
1D wires \cite{auslander05}. The measurements of Ref. \cite{auslander05} also
indicate a localization transition in the wire for low densities which
might be associated with a charge density wave (CDW). Signatures of a CDW have been also observed recently, in 
finite 1D wires coupled to dots, when strong magnetic fields were applied \cite{pepper04}.

A generic model for the situation of a 1D system which is coupled to an impurity,
is a quantum dot (QD) coupled to a lead. If such a dot is controlled by a gate one can change 
its orbital energy, and measure physical quantities which are either on the lead or 
on the dot. For example, the occupation of the dot level is experimentally 
accessible by the charging effect of the impurity on a quantum point contact (QPC)
in its vicinity \cite{johnson04}.

\subsection{Level Broadening vs. Conductivity}
\begin{figure}[ht]\centering
\includegraphics[trim=0mm 0mm 0mm 0mm, clip, width=7cm]{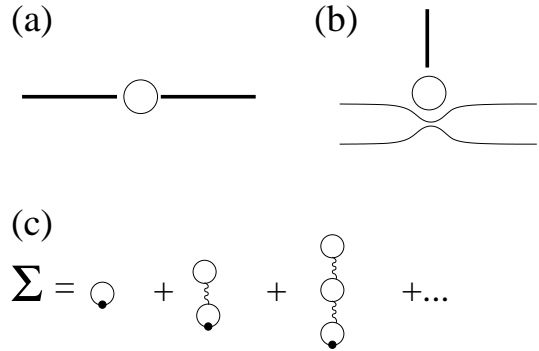}
\caption[Different coupling schemes of a QD and leads]
{\label{fig_1}
Different coupling schemes of a QD and leads. (a) A QD coupled to two 
TLL leads represented by the wide lines. (b) A QD coupled 
electrostatically to a QPC through which the conductance in measured, 
and to an interacting reservoir. (c) The diagrammatic representation of the
RPA approximation of the self energy. The line corresponds to
the lead Green function, the black dot to the hopping into the
dot and the wiggly line to the interaction.}
\end{figure}

The difference between measuring the conductivity through the dot-lead system
or the local density of states at the impurity \cite{grishin04},
and probing the dot occupation using, e.g., a QPC, must be emphasized.
Essentially, as was noted in chapter \ref{cpt:intro}, 
any impurity which is placed in a TLL will lead to an insulating behavior \cite{kane92}. 
The resonance conductance through a QD coupled to a pair of TLL leads 
(see Fig.~\ref{fig_1}a) was found to produce infinitely sharp 
Coulomb blockade peaks at zero temperature \cite{nazarov03}. 
Thus, no level broadening of the dot states is exhibited in the measurement 
of the conduction through that dot. 

Nevertheless, this does not imply that coupling a dot to a TLL
has no effect on the width of the dot filling as a function of the chemical potential.
Consider for example the arrangement depicted in Fig.~\ref{fig_1}b. A dot is connected to
a TLL lead, while its occupation is measured by a QPC.
Thus, the TLL acts as a reservoir for the dot, while the 
QPC is used to probe the dot's level broadening.
In such an arrangement, any additional broadening of the levels due to the
coupling to the reservoir will be seen in the shape of the conductance 
through the QPC.

The conductance through the QPC in the geometry described in Fig.~\ref{fig_1}b
is directly proportional to the occupation of the dot's orbital due to
the capacitive coupling between the charge of the dot and 
the QPC (it is assumed that no tunneling occurs between the dot and
the QPC) \cite{johnson04}. Thus, in principal, the dot occupation ${n}_{\rm dot}$ may be
read off the conductance through the QPC and the effect of coupling 
of the dot to the interacting reservoir can be measured.

The difference between the two arrangements depicted in Figs.~\ref{fig_1}a and \ref{fig_1}b
is that while the first case (Fig.~\ref{fig_1}a) essentially probes the enhancement 
of the backscattering in the TLL in the vicinity of the Fermi energy, the second
(Fig.~\ref{fig_1}b) explores the broadening of the level due
to coupling to states which may be far from the Fermi energy. 
Therefore, one might expect the broadening of the level 
measured in the second arrangement to approach
the conventional Breit-Wigner form, although some signature of
the interactions is anticipated.

\subsection{Chapter's Outline}
The rest of this chapter is organized as follows. In the next section, \ref{sec:ch1_model},
we discuss the Hamiltonian model, and briefly describe the numerical methods we 
use\footnote[1]{A detailed description is given in chapter \ref{cpt:numerics}.}.
We then dedicate section \ref{sec:ch1_nonint} to an analytical calculation in the non-interacting case, 
in which we exactly formulate the physical properties,
which we later calculate numerically for the interacting system,
by using the Green functions techniques.
Next we present the numerical results for interacting systems. 
In section \ref{sec:ch1_notsymH} we show the dot population in the most general TLL phase. 
We continue, in section \ref{sec:ch1_symH}, with a version of the Hamiltonian which conserves 
particle-hole symmetry, and compare the system's properties between the TLL and the CDW
phases. Three physical quantities are obtained: the dot occupation, the total system's 
population and the free energy. 
Focusing on the latter, we compare, in section \ref{sec:ch1_qpt}, its dependence on the
gate voltage to that of the ferromagnetic (FM) phase, and show that in the CDW there is 
a quantum phase transition (QPT) when the dot's energy crosses the Fermi energy. In the FM
phase, on the other hand, a simple level crossing occurs at that point.
We continue the chapter in section \ref{sec:ch1_relation} by exploring another consequence of the 
particle-hole symmetry of the Hamiltonian, which results in a non-trivial integral form.
We finish by a summation of the main findings, and point out a possibility for a future research.

\section{Model}
\label{sec:ch1_model}
\subsection{Hamiltonian}
The system we investigate in this chapter is composed of a QD which is coupled to a 1D
lead. The electrons moving in the system are spinless, the QD is restricted to have a 
single level, and there are interactions between nearest-neighbor (NN) electrons in the lead. 
The Hamiltonian describing such a system can be written as
$\hat H = \hat H_{\rm dot} + \hat H_{\rm dot-lead} + \hat H_{\rm lead}$,
where $\hat H_{\rm dot}$ and $\hat H_{\rm lead}$ are given, as was explained in chapter \ref{cpt:intro}, by 
\begin{eqnarray} \label{eqn:H_dot}
{\hat H_{\rm dot}} &=& \epsilon_{0}{\hat a}^{\dagger}{\hat a}, \\ \nonumber
{\hat H_{\rm lead}^{(1)}} &=& -t
\displaystyle \sum_{j=1}^{L-1}({\hat c}^{\dagger}_{j}{\hat c}_{j+1} + H.c.) 
+ I \displaystyle \sum_{j=1}^{L-1}{\hat c}^{\dagger}_{j}{\hat c}_{j}
{\hat c}^{\dagger}_{j+1}{\hat c}_{j+1},
\end{eqnarray}
and $\hat H_{\rm dot-lead}$, which connects the QD and the lead through hopping and dot-lead interaction terms, is
\begin{eqnarray} \label{eqn:H_lead1}
{\hat H_{\rm dot-lead}} = -V ({\hat a}^{\dagger}{\hat c}_{1} +
{\hat c}^{\dagger}_{1}{\hat a}) + 
I_{\rm dl} {\hat a}^{\dagger}{\hat a} {\hat c}^{\dagger}_{1}{\hat c}_{1}.
\end{eqnarray}
We denote the dot's energy level by $\epsilon_0$,
$V$ ($t$) is the dot-lead (lead) hopping matrix element, and
$I_{\rm dl}$ ($I$) is the dot-lead (lead-lead) NN interaction strength.
Following our convention for the operators notations, 
${\hat a}^{\dagger}$ (${\hat a}$) is the creation (annihilation)
operator of an electron in the dot, and ${\hat c}_j^{\dagger}$ (${\hat c}_j$) is the 
creation (annihilation) operator of an electron at site $j$ in the lead.
The lead hopping matrix element, $t$, is taken as $1$, in order to set the energy scale.

In order to obtain a particle-hole symmetric version of the Hamiltonian,
one should consider a positive background in the two interaction terms. Throughout this chapter,
for the particle-hole symmetric version we will substitute $I_{\rm dl}=0$ in the $\hat H_{\rm dot-lead}$ term,
and we will replace $\hat H_{\rm lead}^{(1)}$ by 
\begin{eqnarray} \label{eqn:H_lead2}
{\hat H_{\rm lead}^{(2)}} = -t
\displaystyle \sum_{j=1}^{L-1}({\hat c}^{\dagger}_{j}{\hat c}_{j+1} + H.c.) 
+ I \displaystyle \sum_{j=1}^{L-1}({\hat c}^{\dagger}_{j}{\hat c}_{j} - \frac{1}{2})
({\hat c}^{\dagger}_{j+1}{\hat c}_{j+1} - \frac{1}{2}).
\end{eqnarray}

In the following we use both forms of these Hamiltonians. 
When the electron-electron interaction is not considered, as in section \ref{sec:ch1_nonint}, 
the two forms $\hat H_{\rm lead}^{(1)}$ and $\hat H_{\rm lead}^{(2)}$ are identical.
For the interacting case, however, they are different.
In section \ref{sec:ch1_notsymH} we examine 
the TLL phase by considering the non-symmetric Hamiltonian 
$\hat H_1 = \hat H_{\rm dot} + \hat H_{\rm dot-lead} + \hat H_{\rm lead}^{(1)}$. 
We explore the interaction effect on the dot population in two cases. The first case
is when an electrostatic potential between the dot and the lead does not exist, i.e., 
for $I_{\rm dl}=0$. In the second case we assume $I_{\rm dl}=I$. 
In section \ref{sec:ch1_symH} we compare between different phases
of the wire by taking the particle-hole symmetric version of the Hamiltonian
$\hat H_2 = \hat H_{\rm dot} + \hat H_{\rm dot-lead} + \hat H_{\rm lead}^{(2)}$
with $I_{\rm dl}=0$.

\subsection{Diagonalization Method}
For each Hamiltonian model we discuss,
the grand canonical potential $\hat \Omega={\hat H}-\mu {\hat N_e}$,
where ${\hat N_e}$ is the particle-number operator and $\mu$ is the chemical potential, is diagonalized using a 
finite-size density-matrix renormalization-group (DMRG) calculation \cite{white93,berkovits03-1}
for different values of $V$, $I$ and $\epsilon_0$ or $\mu$, and with a lead of up to $L=500$ sites.
For the cases in which a particle-hole symmetry is required, 
$\mu$ is set to zero.

As was noted in chapter \ref{cpt:numerics}, the total number of particles in the system is 
not fixed during the DMRG process, so that the results describe the experimental situation of a
finite section of a 1D wire which is coupled to a dot and to an external electron reservoir.
In particular, the total occupation obtained for such a system can be non-integral.

The following ground state properties are calculated as a function of $\epsilon_0$ (or $\mu$):
the system's grand potential $\Omega$, the total number of electrons $N$
and the dot population $n_{\rm dot}$. The population of lead sites are also calculated
in order to differentiate between local effects of the dot population and
global phenomena in the lead.

\section{Non-Interacting Case}
\label{sec:ch1_nonint}
\subsection{Exact Calculation}
In order to estimate the influence of the interactions in the
lead on the different properties of the system, we should start from considering 
the non-interacting case. The coupling of the dot state to the continuum
(akin to the Fano-Anderson model) may be treated using standard
Green function technique \cite{mahan} which leads to:
\begin{eqnarray} \label{eqn:n_mu}
{n}_{\rm dot}(\mu,\epsilon_0) = \frac{1}{\pi} \int_{-\infty}^{\mu} \frac
{\Im \Sigma(\epsilon)}
{(\epsilon-\epsilon_0-\Re \Sigma(\epsilon))^2 + (\Im \Sigma(\epsilon))^2} 
d\epsilon,
\end{eqnarray}
where $\Sigma(\epsilon)$ is the self energy given by:
\begin{eqnarray} \label{eqn:self_energy}
\Sigma(\epsilon) = \sum_k \frac {|V_k|^2}{\epsilon - \epsilon_k - i \delta},
\end{eqnarray}
$\epsilon_k$ are the eigenvalues of the lead, $V_k$ is the coupling
between the eigenstates in the lead and the state in the dot and 
$\delta \rightarrow 0$.

For the idealized case, the density of states in the lead is
constant (i.e., $\epsilon_k=k / L \nu$, where 
$\nu$ is the (constant) local density of states, and $L$ is the leads length). 
The coupling is $V_k=\sqrt{a/L}V$ ($a$ is the distance between NNs), and 
under these conditions
\begin{eqnarray} \label{eqn:self_energy_2}
\Sigma(\epsilon) = \frac{a}{L} 
\int \frac {|V|^2 dk}{\epsilon - k/L\nu - i \delta},
\end{eqnarray}
resulting in $\Im \Sigma(\epsilon) = \pi a \nu |V|^2 = \Gamma/2$ and
$\Re \Sigma(\epsilon)=0$. Thus, one obtains
the Breit-Wigner formula: 
\begin{eqnarray} \label{eqn:breit_wigner}
{n}_{\rm dot}(\mu,\epsilon_0) = \frac{1}{\pi} \int_{-\infty}^{\mu} \frac
{\frac{\Gamma}{2}}
{\left(\epsilon-\epsilon_0 \right)^2 + \left(\frac{\Gamma}{2} \right)^2} d\epsilon.
\end{eqnarray}

For the tight-binding model discussed
$\epsilon_k= -2t \cos(ka)$ and $V_k=\sqrt{2a/L} \sin(ka) V$, resulting in
\begin{eqnarray} \label{eqn:sigma_dot}
\Sigma(\epsilon) = (V/t)^2\epsilon/2 + i (V^2/t) \sqrt{1-(\epsilon/2t)^2}.
\end{eqnarray}
We thus find
\begin{eqnarray} \label{eqn:moshe1}
{n}_{\rm dot}(\mu,\epsilon_0) = \frac{1}{\pi} \int_{-2t}^{\mu} \frac {\frac{V^2}{t} 
\sqrt {1-\frac{\epsilon^2}{4t^2}} }
{\frac{V^4}{t^2} \left(1-\frac{\epsilon^2}{4t^2} \right) + 
\left[ \left(1-\frac{V^2}{2t^2} \right) \epsilon - \epsilon_0 \right]^2} d\epsilon.
\end{eqnarray}

The change in the total density of states in the system,
due to the presence of the dot, can be found in a similar method to be
\begin{eqnarray} \label{eqn:dos}
\Delta \nu(\epsilon) = \frac {1}{\pi}
\Im \frac {\partial} {\partial \epsilon} \ln (\epsilon -\epsilon_{0} - \Sigma(\epsilon)).
\end{eqnarray}

Therefore, the change in the number of electrons in the entire
system is
\begin{eqnarray} \label{eqn:n_miu}
\Delta N(\mu,\epsilon_0) = \int_{-2t}^{\mu} {\Delta \nu(\epsilon) d\epsilon} = \\ \nonumber
\frac {1}{2} + \frac {1}{\pi} & \displaystyle
\tan^{-1}{ \frac {\mu -\epsilon_{0} - \Re \Sigma(\mu)} {\Im \Sigma(\mu)}},
\end{eqnarray}
and the change in the free energy of the system is
\begin{eqnarray} \label{eqn:free_E}
\Delta \Omega(\mu,\epsilon_0) = \Delta E - \mu \Delta N(\mu) = \\ \nonumber
\int_{-2t}^{\mu} (\epsilon-\mu) \Delta \nu(\epsilon) d\epsilon &=& 
- \int_{-2t}^{\mu} \Delta N(\epsilon,\epsilon_0) d\epsilon.
\end{eqnarray}

\subsection{Numerical Results}
\begin{figure}[h]\centering
\includegraphics[trim=0mm 0mm 0mm 0mm, clip, width=7cm]{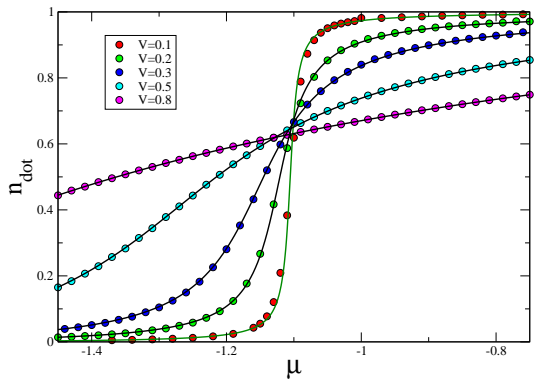}
\caption[Level population in the non-interacting case]
{\label{fig:110}
The dot population as a function of the chemical potential
for $\epsilon_0=-1.1$ is shown for various values of $V$, and without interactions.
The symbols are the DMRG results and the lines are the exact formula, 
Eq.~(\ref{eqn:moshe1}).}
\end{figure}

The non-interacting case, in which the equations given above are relevant, gives an
important possibility to examine the results of the numerical DMRG method. In Fig.~\ref{fig:110} 
the numerical results of the DMRG for the dot population are presented in symbols, and are 
compared to the calculated formula Eq.~(\ref{eqn:moshe1}). A very good 
correspondence\footnote[2]{A good correspondence is obtained for the other two formulas as well, 
i.e., between the DMRG results for the total number of electrons and for the free energy, to the
prediction of Eqs.~(\ref{eqn:n_miu}) and (\ref{eqn:free_E}). These results are shown
in section \ref{sec:ch1_symH}.} is obtained for all values of the dot-lead coupling $V$.

\section{The Dot Population in the TLL Phase}
\label{sec:ch1_notsymH}
In this section we discuss the results obtained for the Hamiltonian
$\hat H_1$, considering a constant $\epsilon_0$ and enumerating over $\mu$.
Alternatively, we will consider a constant $\mu$, and enumerate over $\epsilon_0$.
In both cases the potential $\hat H-\mu \hat N_e$ is diagonalized using the DMRG method 
for a lead of a few hundreds of sites, and the curves of the dot population, whether 
$n_{\rm dot}(\mu)$ or $n_{\rm dot}(\epsilon_0$), are calculated for several coupling strengths, $V$, 
and interaction strengths, $I$. 
 
\subsection{Influence of Interactions in the Lead}
\begin{figure}[ht]\centering
\includegraphics[trim=0mm 0mm 0mm 0mm, clip, width=7cm]{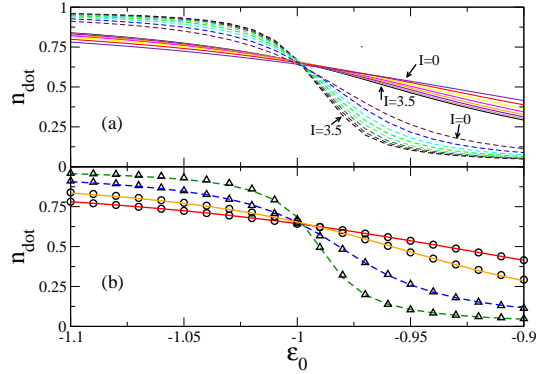}
\caption[Level population as a function of its energy]
{\label{fig:111}
(a) The dot population as a function of the level energy  
for $V=0.2$ (dashed lines) and $0.4$ (solid lines), and a constant
chemical potential $\mu=-1$. 
The interaction strength $I$ takes values between $0$ and $3.5$, in jumps of $0.5$.
(b) The curves for $I=0,3.5$ are redrawn (symbols) together with the fit to the 
Eq.~(\ref{eqn:moshe1}) (lines) with an effective coupling $V_{\rm eff}$.}
\end{figure}

\begin{figure}[ht]\centering
\includegraphics[trim=0mm 0mm 0mm 0mm, clip, width=7cm]{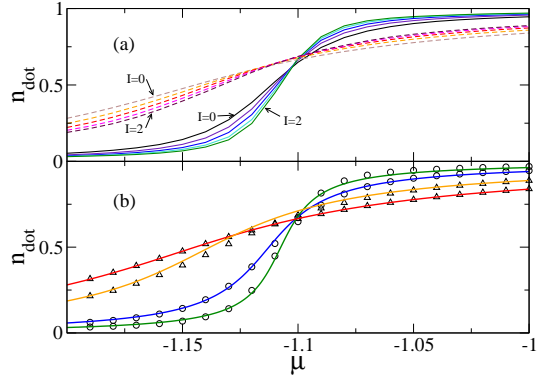}
\caption[Level population as a function of the chemical potential]
{\label{fig:1a}
(a) The dot population as a function of the chemical potential 
for $V=0.15$ (solid lines) and $0.3$ (dashed lines), and a constant
level energy $\epsilon_0=-1.1$.
The interaction strength $I$ takes values between $0$ and $2$, in jumps of $0.5$. 
b) The curves for $I=0,2$ are redrawn (symbols) together with the fit to the 
Eq.~(\ref{eqn:moshe1}) (lines) with an effective coupling $V_{\rm eff}$.
}
\end{figure}

First we discuss the case in which $I_{\rm dl}=0$.
Typical results for $n_{\rm dot}(\epsilon_0$) and $n_{\rm dot}(\mu)$ are shown in Figs.~\ref{fig:111}
and \ref{fig:1a}. As can be seen in both cases, changing the interaction 
strength in the lead ($I$) almost does not affect the level position, which
is centered near the constant $\mu$ in Fig.~\ref{fig:111}, 
and the constant $\epsilon_0$ in Fig.~\ref{fig:1a}. 
The level width, however, decreases with increasing interactions.
Nevertheless, it can be clearly seen that
the shape of the curve is similar for all values of interaction strength shown.

\begin{figure}[ht]\centering
\includegraphics[trim=0mm 0mm 0mm 0mm, clip, width=7cm]{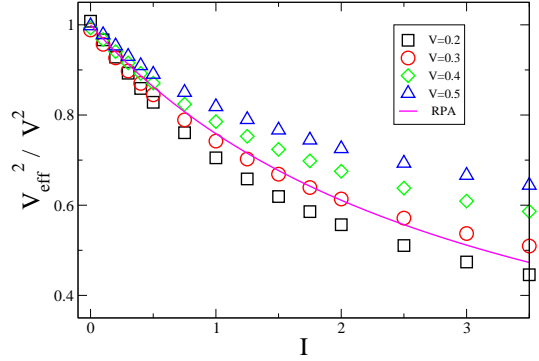}
\caption[Effective dot-lead coupling]
{\label{fig:2}
$V_{\rm eff}^2/V^2$ as a function of $I$ (symbols) as
obtained by fitting the $n_{\rm dot}(\epsilon_0)$
curves of $V=0.2,0.3,0.4$ and $0.5$ to Eq.~(\ref{eqn:moshe1}).
The line corresponds to the dependence according to RPA prediction Eq.~(\ref{eqn:V_eff}),
and is in good accordance with all curves for small $I$.}
\end{figure}

In order to understand the role played by the interactions in the
lead, when it is described by the TLL theory, one must remember
that the dot occupation ${n}_{\rm dot}$ is determined by contributions from all 
energies, and the region around the Fermi energy, which has the special
TLL behavior, does not play a unique role. 
Therefore, one could expect a simple perturbation
description of the interactions in the lead to suffice. Indeed, the
effect of the electron electron interactions in the RPA approximation
on the self energy (see Fig.~\ref{fig_1}c) 
may be written as \cite{berkovits97}:
\begin{eqnarray} \label{eqn:interaction}
\Sigma(\epsilon) = \chi \Sigma^0(\epsilon) 
\end{eqnarray}
where $\Sigma^0(\epsilon)$ is the non-interacting self energy and
\begin{eqnarray} \label{eqn:interaction1}
\chi = \frac{1}{1+a \nu I}.
\end{eqnarray}

Here we assumed a constant local density of states $\nu$ (for the
tight binding lead $\nu = (a \pi t)^{-1}$ which ignores
local density of states  variations), thus one obtains
$\Im \Sigma(\epsilon) = (V^2/[t(1+I/\pi t)]) 
\sqrt{1-(\epsilon/2t)^2}$ and
$\Re \Sigma(\epsilon) = (V^2/[t(1+I/\pi t)])\epsilon/2$, corresponding
to replacing $V^2$ in
Eqs.(\ref{eqn:sigma_dot}) and (\ref{eqn:moshe1}) by an ``effective'' coupling
\begin{eqnarray} \label{eqn:V_eff}
V_{\rm eff}^2=V^2/(1+I/\pi t).
\end{eqnarray}

Returning to the results obtained by the numerical DMRG calculations,
the curves of Figs.~\ref{fig:111} and \ref{fig:1a} can be now fitted to 
Eq.~(\ref{eqn:moshe1}) with $V_{\rm eff}^{2}$ as a fitting parameter.
It is easy to see that the Breit-Wigner form fits quite well
even in the presence of strong interactions, and with the level position 
$\epsilon_0$ remaining constant.
The effect of interactions is indeed limited here to a decrease of 
$V_{\rm eff}^{2}$, i.e., a decrease of the level width $\Gamma$.
The values of $V_{\rm eff}^{2}$ extracted from the fit are plotted
in Fig.~\ref{fig:2} and compared with the RPA predictions
of $V_{\rm eff}^2=V^2/(1+I/\pi t)$. A rather good correspondence,
especially for small values of $I$, is observed for all values of $V$.

Thus, although the TLL has a vanishing local density of states at the end of the lead
in the vicinity of the Fermi energy, the dot level is broadened, 
since all the reservoir states take part in the broadening mechanism.
Nevertheless, as we have seen, the interactions in the reservoir influence the
width of the resonance. One might gain some insight from the following consideration:
For the non-interacting case (for constant density of states in the reservoir)
the width is equal to $\Gamma = 2 \pi a \nu |V|^2$, which
may be rewritten  as $\Gamma = 2 \pi (a/L) |V|^2 \partial N/ \partial \mu$.
The thermodynamic inverse compressibility $\partial N/ \partial \mu$ is 
affected by the interactions \cite{lee82}.
Lets consider the compressibility $\partial \mu/ \partial N$.
In the lowest order approximation \cite{berkovits97}
$\partial \mu/ \partial N = (L \nu)^{-1} + e^2/C$, where $C$ is the capacitance
of the system. For NN interaction $e^2=aI$ and
as usual $C \sim L$. Therefore,  
$\partial N/ \partial \mu = L \nu /(1+a \nu I)$. Inserting this to
the expression for $\Gamma$, we get a result similar to
the RPA approximation results in Eq.~(\ref{eqn:interaction1}).
Although capacitance is proportional to the length $L$ of the lead,
so is the density of states in a 1D system, and therefore
it has an influence even for an infinite lead.

\subsection{Influence of Dot-Lead Interactions}
\begin{figure}[ht]\centering
\includegraphics[trim=0mm 0mm 0mm 0mm, clip, width=7cm]{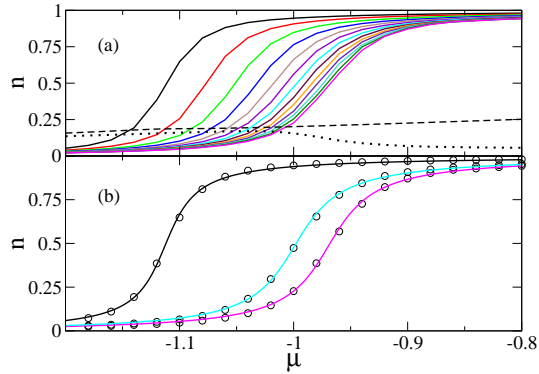}
\caption[Level population in the presence of dot-lead interactions]
{\label{fig:2a}
(a) Population of the dot and of the first site of the lead 
as a function of the chemical potential, for $V=0.15$. 
Dot-lead interaction was included taking $I_{\rm dl}=I$.
The curves shown are for $I$ between $0$ and $3$, in jumps of $0.25$ (full lines) for the dot population,
and for $I=0$ (dashed line) and $I=2$ (dotted line) for the lead population.
(b) The plots for $I=0, 1.5$ and $3$ (symbols)
together with the best fit to Eq.~(\ref{eqn:moshe1}). }
\end{figure}

In order to consider interactions between an electron occupying 
the dot, and the electrons in the lead, we now turn on the dot-lead interaction
term, and we choose to take $I_{\rm dl}=I$.
The corresponding $n(\mu)$ results (Fig.~\ref{fig:2a})
clearly show a change in the resonance width, but also a
change in the level position, which was absent in
the previous case. Nevertheless, these results can still be
fitted to Eq.~(\ref{eqn:moshe1}), using also $\epsilon_{\rm 0,eff}$ as an 
additional fitting parameter, in addition to $V_{\rm eff}^{2}$.
As can be seen, Eq.~(\ref{eqn:moshe1}) describes this system quite well.

\begin{figure}[ht]\centering
\includegraphics[trim=0mm 0mm 0mm 0mm, clip, width=7cm]{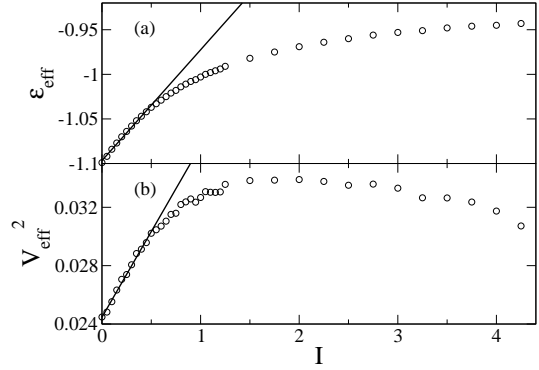}
\caption[Effective level position and dot-lead coupling in the presence of dot-lead interactions]
{\label{fig:2b}
(a) $\epsilon_{\rm 0,eff}$ and (b) $V_{\rm eff}^{2}$ as functions of 
$I_{\rm dl}=I$ 
(symbols) and linear fits for the region $I\le0.5$ (lines). }
\end{figure}

The movement of the resonance center, $\epsilon_{\rm 0,eff}$, as well as the
width, $V_{\rm eff}^{2}$, that were obtained from the fit are shown
in Fig.~\ref{fig:2b}. For small values of interaction
both grow linearly.  First lets try to explain the shift in the
resonance center. As noted, almost no shift was seen for $I_{\rm dl}=0$.
In the presence of weak dot-lead interactions, one may approximate
the dot-lead interaction operator by
${\hat H_{\rm dl}} \approx I_{\rm dl} n_1 {\hat a}^{\dagger}{\hat a}$, 
where $n_{1}$ represents the average occupation of the first
site in the lead. As can be seen in Fig.~\ref{fig:2a}a, $n_1$ is not
very sensitive to the occupation of the dot and may be replaced
by its typical value. Thus, the energy of the orbital in 
Eq.~(\ref{eqn:H_dot}) may be rewritten as
$\epsilon_{\rm 0,eff} = \epsilon_{0}+ n_1 I_{\rm dl}$. Indeed, 
this formula fits well the numerical results for small 
values of $I_{\rm dl}$ (as can be seen
in Fig.~\ref{fig:2b}(a)), for $n_{1} = 0.14$. This result agrees well
with the value $n_{1} \approx 0.15$ in the region of the resonance,
taken from the data of Fig.~\ref{fig:2a}(a).

A more striking feature is the behavior of $V_{\rm eff}^{2}$, i.e,
the width of the resonance. There is a distinct qualitative 
change in the width behavior,
compared to the case without dot-lead interactions.
As opposed to the monotonic decrease of $V_{\rm eff}^{2}$, which 
was demonstrated in Fig.~\ref{fig:2}, 
Fig.~\ref{fig:2b}b (symbols) shows that $V_{\rm eff}^{2}$ increases with $I$, 
until a maximal value is achieved around $I=2$. For larger values of
interaction a decrease in the width is observed.

This enhancement of $V_{\rm eff}^{2}$ is associated to the interplay
between the population of the dot level to the depopulation
of the first site in the lead, ignored in our treatment of $\epsilon_{\rm 0,eff}$.
This leads to a reduction in the effect of the dot-lead interaction
which results in an increase in the width as depicted 
in Fig.~\ref{fig:2b}b. For weak interactions the enhancement 
of  $V_{\rm eff}^{2}$ is linear.

\section{A Comparison Between the TLL and the CDW Phases}
\label{sec:ch1_symH}
The 1D wire can be, as noted in chapter \ref{cpt:intro}, in other phases than the TLL, 
e.g. the CDW phase. The cleanest appearance of the CDW phase occurs at $\mu=0$, 
when the system is half-filled and particle-hole symmetry exists. 
Therefore, in order to compare the results for these two phases, we use, 
here, the symmetric version of the Hamiltonian, $\hat H_2$. 

In order not to increase the number of free parameters in this section, 
the model discussed here does not contain a dot-lead interaction term. 
Nevertheless, as the results of the previous section have pointed out, 
and according to the results we are going to present, we note that
such a term is not expected to influence qualitatively our results.

\subsection{Level Occupation}
\begin{figure}[htb]\centering
\includegraphics[trim=0mm 0mm 0mm 0mm, clip, width=6cm]{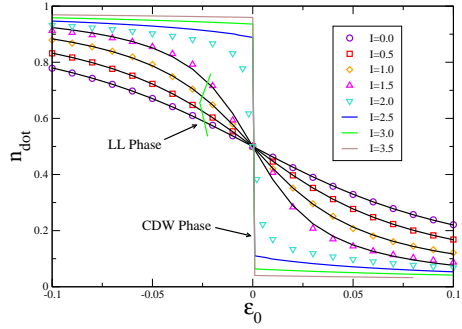}
\caption[Level population in the TLL and CDW phases]
{\label{fig:n_dot}
The dot population $n_{\rm dot}$ as a function of the level energy $\epsilon_0$.
For $I<2t$ (TLL phase) the curves fit the non-interacting
formula with an effective coupling
constant $V_{\rm eff}$ (lines - DMRG results, symbols - fit to Eq.~(\ref{eqn:moshe1})),
while for $I>2t$ (CDW phase) the width is zero.
Exactly at the transition point ($I=2t$) $n_{\rm dot}$ is continuous
but doesn't fit the non-interacting formula.}
\end{figure}

We begin by comparing the behavior of $n_{\rm dot}$ in the two phases of the interacting lead.
As pointed out in section \ref{sec:int_1d}, the system experiences a phase transition 
(of a Kosterlitz-Thouless type \cite{mikeska04}), between the TLL and the CDW 
phases, at $I=2t$. 
Indeed, the results presented in Fig.~\ref{fig:n_dot} show a qualitative difference
between $n_{\rm dot}(\epsilon_0)$ in these two regimes. 
We have shown in the previous section that in the TLL phase
the curves fit quite well the Breit-Wigner formula with an effective coupling $V_{\rm eff}$.
In the CDW phase, however, the width of the level becomes zero, and a jump in $n_{\rm dot}$ occurs
at $\epsilon_0=0$. This jump is associated with the degeneracy of the CDW ground state.

In order to support this argument, we first calculate the
properties of the ground state for a lead not coupled to a dot,
by taking $V=t$ and the limit $\epsilon_0 \rightarrow 0$.
The CDW order parameter can be defined as \cite{pang_93}
\begin{eqnarray} \label{eqn:P_cdw}
{P(i) = \frac {1}{2} (-1)^i [ 2n(i)-n(i-1) - n(i+1)]},
\end{eqnarray}
where $n(i)= \langle {\hat c}^{\dagger}_{i}{\hat c}_{i} \rangle$
is the occupation of the i'th lead site in the ground state.
The value of $P(i)$ does not change much as a function of the spatial
coordinate, except in the vicinity of the lead's edges.
We thus define $P=P(L/2)$.

\begin{figure}[htb]\centering
\includegraphics[trim=0mm 0mm 0mm 0mm, clip, width=5.5cm]{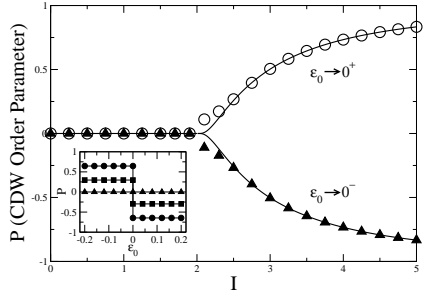}
\caption[CDW order parameter as a function of the interaction strength]
{\label{fig:cdw_order}
The CDW order parameter $P$ as a function of the interaction $I$
for a 300-sites lead (DMRG - symbols, theory - lines).
For the CDW phase $P$ is inverted
between the cases $\epsilon_0 \rightarrow 0^+$ and $\epsilon_0 \rightarrow 0^-$,
while in the TLL phase both cases result in $P=0$.
Inset: $P$ as a function of $\epsilon_0$ for
$I=1.5$ (triangles), $2.5$ (squares) and $3.5$ (circles). The TLL
case results in a constant $P=0$, while an inversion of
$P$ occurs at $\epsilon_0=0$ for the two CDW cases.}
\end{figure}

In Fig.~\ref{fig:cdw_order} the dependence of $P$
on the interaction strength $I$, as obtained by the DMRG method, is shown, and
compared to the exact results \cite{baxter73}.
For $I<2t$, the system is in the TLL phase and indeed $P=0 \pm 10^{-4}$.
In this case the population of each lead site is $1/2$ 
and the lead is half filled; i.e., $N=L/2$.

As expected, the metal-insulator transition occurs at $I=2t$.
For $I>2t$, in which the system is in the CDW phase,
the value of $P$ is finite. For values of $I$ which are far
from the transition point ($I > 2.5t$), we get a very good fit to the theory.
Near the transition point the numerics tend to emphasize the charge oscillations,
resulting in too large values of $|P|$. This tendency, however, does not
affect the following qualitative conclusions.

As can be seen, the CDW ground state is different for the cases
$\epsilon_0 \rightarrow 0^+$ and $\epsilon_0 \rightarrow 0^-$,
resulting in two values of $P$ (i.e., $\pm |P|$) for any value of $I$ in the CDW regime.
For $\epsilon_0 = 0$ the ground state is two-fold degenerate in the
thermodynamic limit.

In the CDW case, special care should be devoted to
the number of electrons in the system, and
to the difference between even and odd lead length $L$.
For $\epsilon_0 \rightarrow 0$ and $V=t$, the lead length is
effectively $L+1$.
Denoting the ground state for $\epsilon_0 \rightarrow 0^{\pm}$
by $\psi^{(\pm)}_0$,
and its population in each site as $n^{(\pm)}_j$, where $j=0$ for the dot
and $1 \le j \le L$ for
the lead sites, the particle-hole symmetry implies that
$n^{(-)}_j=1-n^{(+)}_j$ everywhere.
For odd $L$ (even $L+1$) one has the requirement
$n^{(-)}_j=n^{(+)}_{L-j}$, and the number of electrons in both
$\psi^{(+)}_0$ and $\psi^{(-)}_0$ is $N=(L+1)/2$.
When $L$ is even ($L+1$ is odd), however, one has an additional symmetry requirement,
$n_j = n_{L-j}$, for both $\psi^{(+)}_0$ and $\psi^{(-)}_0$,
and the states differ in their electrons number:
$\psi^{(+)}_0$ contains $N=L/2$ electrons, for any $I>2$,
while $\psi^{(-)}_0$ has $N=L/2+1$.
It is worth noting that $n_1^{(+)} > 1/2$,
while $n_1^{(-)} < 1/2$.

\begin{figure}[htb]\centering
\includegraphics[trim=0mm 0mm 0mm 0mm, clip, width=6cm]{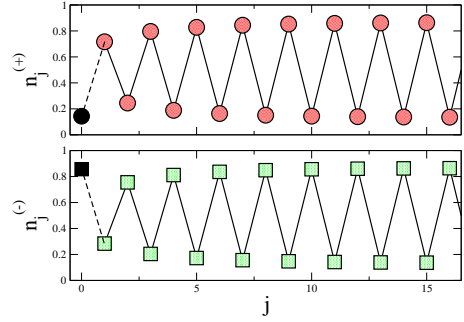}
\caption[Doubly degenerate CDW ground state]
{\label{fig:cdw_dot_scheme}
The occupation of the dot and the first lead sites of the
CDW states $\psi_0^{(+)}$ (circles)
and $\psi_0^{(-)}$
(squares), of a lead with 300 sites and $I=4$ which is
coupled to the dot (filled symbols) with $V=0.8$.
For $\epsilon_0=0$ these two states are degenerate in the thermodynamic limit.
Lines are guide to the eye.}
\end{figure}

When $\epsilon_0 \ne 0$, i.e., a dot with a finite on-site energy is connected to the lead, it
has some local influence\footnote[3]{The investigation of this influence takes a major part of 
chapter \ref{cpt:ch2}.}~on the ground state in its vicinity.
For a lead in the CDW phase, the two states
described above are slightly modified, and can be denoted by $\psi^{(+)}$
and $\psi^{(-)}$. Nevertheless, the
main influence of the dot is to lift the degeneracy between those states.
If $\epsilon_0<0$, the dot state population is high and $\psi^{(-)}$
is energetically preferable due to the dot-lead hopping term.
For $\epsilon_0>0$ the opposite happens, resulting
in a preference of $\psi^{(+)}$.
The occupation of the dot-lead system in the CDW phase is shown
in Fig.~\ref{fig:cdw_dot_scheme}.
When the dot's orbital energy changes from a negative
to a positive value, the system switches from $\psi^{(-)}$ to
$\psi^{(+)}$. In the following section we show, by a
calculation of the free energy, that this is indeed a
first order QPT. The resulting phase diagram is drawn in the inset of
Fig.~\ref{fig:n_tot}.

\subsection{Total Number of Electrons}
\begin{figure}[htb]\centering
\includegraphics[trim=0mm 0mm 0mm 0mm, clip, width=6cm]{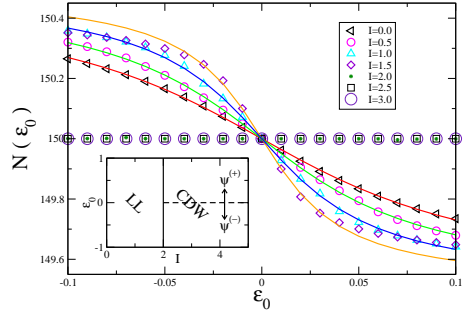}
\caption[Total number of electrons in a dot-lead system]
{\label{fig:n_tot}
The total number of electrons in the system,
as a function of $\epsilon_0$, for $L=299$, $\mu=0$, $V=0.3$, and
different values of $I$. Symbols - DMRG results.
Lines - fit to Eq.~(\ref{eqn:n_miu}).
Inset: the phase diagram: a Kosterlitz-Thouless transition (solid line)
occurs at $I=2$, and a first order phase transition (dashed line) at 
$\epsilon_0=0$ for $I>2$.}
\end{figure}

In Fig.~\ref{fig:n_tot} the total number of electrons
in the system is presented. For $I=0$,
$N$ fits the predicted formula, Eq.~(\ref{eqn:n_miu}), well.
As noted above, the agreement between the numerical results for a
finite lead and the exact result for a semi-infinite lead, is obtained
due to the fact that the number of particles can vary during the DMRG process.
For the TLL phase ($I<2t$),
$N(\epsilon_0)$ looks quite similar, varying between
$L/2+1$ (at $\epsilon_{0} \rightarrow -\infty$)
and $L/2$ (at $\epsilon_{0} \rightarrow \infty$),
taking the average value $(L+1)/2$ at $\epsilon_{0}=0$. As in the case of
$n_{\rm dot}$, the results fit Eq.~(\ref{eqn:n_miu}) with a
renormalized dot-lead coupling for small values of $I$. Increasing $I$
towards the transition point ($I > 1.5t$), results in a less accurate fit.

The values of $V_{\rm eff}(I)$ obtained by the fit of the $N$ curves
to Eq.~(\ref{eqn:n_miu}),
are in good agreement with the values obtained by the fit of $n_{\rm dot}$
to Eq.~(\ref{eqn:moshe1}). We find that $V_{\rm eff}$ decreases
monotonically with increasing $I$, exhibiting the RPA like
behavior described in section \ref{sec:ch1_notsymH}.

The CDW phase ($I>2t$), however,
is qualitatively different: for an odd lead length $N(\epsilon_0)$
remains a constant integer ($N=(L+1)/2$) which
does not depend on $\epsilon_0$ at all.
For even $L$, $N=L/2+1$ for $\epsilon_0<0$, and
$N=L/2$ for $\epsilon_0>0$.
This is a direct result from the
switch of the system ground state from
$\psi^{(-)}$ to $\psi^{(+)}$
at $\epsilon_0=0$.
For an odd lead length, the total number of
sites in the system ($L+1$) is even, so that $N$ is equal for both states.
For an even lead length, $L+1$ is odd, and the total number of
electrons is changed by one when $\epsilon_0$ passes $0$.
Except for the decrease of one electron at $\epsilon_0=0$ for an even lead,
$N$ remains constant in the CDW phase, independent of $\epsilon_0$.
Thus even with the continuous change in the population of the
dot as a function of the dot's level for $\epsilon_0 \ne 0$,
the number of electrons in the entire system remains constant.
The change in the occupation of the dot as a function of $\epsilon_0$
is compensated by the lead.

This difference in the behavior of $N$ as a function of $\epsilon_0$
between the two phases (i.e., constant for
the CDW phase, compared to a continuous decrease
for the TLL phase) is a direct manifestation of their transport properties.
The TLL phase is metallic, and therefore compressible. Hence, infinitesimal
changes of the electrons number are possible. On the other hand, the CDW phase
is insulating and thus incompressible, which results in a constant $N$.

\subsection{Free Energy}
\begin{figure}[htb]\centering
\includegraphics[trim=0mm 0mm 0mm 0mm, clip, width=6cm]{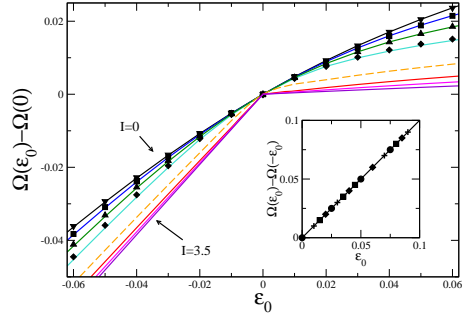}
\caption[Ground state free energy in a dot-lead system]
{\label{fig:free_energy}
The ground state free energy as a function of $\epsilon_0$ for different
interaction strengths. The lead-dot coupling was taken as
$V=0.3$, and the interaction strength
$I$ takes values between $0$ and $3.5$, in jumps of $0.5$.
The lines represent the DMRG results (dashed line - $I=2$), while the symbols show the
fit to Eq.~(\ref{eqn:free_E}) for values of $I$ lower than $2$.
Inset: $\Omega(\epsilon_0)-\Omega(-\epsilon_0)$
as a function of $\epsilon_0$ for $I=0,1,2,3$ (symbols) 
and $\Omega(\epsilon_0)-\Omega(-\epsilon_0) = \epsilon_0$ (line).
For all values of interaction we get
$\left| \Omega(\epsilon_0)-\Omega(-\epsilon_0) - \epsilon_0 \right| < 10^{-6}$.
This result is discussed in section \ref{sec:ch1_relation}.
}
\end{figure}

In Fig.~\ref{fig:free_energy}, typical numerical results for
the free energy $\Omega(\epsilon_0)-\Omega(0)$ as
a function of $\epsilon_0$ are shown.
A perfect fit between the DMRG results
and the exact formula Eq.~(\ref{eqn:free_E}) for $I=0$ is obtained.
In the TLL phase ($I<2t$) our DMRG calculations show
that the effect of interactions on $\Omega$ can be fitted by
replacing $V$ in Eq.~(\ref{eqn:free_E}) by the same
effective coupling $V_{\rm eff}$ obtained for the behavior
of $n_{\rm dot}$ and of $N$ which were discussed above.
For the CDW phase ($I>2t$), however,
there is obviously a qualitative change in the energy curve:
the dependence of $\Omega$ on $\epsilon_0$ is linear
both below and above $\epsilon_0=0$, with an abrupt
change of $\frac {d\Omega}{d\epsilon_0}$ at $\epsilon_{0}=0$.

These results point out that the single impurity,
connected at one end of a long interacting lead which is in a CDW state, has a well defined
influence on the ground state of the entire coupled system. As
discussed above, when the dot level passes through $\epsilon_0=0$,
the lead's population is inverted at every site, leading to
an inversion of the CDW order parameter,
as presented in Fig.~\ref{fig:cdw_order}(inset).
The dot population is inverted as well. As a result,
a dramatic change in the dependence of the free energy of the system
on $\epsilon_{0}$ occurs.

\section{Quantum Phase Transition}
\label{sec:ch1_qpt}
\subsection{QPT or a Simple Level Crossing?}
Since $\Omega$ is the free energy of the system,
the jump in its first derivative might be a sign of a first order QPT.
In order to see whether this non-analyticity of the free
energy is not just a trivial level crossing (LC) of two levels in the system,
the dependence of the transition shape on the system size $L$ is explored.
For the case that the sharp transition in the energy results from the fact
that the external field (in our case the gate voltage) commutes with the
Hamiltonian, and thus a LC is possible, no size dependence of the sharpness
of the transition is expected. On the other hand, a real QPT will become
sharp only in the thermodynamic limit (i.e., semi-infinite lead).

In order to compare the two scenarios (LC vs. QPT) we solve
the same Hamiltonian with strong
attractive interactions, i.e., $I<-2t$. As discussed in chapter \ref{cpt:intro}, 
such strong attractive interactions regime corresponds to the FM
phase, while the CDW is equivalent to the N\'eel phase.
In the fermionic case the FM phase yields a two-fold
degenerate ground state, composed of states in which the sites are either 
entirely occupied or empty\footnote[4]{No explicit restriction to a certain filling factor is assumed.}.
Pinning by the dot lifts the degeneracy, because an occupied dot
causes a preference of an empty lead, and vice versa.
Thus the influence of the dot on the coupled wire is superficially similar
to the CDW case in so far as in both cases the ground state
of the entire system is determined by the dot orbital.
Nevertheless, the FM system is
clearly a LC system, and no dependence on the system size is expected.

\begin{figure}[htb]\centering
\includegraphics[trim=0mm 0mm 0mm 0mm, clip, width=8cm]{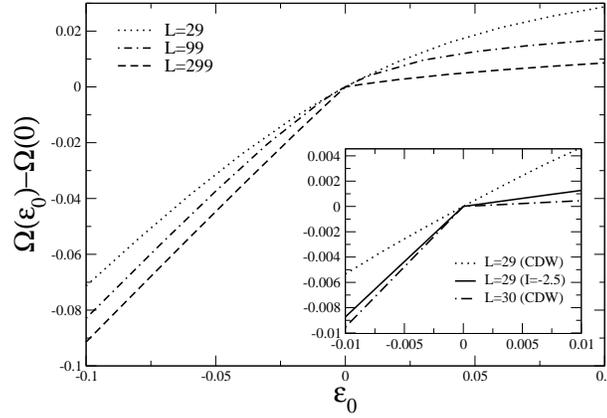}
\caption[Free energy for different models and system sizes]
{\label{fig:qpt_L}
The ground state free energy as a function of $\epsilon_0$ for different
CDW system sizes, with $V=0.3$ and $I=2.5t$ (dotted, dashed-dotted and dashed lines).
Inset: the results for $L=29$ are compared to the case of $I=-2.5t$,
which is inside the FM phase, and to the CDW case with $L=30$.
}
\end{figure}

The dependence of $\Omega$ on $\epsilon_0$ was calculated for
$L=29,99,299$ in the CDW case, and for $L=29$ in the FM case.
From the results shown in Fig.~\ref{fig:qpt_L} (and its inset)
it is clear that these two cases are different.
The FM system is indeed a trivial LC system, with two
competing states whose energies cross each other for $\epsilon_0=0$.
Since these two states are eigenstates of the Hamiltonian even for
a finite $L$, the size does not play a role, so that even for $L=29$
one can see a sharp transition between the two ground states.

On the other hand, for a small (but with an odd $L$) CDW system $\Omega$
shows a smooth dependence on $\epsilon_0$, without any non-analyticity.
As a matter of fact, for any finite system the electron levels are expected
to be mixed, resulting in avoided crossing of the two lowest many-body
levels \cite{qpt_book}. In other words, the CDW states are not true
eigenstates of the Hamiltonian for a finite system. Indeed,
indications for the sudden jump of $\frac {d\Omega}{d\epsilon_0}$ are seen
only for larger system sizes (i.e., for $L \gtrsim 200$).

The dot-dashed line in Fig.~\ref{fig:qpt_L}(inset) shows that for a short lead
of an even size ($L=30$ in that case) there is, however, a non-analyticity
in $\Omega$ as a function of $\epsilon_0$. The comparison between even and odd
lead sizes emphasizes our conclusion stated above.
While for an odd lead size $\psi^{(-)}$ and $\psi^{(+)}$ have the
same number of electrons, for an even size of the lead, the transition
from $\psi^{(-)}$ to $\psi^{(+)}$ at $\epsilon_0=0$ involves a decrease
of the electrons number by one.
The Hamiltonian $\hat H$ conserves the number of particles, so that in
the case of $L$ even, these two states are not coupled, and the 
transition between them is a simple LC, not showing a size dependence.
For an odd $L$, however, $\psi^{(-)}$ and $\psi^{(+)}$ are coupled
by $\hat H$, so that for a finite $L$ they are actually mixtures of the
CDW states, thus presenting sharper $\Omega(\epsilon_0)$ dependence
for larger systems.
We thus conclude, based on these two comparisons, that in the case of a CDW
with an odd $L$ this transition is a QPT, which happens for
$L \rightarrow \infty$, i.e., for a semi-infinite lead.

\subsection{Scaling Results of Small Systems}
As pointed out above, in the numerical results for large lead sizes, the
non-analyticity of the free energy is seen clearly, so that it is obviously
a first order transition. However, a similar conclusion can be drawn from
the results of short leads, by scaling the results with the lead size $L$.
Although for finite size systems the order parameter changes continuously,
the slope of this change grows with the system size $L$.
For a first order transition in $d$ dimensions one expects a power law dependence
of the slope as $L^d$, while for a second order transition
the power law should be fixed by some universal exponents \cite{binder84}.

\begin{figure}[htb]\centering
\includegraphics[trim=0mm 0mm 0mm 0mm, clip, width=8cm]{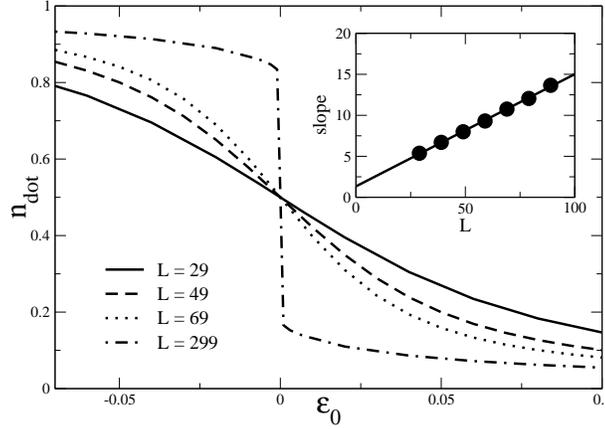}
\caption[Level occupation for different CDW system sizes]
{\label{fig:scaling_L}The dependence of the dot occupation on 
$\epsilon_0$ for different CDW system sizes, with $V=0.3$ and $I=2.5$.
Inset: the absolute value of the slope in the limit $\epsilon_0 \rightarrow 0$
as a function of $L$ for short leads, together with a linear fit.
}
\end{figure}

For the CDW model, the population of the dot plays the role
of an order parameter, and one can check its behavior near $\epsilon_0=0$ for different
lead sizes. In Fig.~\ref{fig:scaling_L} the occupation as a function of $\epsilon_0$
is shown for some short lead sizes ($L=29,49,69$) and for a large one ($L=299$).
It is indeed seen that while for a long enough lead there is a jump in the
dot population, for short leads there is a continuous change in the occupation,
with a larger slope for larger lengths. Zooming into the regime of $\epsilon_0=0$
we find that although $L \gtrsim 200$ is required in order to see a clear discontinuity
in $\Omega$,
$L \gtrsim 100$ is long enough for showing a jump in $n_{\rm dot}$.
The dependence of the slope on $L$, for $L<100$,
is shown in the inset of Fig.~\ref{fig:scaling_L}. As one expects from
finite size scaling predictions for first order phase transition,
there is a very good linear fit of the the order parameter on the lead size.

\section{Another Implication of Particle-Hole Symmetry}
\label{sec:ch1_relation}
We now turn to another consequence of the particle-hole symmetry.
For $\epsilon_0 \ne 0$ the symmetry results in the fact that the
ground states for $+\epsilon_0$ and for $-\epsilon_0$
($\psi^{(+)}$ and $\psi^{(-)}$ respectively)
have an inverted population everywhere.
It is thus clear that $\langle \psi^{(+)}|\hat H_{\rm lead}|\psi^{(+)} \rangle =
\langle \psi^{(-)}|\hat H_{\rm lead}|\psi^{(-)} \rangle$, and similarly
$\langle \psi^{(+)}|\hat H_{\rm dot-lead}|\psi^{(+)} \rangle =
\langle \psi^{(-)}|\hat H_{\rm dot-lead}|\psi^{(-)} \rangle$. This implies
\begin{eqnarray} \label{eqn:E_diff}
\Omega(\epsilon_0)-\Omega(-\epsilon_0) &=& \\ \nonumber
\langle \psi^{(+)}|\hat H_{\rm dot}|\psi^{(+)} \rangle &-&
\langle \psi^{(-)}|\hat H_{\rm dot}|\psi^{(-)} \rangle = \\ \nonumber
&& \epsilon_0 n_{\rm dot}^+ - (-\epsilon_0) n_{\rm dot}^- = \epsilon_0,
\end{eqnarray}
where $n_{\rm dot}^+$ ($n_{\rm dot}^-$) represents the dot populations of
$\psi^{(+)}$ ($\psi^{(-)}$).
The last equality results from the symmetric population
$n_{\rm dot}^- = 1 - n_{\rm dot}^+$.

The relation Eq.~(\ref{eqn:E_diff}) does not depend on the interaction,
and should exist for both the TLL and the CDW phases.
Obviously it should be obeyed for the non-interacting case,
where the energy is given by Eq.~(\ref{eqn:free_E}), thus
leading to the following nontrivial relation
\begin{eqnarray} \label{eqn:E_diff_I0}
{
- \frac {1}{\pi} \int_{-2t}^{2t}
\tan^{-1} \frac {  (1- \frac {V^2} {2t^2}) \epsilon -\epsilon_{0} }
{ \frac {V^2}{t} \sqrt {1- \frac {\epsilon^2}{4t^2}} }
 d\epsilon
= \epsilon_{0}},
\end{eqnarray}
in which the result of the integral on the LHS does not depend on the parameters
$V$ and $t$.

A physical insight into Eq.~(\ref{eqn:E_diff_I0}) can be gained by
taking the derivative of both sides with respect to $\epsilon_0$, and using the
definition of the self energy given above in Eq.~(\ref{eqn:sigma_dot}). This yields
\begin{eqnarray} \label{eqn:E_tag_diff}
{\frac {1}{\pi} \int_{-2t}^{2t}
\frac {\Im \Sigma(\epsilon)} {(\epsilon -\epsilon_{0} - \Re \Sigma(\epsilon))^2 +
(\Im \Sigma(\epsilon))^2} d\epsilon = 1},
\end{eqnarray}
which is evident since the LHS is the occupation of the dot when $\mu > 2t$.

In Fig.~\ref{fig:free_energy}(inset) a plot of
$\Omega(\epsilon_0)-\Omega(-\epsilon_0)$ 
as a function of $\epsilon_0$ is shown, for values of $I$ between $0$ and $3$.
As can be seen, Eq.~(\ref{eqn:E_diff}) is valid 
for all values of $I$.

\section{Conclusions and Future Prospects}
In conclusion, we have shown in this chapter that the occupation of the dot level 
can be used to identify the different phases in the wire. When the reservoir is 
in the TLL phase, the occupation of the QD does not show a jump when the 
dot level crosses the Fermi energy. The interactions in the reservoir, in that case,
leave clear fingerprints on the width of the resonance. 
The main influence is a decrease in the resonance width due to a change in the 
inverse compressibility of the reservoir. On the other hand, a dot-lead interaction 
shifts the resonance position and may also enhance the width.

In the two other phases, the CDW and the FM phases, the dot level splits 
the ground state degeneracy by favoring one of the two states depending on whether 
the QD level is empty or filled. Nevertheless, we have shown that the physics in 
these cases is different. While for the FM phase a simple LC
occurs, with a sharp jump in the occupation of the level for any length of the wire,
in the CDW phase the position of the dot level drives a first order QPT in the 
thermodynamical limit between the two CDW states, while for a finite wire the jump 
is smeared. This phase transition shows all the hallmarks of a first order QPT, 
such as a size dependence, a jump in the order parameter and a discontinuity of the 
derivative of the grand canonical potential.

\begin{center}
*~~~~~*~~~~~*
\end{center}

One of the results discussed in this chapter is the abrupt jump of $n_{dot}$ in the 
FM and the CDW phases, which results from the degeneracy of the ground state at 
$\epsilon_0=0$. In the TLL phase, as noted, no such a jump was demonstrated, and the
dot occupation has a smooth analytic form.
Nevertheless, Furusaki and Matveev have predicted \cite{matveev} that a jump in the 
dot population should appear in the TLL phase as well, if the parameter $\kappa$, 
which describes the interaction in the TLL (see section \ref{sec:int_TLL}), is 
smaller than $\frac{1}{2}$.

In the model for a 1D lead which we treat in this chapter, that of spinless fermions 
with a NNs interaction term, the TLL phase is restricted to the regime 
$-2t < I < 2t$, and the relation between the interaction strength $I$ and the TLL 
interaction parameter $\kappa$ is given, for a half filled lead, by \cite{g_formula}
$\kappa=\frac {\pi} {2 \cos ^{-1} [-I/(2t)]}$.
Therefore, the values of $\kappa$ are limited to the regime $\kappa > \frac{1}{2}$. 
At the point in which $k$ approaches $\frac{1}{2}$ the lead forms a CDW, and is no longer
described by the TLL theory. Therefore, by using the NN interactions model of spinless 
fermions, with a half filled lead, one cannot get the Furusaki-Matveev jump.

When the lead is not half filled, however, a CDW phase is usually not formed, even for strong 
interactions, since such a transition is based on umklapp processes which occur only 
for commensurate fillings. Whereas there is no analytic relation between $\kappa$ and 
$I$ for systems which are not half filled, yet numerical studies of Haldane \cite{haldane80} 
point out that a regime in which $\kappa < \frac{1}{2}$ exists, when the interactions are 
strong enough, and the lead occupation is not too far from half filling. 

Using a numerical method, however, it is a bit difficult to obtain this regime. 
The ground state of the TLL phase, with strong interactions ($I>2t$) and near half filling, 
is basically built upon an half filled CDW, with some extra electrons which convert its
behavior to a TLL. The number of these extra electrons is very important, since the TLL description 
assumes a linearization of the energy curve near the Fermi points (see chapter \ref{cpt:intro}), and for a small number of 
particles it is not a good enough approximation. Therefore, unless one uses a very long lead 
(which might be computationally difficult), the strong interactions can cause the CDW ordering 
to be more favorable than a non-accurate description of the lead as a finite TLL.
As a result, the TLL description, which is true for an infinite lead, is replaced
by a finite-size CDW. 

There are also some modifications of the 1D lead model which may be used to explore the 
$\kappa < \frac{1}{2}$ regime. For example, one can enlarge the interaction range, e.g. 
by considering also next-nearest-neighbor (NNN) interactions. Such a model was used in some 
previous studies, especially at half filling \cite{poilblanc97,tsiper97,zhuravlev99}, 
in which it results in three different phases 
as functions of $V_1$ and $V_2$ (denoting the NN and the NNN interaction strengths, accordingly). 
For $V_1 \gg V_2$ the ground state is a two-fold degenerate $2k_F$ CDW, i.e., the occupation seems like 
$\cdots \bullet \circ \bullet \circ \cdots$. The other limit, in which $V_2 \gg V_1$, 
results in a four-fold degenerate $k_F$ CDW creation, with sites occupancy such as 
$\cdots \bullet \bullet \circ \circ \bullet \bullet \circ \circ \cdots$. The intermediate 
regime, including the frustration line $V_1 = 2V_2$, is described by a TLL, 
at least for not very strong interactions \cite{zhuravlev99}.

As in the NN model, the half filled case in the NNN model might also be limited to 
$\kappa > \frac{1}{2}$, as the numerical studies suggest. Nevertheless, when the system is not 
half filled, the case of $\kappa < \frac{1}{2}$ can be reached by this model with 
much smaller values of interaction \cite{zhuravlev01}, in which the CDW excitations 
are weaker, thus shorter leads might suffice.

Another possibility, which seems the most promising, is to consider spin $1/2$ electrons. In this 
case the interactions parameter has the analytic form $\kappa^{{\rm spin} 1/2} = \kappa^{\rm spinless}/2$, 
resulting, for a half filled lead, in the regime $\kappa > \frac{1}{4}$. With this model any 
non-vanishing value of repulsive interactions result in $\kappa < \frac{1}{2}$, and can 
demonstrate the Furusaki-Matveev jump.

The required modifications of the DMRG process for NNN model or for spin $1/2$ electrons are discussed
in section \ref{sec:num_nnn}. Here we just note that whereas in the spinless NN model there is an 
additional one state in each iteration, these modifications demand an iterative addition of a couple of states,
whether they are two sites with spinless electrons, or a single site which might be doubly occupied.
Therefore, in order to obtain the same accuracy, the DMRG block sizes, and thus the computational 
resources which are required, are larger. The search for the Furusaki-Matveev jump is thus left
for a future study.

\cleardoublepage
\chapter[A QD coupled to a {\textit disordered} 1D lead]
{A level coupled to a {\it disordered} 1D interacting reservoir}
\label{cpt:ch2}

In this chapter we calculate numerically the Friedel oscillations caused due to an impurity 
located at one edge of a disordered interacting quantum wire.
The electron density in the system's ground state is determined using the 
density-matrix renormalization-group method,
and the Friedel oscillations data is extracted using the density difference
between the case in which the wire is coupled to an impurity and
the case where the impurity is uncoupled.

For the non-interacting case, we develop an exact formula of the Friedel oscillations
in the one-dimensional tight-binding model. The excellent fit to
the numerical results serves as a proof for the accuracy of the method.

For a one-dimensional wire which is described by Tomonaga-Luttinger liquid theory the 
oscillations of a clean interacting sample decay as a power law. We show that 
once the wire moves into an Anderson insulator phase, which happens due to the 
introduction of a disorder, the power law decay 
is multiplied by an exponential decay term due to the disorder.
Scaling of the average Friedel oscillations by this exponential term collapses the
disordered samples data on the clean results. The decay length is shown 
to decrease as a function of the interaction strength.

However, when a short enough mesoscopic wire is in a charge density wave phase, 
the presence of weak disorder may not destroy the long range order, so that the wire
will be described by a disordered Mott insulator phase. In this case we find that 
the effect of interactions is the opposite, and the disorder significance
decreases as the interactions strength increases.

We prove that the length scale governing the exponential decay, in the Anderson insulator phase, may be
associated with the Anderson localization length and thus be used as a convenient
way to determine the dependence of the localization length on disorder and
interactions. Our results, which show a decrease of the localization length as a function 
of the interaction strength, are in accordance with previous predictions.

\section{Introduction}
The interplay between repulsive interactions and disorder in low-dimensional systems,
and their influence on the conductivity, were the subjects of many studies in recent years.
Some of this interest was
motivated by the experimental observations of a crossover in
the temperature dependence of the conductance of low density two-dimensional
electrons from an insulating-like dependence
at low densities to a metallic one at higher densities \cite{reviews}.
Nowadays it is generally accepted that even if such a two-dimensional metal-insulator
transition exists, it must be related to the spin degree of freedom \cite{punnoose05}
and therefore absent for spinless electrons.

It seems therefore clear that for spinless one-dimensional (1D) systems
no metal-insulator transition is expected for repulsive interactions, although
for a certain range of attractive interactions a delocalized regime was found in
several studies \cite{schmitt1}.
Nevertheless, it was shown that there might be a certain strong disorder
and interaction regime, in which there is an increase of the localization length $\xi$, 
defined through the dependence of the zero temperature conductance on the
system size: $g(L) \sim \exp(-L/\xi)$. Similar behavior was demonstrated
also in properties which are usually related to $\xi$ such as the persistent current
\cite{abraham93}. A sample dependent increase 
in the localization length was also reported for weaker values of disorder and
repulsive interactions for longer (of order of $100$ sites) wires \cite{pichard}.

On the other hand, several analytic studies which were performed in the Tomonaga-Luttinger liquid (TLL) framework \cite{giamarchi03} 
have concluded that the localization length of a 1D wire decreases monotonically with increasing 
repulsive interaction. Using either renormalization group \cite{apel_82} or
self consistent Hartree-Fock \cite{suzumura_83} methods it was shown that 
the localization length, renormalized by the interaction, scales as
\begin{eqnarray} \label{eqn:xi_I}
\xi(\kappa) \sim (\xi_0)^{1/(3-2\kappa)},
\end{eqnarray} 
where $\xi_0$ is the localization length of the free electron system, and $\kappa$ 
is the TLL interactions parameter (see chapter \ref{cpt:intro}) with $\kappa=1$ for 
non-interacting electrons. Since for repulsive interactions $\kappa$ decreases as a 
function of the interaction strength, one finds that the localization length 
always decreases as a function of the interaction strength.

One must be careful though to differentiate between weak and strong interaction strength. 
A clean 1D system of spinless fermions on a lattice undergoes a metal-insulator 
phase transition between a TLL and a Mott Insulator (MI) as a function of the interaction strength.
The MI phase, for strong interactions, appears for spinless 1D electrons as a $2k_F$ charge density wave (CDW). 
This phase transition, caused by umklapp processes, is exhibited for commensurate fillings.
Once disorder is turned on, the TLL transport properties change drastically.
For more than a decade it is well known \cite{kane92} that the
conductivity of a TLL wire vanishes in the presence of impurities, 
thus a metal-insulator transition as a function of interaction strength no longer exists.

Nevertheless, a difference between the TLL and the CDW phases may still exist, even in the presence of disorder.
When the interactions are weak, so that the wire is in a TLL phase, the addition of disorder 
turns the metallic system into an Anderson insulator (AI). However, for strong interactions (i.e., when the clean 
system is a CDW, which is a MI) the exact effect of disorder depends on its strength, and 
in general is not completely understood. While for clean systems the MI phase is a well 
studied problem \cite{mott90}, the addition of disorder opens a few questions, which have 
attracted several studies in the last decade \cite{pang93,fujimoto96,mori96,orignac99,giamarchi01}.

When the disorder is strong, i.e., when the random potential felt by the electron 
is much larger than any other energy scale in the problem, the MI state is destroyed,
and an AI phase emerges. For weak disorder, however, it was shown in several studies that 
the Mott energy gap vanishes only when a finite disorder is introduced, so that
below this critical disorder the MI phase is stable \cite{ma,sandvik94}. 
Usually this is not the case for a MI consisting of spinless particles, such as the CDW we study,
since an Imry-Ma type of argument \cite{imry_ma} shows that the long range order is destroyed 
even for an infinitesimal disorder \cite{shankar}. Yet, for a finite sized mesoscopic sample, 
the Imry-Ma length scale might be a few orders of magnitude larger than the sample's size,
so that the effective ground state for a weak enough disorder remains a MI one.
Increasing the disorder above a critical strength changes the MI state either
to a Mott glass\footnote[1]{This issue will be addressed at the last section of this chapter.} 
or to an AI \cite{orignac99,giamarchi01}. 

Therefore, a finite size CDW state is expected to remain stable against the application 
of a weak enough disorder, i.e., to remain a MI state. For example, previous numerical 
simulations have presented the long range order of such a weakly disordered CDW \cite{pang93}. 
Furthermore, such finite 1D wires coupled to dots have been recently
manufactured, and signatures of a CDW in strong magnetic fields have been observed \cite{pepper04}. 
In order to verify numerically the existence of a CDW order in the presence of disorder 
for the length scales considered, one can check the electron density of the entire system.

Most studies on disordered 1D wires concentrate on either the AI or the MI phases, 
thus a full comparison between the two regimes is still lacking. 
Nevertheless, a qualitatively different behavior between these two regimes was 
demonstrated in a few cases. For example, the 
effect of interactions on the persistent current in 1D disordered rings
was calculated in previous works \cite{abraham93,bouzerar94}, and an important
difference between the AI and MI phases was found. 
While for strong interactions and weak disorder (MI phase) 
the persistent current was reduced, for strong disorder (AI phase) an increase
of the current was found. However, the exact diagonalization techniques which 
were used in these studies, are applicable only for very small system sizes.

In this chapter we investigate the influence of interactions on the Friedel oscillations (FO) 
in a disordered 1D wire, and compare this behavior between the AI and the MI regimes.
We study interacting spinless electrons confined to a 1D wire which can be in either 
its AI or MI phases.
The entire parameter range is very relevant to the explanation of 
recent transport measurements in various 1D or quasi 1D systems, such 
as single-wall carbon nanotubes \cite{swnt1,swnt2,swnt3}, multi-wall carbon nanotubes \cite{mwnt}, 
polymer nanofibers \cite{polyfiber} and MoSe nanowires \cite{mose}, which have led
to many theoretical works \cite{malinin04,mirlin05,mirlin052,mora06,kwapinski06}.
With this in mind, we study the effect of the interplay of interactions 
and disorder on the behavior of the FO in a wire due to its coupling to an impurity at its edge.

\subsection{Friedel Oscillations}
Once a single-level impurity (dot) is coupled to a clean metallic system 
the density of electrons in its vicinity oscillates with a $2k_F$ period, and 
the envelope of the oscillations decays as a power law of $r$, the distance from the impurity \cite{friedel}.
For non-interacting systems the perturbation of the density in the vicinity of the
impurity depends on the dimensionality, $d$, of the system, and can be expressed as
\begin{eqnarray} \label{eqn:rho_frid}
\delta \rho(r) = A \frac {\cos(2k_F r+\eta)} {|r|^d},
\end{eqnarray}
where the coefficient $A$ and the phase shift $\eta$ do not depend on $r$. These oscillations
are the famous Friedel oscillations, which have been observed
experimentally during the last decade using various techniques, such as
scanning tunneling microscopy in low temperatures \cite{wielen}
and X-Ray diffraction \cite{brazovskii}.

Whereas for higher dimensions ($d \ge 2$) Eq.~(\ref{eqn:rho_frid}) is in 
general true even in the presence of interactions, this is not the case for 1D systems.
For the TLL phase, using field theoretical approaches, it was shown \cite{frid_LL}
that the $x^{-1}$ dependence is replaced by a different power law, $x^{-\kappa}$.
For the non-interacting case $\kappa=1$, it leads to the expected $x^{-1}$ decay,
while for repulsive interactions $\kappa<1$ and thus a slower decay of the FO envelope
is expected.

Therefore, the density change in the TLL phase, which is metallic, shows FO with a $2k_F$ wave vector 
and a power law decay. In the CDW phase the picture is different, since the CDW phase is insulating,
and thus the power law is replaced by an exponential decay. 
The length scale of this exponential decay is related to
the CDW correlation length \cite{mikeska04}, $\zeta$. 

In the presence of a weak disorder, one can expect to find a similar decay to that of the 
clean case, with an additional exponential decay due to the disorder. 
By calculating this exponential decay we are able to present a clear picture of the dependence 
of the decay length due to disorder on interactions, in both the AI and MI regimes.
In the following sections we show a different behavior of this decay length in the two 
phases, and the origin of this difference is explained.

From Eq.~(\ref{eqn:rho_frid}) it is clear that the observation of the density
fluctuations, either experimentally or numerically, is easier 
in the vicinity of the impurity. 
When disorder is also introduced, this distance becomes
even shorter since there are also density fluctuations caused by the disorder.
Yet, in common experimental 1D situations disorder is usually present.
Therefore, although the presence of disorder hampers observing the FO, it
is beneficial to develop a method to tease the FO out of the density
fluctuations of a disordered system.

\subsection{Localization Length in the TLL Phase}

In general, the study of a dot (or impurity)
coupled to a 1D lead, has been shown to shed  
light over the physics of the lead. Certain thermodynamic observables, such as the 
occupation of the impurity level (see for example chapter \ref{cpt:ch1} and Ref.~\cite{matveev})
and the corresponding electron density changes in the lead \cite{schmitt2,andergassen}, 
were recently used to analyze different wire properties, such as the strength and form 
of the interactions, and even the wire's phase (e.g., TLL vs. CDW). In a similar fashion, 
we show in this chapter how the electron density of a disordered wire in the TLL phase, 
coupled to an impurity level, can be used in order to extract its localization length.

In the TLL case, for weak enough forward scattering we will show in the following that 
the decay length of the density-density correlations is equivalent to the localization length. 
It is important to note that the extraction of the localization length for
interacting systems is plagued with difficulties. The straightforward
method of measuring the decay length of the envelope of the single-electron
state has no direct translation to a many-electron state.
Nevertheless, one would prefer to stick to a ground-state property of the
system, since the calculation of excited state dependent properties such as the 
conductance is computationally taxing. The sensitivity to boundary conditions
(i.e., persistent current) which is the natural candidate for a ground-state property 
is problematic since it incorporates both interaction corrections to the localization
length as well as interaction corrections to the inverse compressibility of the
system \cite{berkovits96}. Separating the two is not easy, while computationally
it requires both a calculation of the sensitivity of the ground state to flux,
as well as the dependence of the number of electrons in the system on the chemical 
potential.

Therefore, the study of the influence of interaction on the FO in the
Anderson phase is not only interesting on its own account, but it establishes a new
numerical method using a ground-state property which is convenient for a 
direct evaluation of the localization length for not too strong disorder. Using this method 
we show that the localization length decreases as a function of the interaction strength, 
in correspondence to Eq.~(\ref{eqn:xi_I}).

\subsection{Chapter's Outline}
The rest of this chapter is organized as follows. In section \ref{sec:FO_model} the model
Hamiltonian and its diagonalization method are presented, followed by a description of the
method used to extract the FO data from the ground-state wave functions.
In section \ref{sec:FO_I0} an analytic calculation is preformed for the non-interacting case,
resulting in an exact formula for the FO for a 1D tight-binding model. 
Next we present the results for the two phases. The results for the TLL phase, which
is replaced by an AI phase once disorder is introduced, are shown in section \ref{sec:FO_TLL}.
The fact that the decay length can be used to approximate the localization length in the AI regime 
is also proven in this section.
The results for the MI regime are shown in section \ref{sec:FO_CDW}, where we also provide an
explanation to the difference found between the two phases. We conclude in section \ref{sec:FO_con},
in which we point out some of our planned future studies.

\section{Model}
\label{sec:FO_model}
\subsection{Hamiltonian}
The system under investigation is composed of spinless  
electrons on a 1D lattice coupled to an impurity in one end. 
We model the 1D wire by a lattice
of size $L$ with repulsive nearest neighbor (NN) interactions and with an on-site disorder. 
The system's Hamiltonian is thus given by
\begin{eqnarray} \label{eqn:H_wire}
{\hat H_{wire}} &=& 
\displaystyle \sum_{j=1}^{L} \epsilon_j {\hat c}^{\dagger}_{j}{\hat c}_{j}
-t \displaystyle \sum_{j=1}^{L-1}({\hat c}^{\dagger}_{j}{\hat c}_{j+1} + H.c.) \\ \nonumber
&+& I \displaystyle \sum_{j=1}^{L-1}({\hat c}^{\dagger}_{j}{\hat c}_{j} - \frac{1}{2})
({\hat c}^{\dagger}_{j+1}{\hat c}_{j+1} - \frac{1}{2}),
\end{eqnarray}
where $\epsilon_j$ are the random on-site energies, taken from a uniform 
distribution in the range $[-W/2,W/2]$,
$I$ is the NN interaction strength ($I \ge 0$), and $t$ is the
hopping matrix element between NNs, henceforth taken as unity.
${\hat c}_j^{\dagger}$ (${\hat c}_j$) is the creation (annihilation)
operator of a spinless electron at site $j$ in the lattice, and
a positive background is included in the interaction term.

Without the disorder term, a similar system - in the limit $L \rightarrow \infty$ and with
periodic boundary conditions - has a well known exact solution. Depending 
on the interaction strength, the wire can be either metallic or insulating. As was detailed in chapter
\ref{cpt:intro}, the metallic
phase is described by TLL, occurring for $I<2t$, and the insulating
phase, in which $I>2t$, is a CDW. In chapter \ref{cpt:ch1} we have shown that
wires of the order of a few hundreds sites lead to a
similar phase diagram, even when employing open boundary conditions.

Introducing an impurity at one end of the wire results in adding 
the following term to the Hamiltonian:
\begin{eqnarray} \label{eqn:H_imp}
{\hat H_{imp}} = \epsilon_0 {\hat c}^{\dagger}_{0}{\hat c}_{0} 
-V ({\hat c}^{\dagger}_{0}{\hat c}_{1} + H.c.) 
+ I ({\hat c}^{\dagger}_{0}{\hat c}_{0} - \frac{1}{2})
({\hat c}^{\dagger}_{1}{\hat c}_{1} - \frac{1}{2}),
\end{eqnarray}
where $\epsilon_0$ describes the impurity strength, 
and $V$ is the hopping matrix element between the impurity and the first lead site.
Along this chapter we use $\epsilon_0 \gg W$ and $V=t$.

The resulting Hamiltonian $\hat H = \hat H_{wire} + \hat H_{imp}$
describes a disordered 1D wire of length $L$ ($1 \le j \le L$), 
which is coupled to a single level at one of its edges ($j=0$).
Practically the $j=0$ site is equivalent to any other site, except for having
a constant on-site energy, whereas the other sites have energies drawn from a
distribution with a zero average over different realizations.

\subsection{Diagonalization Method}
The Hamiltonian $\hat H$ was diagonalized using the finite-size density-matrix renormalization group (DMRG) method
\cite{white93}, and the occupation of the lattice sites were calculated,
for different values of $\epsilon_0$, $W$ and $I$. The size of the wire was up to $L=500$ sites.
During the renormalization process the number of particles in the system is not fixed, 
so that the results describe the experimentally realizable situation of a
finite section of a 1D wire which is coupled to a dot and to an external electron reservoir.
Yet, the calculated density remains close to half filling in all the calculated
scenarios (even in the presence of disorder)
since the interaction term contains a positive background, 
and the calculation is done for $\mu=0$.

\subsection{Extracting the Friedel Oscillations Decay}
\begin{figure}[htb]\centering
\includegraphics[trim=0mm 0mm 0mm 0mm, clip, width=8cm]{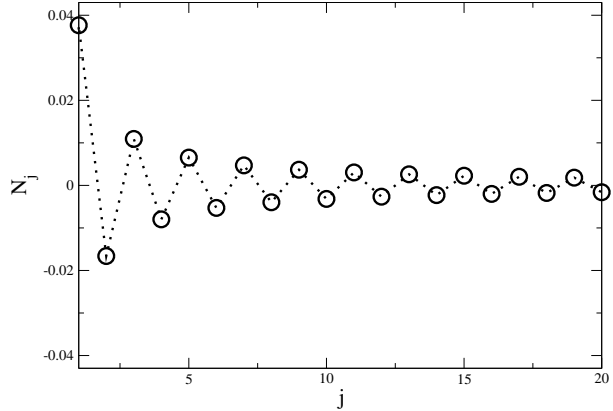}
\caption[A typical form of Friedel oscillations]
{\label{fig:fo_sample}
A typical form of Friedel oscillations, without disorder and without interactions.
The results shown are for $\epsilon_0=10$ and $L=280$. 
The impurity is located at $j=0$ and the population 
of the first 20 lead sites is shown.
}
\end{figure}

In the TLL phase, when no disorder is present ($W=0$), $\hat H_{wire}$ has a
particle-hole symmetry, and the particle density of the wire's 
ground state is flat, with filling factor $n=1/2$. In this case $2k_F = \pi$ and
the oscillating part of Eq.~(\ref{eqn:rho_frid}) alternates according to $(-1)^j$.
Denoting by $n^{wire+imp}_j$ ($n^{wire}_j$) the electron density at site $j$ of the wire
when coupled (not coupled) to the dot, one has $n^{wire}_j=n=1/2$ for any $j$.
Clearly this is not the case in the presence of the impurity, and
the effect of the impurity is measured by $N_j \equiv n^{wire+imp}_j - n$.
A typical result of $N_j$, showing the $2k_F$ oscillations caused by the impurity at $j=0$,
is presented in Fig.~\ref{fig:fo_sample}. 

\begin{figure}[htb]\centering
\includegraphics[trim=0mm 0mm 0mm 0mm, clip, width=8cm]{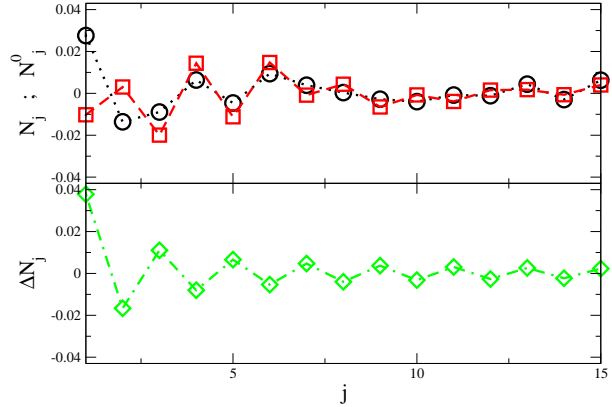}
\caption[Typical Friedel oscillations with a disorder (AI)]
{\label{fig:frid_dis_sample}
Typical FO for a disordered sample with $L=280$, $W=0.1$ and $\epsilon_0=10$
(without interactions, TLL phase). 
The upper panel shows $N_j$ (circles) and $N^0_j$ (squares),
and the lower panel presents the difference between them ($\Delta N_j$).
The FO are observed much better using $\Delta N_j$ instead of $N_j$.
}
\end{figure}

When $W \ne 0$, on the other hand, although the average
filling factor is still $n \approx 1/2$,
there is no local symmetry between particles and holes, and 
the disorder effects are seen in the fluctuations of the electron density.
The density oscillations generated by the additional impurity are then difficult to discern,
since in a distance of a few lattice sites from the impurity the disorder fluctuations are dominant. 
A typical result of $N_j$ together with $N^0_j = n^{wire}_j - n$ 
(the electron density of the disordered wire without an impurity), 
is shown in the upper panel of Fig.~\ref{fig:frid_dis_sample}.

However, the influence of the impurity can be observed by isolating the density
fluctuations created by the disorder. This is achieved by 
comparing the electron density of the two cases shown in the upper panel of Fig.~\ref{fig:frid_dis_sample}, 
i.e., one with the additional impurity and the other without it, for every disorder realization.
Averaging over realizations is thus done for 
\begin{eqnarray} \label{eqn:Y_j_1}
\Delta N_j \equiv N_j - N^0_j = n^{wire+imp}_j - n^{wire}_j,
\end{eqnarray}
instead of just averaging over $N_j$. 
The curve of $\Delta N_j$ in the lower panel of Fig.~\ref{fig:frid_dis_sample} is for the 
same realization as in the upper panel.
It is obvious that the FO which were hardly seen for $N_j$ become clear 
once $\Delta N_j$ is considered.

In the CDW phase, when no disorder is considered ($W=0$), the ground state of the system 
is twofold degenerate. This degeneracy is broken, however, once a pinning impurity, denoted 
by $\epsilon^{(0)}_0 \rightarrow 0^+$, is coupled to one end of the wire, and the particle 
density ($N^0_j$, for $j=1 \dots L$) shows a $2k_F$ modulation (see chapter \ref{cpt:ch1}). 
This is different from the TLL phase, in which the density without the dot is flat.

When the pinning impurity is replaced by a dot level with $\epsilon_0 \gg \epsilon^{(0)}_0$, 
the particle density in the wire ($N_j$) is changed by an additional oscillating $2k_F$ term.
One should notice that once the dot is coupled, a new CDW state emerges, having also
$2k_F$ oscillations, but with a different
amplitude. The difference between these two states has a  
$2k_F$ oscillation, with an exponential decay from its value at the edge of
the wire. 

\begin{figure}[htb]\centering
\includegraphics[trim=0mm 0mm 0mm 0mm, clip, width=8cm]{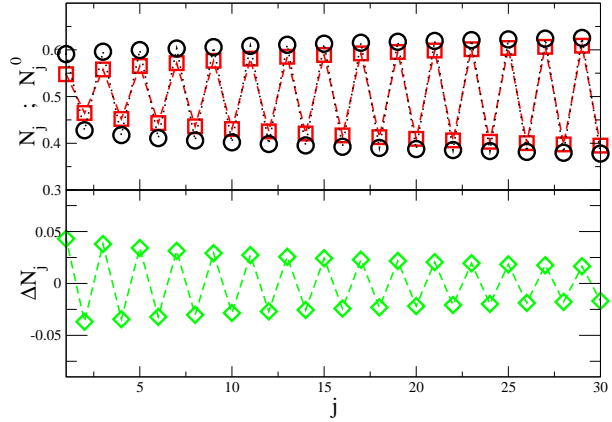}
\caption[Typical Friedel oscillations (MI)]
{\label{fig:fo_sample_W0}
Typical FO for a clean sample with $L=280$ 
for a CDW with $\epsilon_0=10$ and $I=2.5$. 
The upper panel shows $N_j$ (circles) and $N^0_j$ (squares),
and the lower panel presents their difference $\Delta N_j$.
}
\end{figure}

In order to calculate the density difference between the cases when the
quantum dot is coupled or uncoupled to the wire, it is easy to convince oneself that the
definition of $\Delta N_j$ in Eq.~(\ref{eqn:Y_j_1}) should work also in the CDW case.
A typical result of $N_j$ vs. $N^0_j$ for a CDW state, and the resulting $\Delta N_j$, 
showing the $2k_F$ oscillations caused by the dot orbital at $j=0$,
is presented in Fig.~\ref{fig:fo_sample_W0}. 

When $W \ne 0$, the CDW ground state is no longer degenerate, and the infinitesimal pinning impurity is
not required. The disorder itself pins the CDW to different places on the lattice,
with the ability to break the long range order of the clean CDW by localized solitons, 
with a density which depends on the disorder strength \cite{pang93}.
Yet, when a dot level with $\epsilon_0 \gg W$ is connected to one side 
of the wire, the local effect in its vicinity is stronger than the 
pinning caused by the disorder. This results in a change of the particle 
density near the dot, and this change decreases with distance. It 
turns out that the definition of $\Delta N_j$ in Eq.~(\ref{eqn:Y_j_1})
is suitable for the disordered case as well, since it cancels out the 
disorder pinning effects which are the same for the two cases, isolating 
the density fluctuations created by the dot.

\begin{figure}[htb]\centering
\includegraphics[trim=0mm 0mm 0mm 0mm, clip, width=8cm]{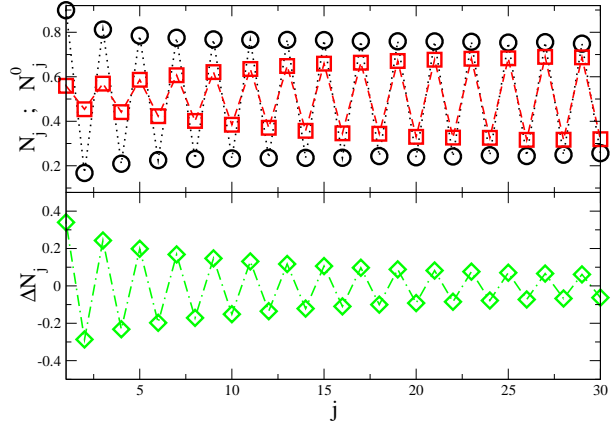}
\caption[Typical Friedel oscillations with a disorder (MI)]
{\label{fig:CDW_frid_sample}
Typical FO for a single disordered sample with $L=280$, $W=0.1$ and $\epsilon_0=10$, 
for a CDW with $I=3$. 
The upper panel shows $N_j$ (circles) and $N^0_j$ (squares),
and the lower panel presents their difference $\Delta N_j$.
}
\end{figure}

A typical picture of $\Delta N_j$ for a disordered CDW sample is presented in 
Fig.~\ref{fig:CDW_frid_sample}. Whereas the upper panel shows the density of the
two similar systems, one which is coupled to the quantum dot and the other is not, the lower panel presents
the difference between these two densities, and the decay of the oscillations can be clearly seen.

\section{Non-Interacting Case}
\label{sec:FO_I0}
\subsection{Exact Calculation of Friedel Oscillations}
We now wish to exactly calculate 
$N(m)$, the density of electrons in site $m$, of a half filled 1D 
tight-binding lead, which is coupled to an impurity, in the asymptotic 
($m \gg 1$) limit. The system is described by the Hamiltonian
\begin{eqnarray} \label{eqn:H_app}
{\hat H} = 
\displaystyle 
\epsilon_0 {\hat c}^{\dagger}_{0}{\hat c}_{0} 
-V ({\hat c}^{\dagger}_{0}{\hat c}_{1} + H.c.) 
-t \displaystyle \sum_{j=1}^{L-1}({\hat c}^{\dagger}_{j}{\hat c}_{j+1} + H.c.).
\end{eqnarray}

$N(m)$ can be calculated using the
retarded Green function of an electron in the m'th site $G^R (\omega;m,m)$
\cite{mahan}, and the relation (for a half filled band)
\begin{eqnarray} \label{eqn:N_G}
N(m) = - \frac {1}{\pi} \Im \int_{-2t}^0 G^R(\omega;m,m)d\omega,
\end{eqnarray}
where we are possibly neglecting bound states with energy lower than $-2t$, which give exponentially
small contributions for large m.
The Green function itself is determined by 
\begin{eqnarray} \label{eqn:G_mm}
G^R(\omega;m,m) &=& G_0^R(\omega;m,m) + \\ \nonumber
 G_0^R(\omega;m,1) ~~ \cdot &V& \cdot ~~ G(\omega;0,0) ~ \cdot ~ V ~ \cdot ~ G_0^R(\omega;1,m).
\end{eqnarray}

In this expression $G_0^R(\omega;m,l)$ is the bare (i.e., without dot) 
lead Green function, while
$G(\omega;0,0)=(\omega-\epsilon_0-\Sigma(\omega))^{-1}$ is the dot's Green function,
where $\Sigma(\omega) = \frac {V^2}{t} ( \frac {\omega}{2t} - i \sqrt{1-(\frac {\omega}{2t})^2})$
is the self energy of the dot, which was calculated in chapter \ref{cpt:ch1}. 
The first term in the RHS of the equation simply gives the constant $n=1/2$ occupation in the 
absence of the dot.

Substituting the known wave functions and energies of the tight-binding Hamiltonian
one finds
\begin{eqnarray} \label{eqn:G0_ml}
G_0^R(\omega;m,l) = 
\frac {a}{L} &\displaystyle \sum _{k > 0}& \frac {\cos(ka(m-l)) - \cos(ka(m+l))} {\omega+2t \cos(ka)},
\end{eqnarray}
where $k=\frac{\pi}{L}n_k$, for integer $n_k$.
Transforming to integration over unit circle in the complex plane leads to
\begin{eqnarray} \label{eqn:G0_m1}
G_0^R(\omega;m,1) = -\frac{1}{t} \left[ -\frac{\omega}{2t}+i \sqrt{1-(\frac{\omega}{2t})^2} ~ \right] ^m.
\end{eqnarray}

Combining Eqs.~(\ref{eqn:N_G}), (\ref{eqn:G_mm}) and (\ref{eqn:G0_m1}), one can get,
\begin{eqnarray} \label{eqn:DN_m}
\Delta N(m) = N(m) - 1/2 = 
-\frac{V^2}{\pi t^2} \Im \int_{-2t}^0 d\omega
\frac {\big{(} -\frac{\omega}{2t}+i \sqrt{1-(\frac{\omega}{2t})^2} \big{)} ^{2m} }
{ \omega - \epsilon_0 - \frac{V^2}{t} 
\big{(} \frac{\omega}{2t} - i \sqrt{1-(\frac{\omega}{2t})^2} \big{)} },
\end{eqnarray}
and by substituting $\omega = -2t \cos \theta$, we find
\begin{eqnarray} \label{eqn:DN_m2}
\Delta N(m) = \frac{V^2}{\pi t i} \int_{-\pi/2}^{\pi/2} d\theta
\frac {\sin(\theta) e^{i2m\theta}}
{2t \cos (\theta) + \epsilon_0 - \frac{V^2}{t} e^{i\theta}}.
\end{eqnarray}

\begin{figure}[htb]\centering
\includegraphics[trim=0mm 0mm 0mm 0mm, clip, width=3cm]{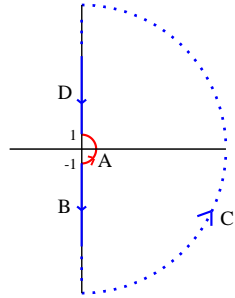}
\caption[Integration contour for the calculation of Friedel oscillations]
{\label{fig:contour}
The integration contours A and B-C-D which connect the points $[0,-1]$ and $[0,1]$.
}
\end{figure}

One now defines $z=e^{-i\theta}$ in order to get 
\begin{eqnarray} \label{eqn:DN_m3}
\Delta N(m) = -\frac{V^2 i}{2 \pi t^2} \int_A \frac {dz}{z}
\frac {(z^2-1)z^{-2m}}
{z^2+\epsilon_0 z/t + 1-V^2/t^2},
\end{eqnarray}
where the integration is over the right half of the unit circle, between the points 
$\pm 1$ on the imaginary axis (contour A in Fig.~\ref{fig:contour}).

Next we deform our contour to the contour B-C-D in Fig.~\ref{fig:contour}.
In doing so we neglect the contribution of poles which may occur inside the closed 
line A-B-C-D. These represent states bound at the impurity, and as we have mentioned above,
contribute exponentially small terms for large $m$.
The integration in parts B and D is done by defining $z=\pm ix$, respectively, $x \in [1,\infty)$,
while the contribution of the semicircle C vanishes as its radius goes to infinity.
Therefore we get
\begin{eqnarray} \label{eqn:DN_m4}
\Delta N(m) = 
\frac{V^2}{\pi t^2} (-1)^m \Im \int_1^\infty 
\frac {(x^2+1)}
{x^2+i\epsilon_0 x/t - 1+V^2/t^2}
x^{-2m-1} dx.
\end{eqnarray}

For $m \gg 1$ the term $x^{-2m-1}$ varies much faster than the other terms,
and the rest of the integrand can be
evaluated at $x \approx 1$ to give $\frac {2}{V^2/t^2+i\epsilon_0/t}$. One thus gets the final form 
\begin{eqnarray} \label{eqn:N_j_exact2}
\Delta N(m) = \frac {(-1)^{m+1}}{\pi m} \big ( \frac {\epsilon_0 t}{V^2} + \frac {V^2}{\epsilon_0 t} \big ) ^{-1}.
\end{eqnarray}

\subsection{Numerical Results}
In order to check the accuracy of the FO data obtained by the DMRG method, we
begin by a comparison of the numerical results of the clean non-interacting case, to the
exact formula we've just calculated. For our model, substituting $V=t$ in Eq.~(\ref{eqn:N_j_exact2}), 
one gets the amplitude of the FO as
\begin{eqnarray} \label{eqn:A_exact}
A(\epsilon_0) = - \frac {1}{\pi} \left( \epsilon_0/t + 
\frac {1}{\epsilon_0/t} \right) ^{-1}.
\end{eqnarray}

\begin{figure}[htb]\centering
\includegraphics[trim=0mm 0mm 0mm 0mm, clip, width=7cm]{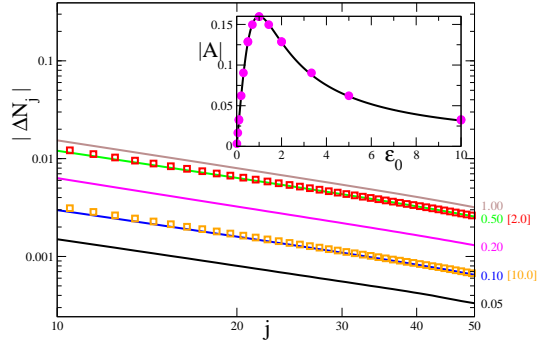}
\caption[Friedel oscillations decay for clean samples (without interactions)]
{\label{fig:A_E}
The FO decay (in log-log scale) for different values of $\epsilon_0$ (shown next
to the curves) for clean samples without interaction. 
The slope - representing the decay exponent  - is constant, 
and the only effect of $\epsilon_0$ is a change of the amplitude $A$.
The curves for $\epsilon_0=2.0,10.0$ are drawn with symbols.
Inset: the dependence of $A$ on $\epsilon_0$ for $V=t$
together with the exact formula Eq.~(\ref{eqn:A_exact}). 
}
\end{figure}

As expected, the amplitude $A$ does depend on $\epsilon_0$, and this dependence is presented 
in Fig.~\ref{fig:A_E}. Curves obtained for different values of $\epsilon_0$ have the same slope,
but not the same amplitude. The limits of $\epsilon_0 \rightarrow 0$ and $\epsilon_0 \rightarrow \infty$ 
are well understood, because in both of them the impurity does not play any role, the lead has a 
hard wall boundary, and the particle-hole symmetry imposes that the FO amplitude goes to zero.
For finite values of $\epsilon_0$, the behavior of the amplitude shown in the inset of Fig.~\ref{fig:A_E} 
is compared to the exact relation Eq.~(\ref{eqn:A_exact}).
The correspondence between the numerical results and this formula is excellent.
\section{From the TLL Phase Towards Anderson Insulator}
\label{sec:FO_TLL}
\subsection{A Clean Lead: Extracting the TLL Parameter}
\begin{figure}[htb]\centering
\includegraphics[trim=0mm 0mm 0mm 0mm, clip, width=8cm,height=!]{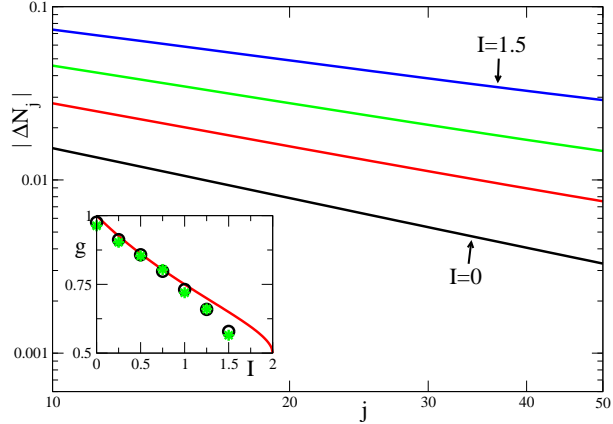}
\caption[Friedel oscillations decay for clean samples (TLL)]
{\label{fig:I_slow_decay}
The DMRG results for the FO decay in log-log scale for $I=0,0.5,1$ and $1.5$ (bottom to top), 
which correspond to the TLL phase,
with $\epsilon_0=10$ and $L=280$ and without disorder. As $I$ increases, the decay gets slower.
Inset: The interaction parameter $\kappa$ as found by fitting the FO decay to $x^{-\kappa}$ (symbols),
together with the theory prediction Eq.~(\ref{eqn:g_theory}) (line). The DMRG results were
obtained for $L=500$ with $\epsilon_0=1$ (circles) and $\epsilon_0=10$ (stars).
}
\end{figure}

We now move to the interacting case.
For $0 \le I<2t$, i.e., when the fermions in the lattice 
are described by the TLL theory, the decay is expected to be proportional \cite{frid_LL} 
to $j^{-\kappa}$.
In our model the TLL parameter $\kappa$ is given by \cite{g_formula}
\begin{eqnarray} \label{eqn:g_theory}
{\kappa=\frac {\pi} {2 \cos ^{-1} [-I/(2t)]}}.
\end{eqnarray}
For non-interacting particles one gets $\kappa=1$ so that the oscillations decay as $j^{-1}$, while in
the interacting regime a monotonic decrease of $\kappa$ toward the limit $\kappa=1/2$
occurs as a function of interaction strength. 
Thus, as $I$ becomes stronger, $\kappa$ decreases, and a slower decay is predicted. This trend
is seen in the DMRG results presented in Fig.~\ref{fig:I_slow_decay}.

In the inset of Fig.~\ref{fig:I_slow_decay}, the results obtained for $\kappa$
by fitting the FO decay of a $500$ sites wire, to the predicted decay of $x^{-\kappa}$, 
are presented together with the theory prediction for $\kappa(I)$ of Eq.~(\ref{eqn:g_theory}).
As can be seen, the results are in good accordance with the theory for
interaction strength $I/t \lesssim 1$. Similar results were obtained using 
other implementation of the DMRG method
(with a constant number of particles) \cite{schmitt2}, and by functional 
renormalization-group studies \cite{andergassen}.
In these works it was argued that for the system sizes treated, the asymptotic regime 
in which the $x^{-\kappa}$ behavior is predicted is not yet reached. In Ref.~\cite{andergassen}
it was shown that using the functional renormalization-group method, which is argued to be as accurate 
as the DMRG method, even $L$ of the order of $10^6$
is not sufficient to obtain the values of $\kappa$ of Eq.~(\ref{eqn:g_theory})
for $I/t \gtrsim 1$.

\subsection{A Disordered Lead: Extracting the Localization Length}
\begin{figure}[htb]\centering
\includegraphics[trim=0mm 0mm 0mm 0mm, clip, width=8cm]{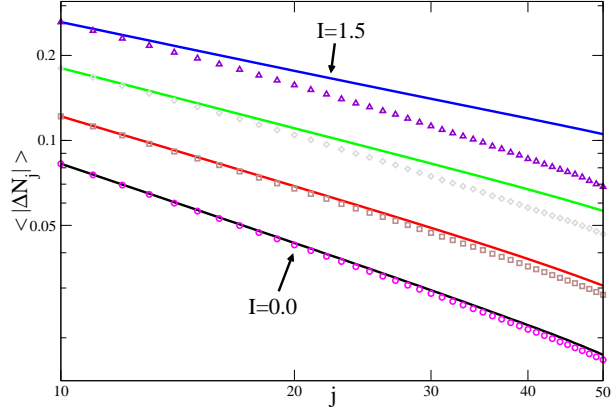}
\caption[Friedel oscillations decay for disordered samples (AI)]
{\label{fig:decay_I}
The decay of FO for $I=0.0,0.5,1.0$ and $1.5$ (bottom to top)
with $L=500$ and $\epsilon_0=10$. 
The symbols are for $W=0.1$ (AI phase) and the lines are for the clean ($W=0$) case (TLL).
The disorder effect becomes significant for large values of $I$ where the
localization length is small. The average was done over 100 realizations.
}
\end{figure}

We now turn on the disorder by taking $W \ne 0$. In this case the results of 
$\Delta N_j$ are averaged over $100$ different realizations of disorder,
which are sufficient to reduce the sampling error to less than one percent.
In Fig.~\ref{fig:decay_I} the averaged particle density for $W=0.1$
is shown and compared to the $W=0$ case for various interaction strengths. 
As can be seen, for small values of the interaction the effect of disorder is very weak,
while for large values of $I$, the FO decay faster in the presence of disorder.
Zooming into these curves, it can be shown that the effect of disorder is 
to multiply the clean FO decay by an exponential factor $e^{-x/\xi}$, 
where $\xi$ is a characteristic decay length. 

\begin{figure}[htb]\centering
\includegraphics[trim=0mm 0mm 0mm 0mm, clip, width=8cm]{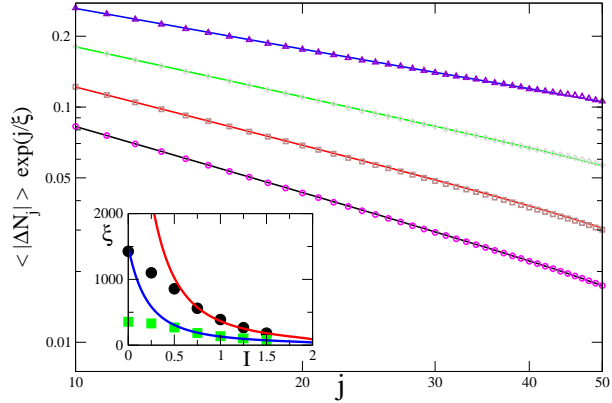}
\caption[Scaling of the Friedel oscillations decay (AI)]
{\label{fig:LL_scaling_I}
The rescaled decay of FO for the $W=0.1$ curves over the clean curves of the TLL phase 
with different interaction strengths. 
Inset: the localization length found by the best fit for each value of $I$ for $W=0.1$ (circles)
and $W=0.2$ (squares). The lines correspond to the theory prediction of Eq.~(\ref{eqn:xi_I}) with 
$\xi_0=7000$ and $1500$ respectively.
}
\end{figure}

For each strength of the interaction, 
one can rescale the disordered $W \ne 0$ curves, to the clean 
$W=0$ one by simply multiplying it by $e^{x / \xi}$, using $\xi$ as a fitting parameter. 
As can be seen in Fig.~\ref{fig:LL_scaling_I}, 
by using this rescaling method, the averaged disordered data collapses on the 
curves of the clean sample. 

The dependence of the decay length $\xi$ on the interaction strength $I$
is shown in the inset of Fig.~\ref{fig:LL_scaling_I}. 
We shall now show that this quantity
$\xi$ is effectively the mobility localization length.

The effect of disorder in the continuum limit can be divided to the 
forward and backward scattering terms.
Whereas the backward scattering term is related to the conductance and thus to
the localization length of the electrons, forward scattering processes
contribute only to the decay length of the FO, but not to localization. Thus, at first
sight $\xi$ does not necessarily correspond to the localization length. Nevertheless,
in this case one can argue that the contribution of the forward scattering
process to $\xi$ is small and therefore $\xi$ is a good measure of the localization
length.

Using standard bosonization technique it can be shown that the forward scattering 
processes result in the following term in the Hamiltonian:
\begin{eqnarray} \label{eqn:forward}
{H_{\rm fs} = -\int {dx \eta(x)\frac {1} {\pi} \bigtriangledown \phi}},
\end{eqnarray}
where $\phi$ is the TLL field which is related to the density operator by 
$\rho(x) = -\frac{1}{\pi} \bigtriangledown \phi(x)$ and $\eta(x)$ is the $q \approx 0$ 
component of the random potential. Since the TLL Hamiltonian 
(see chapter \ref{cpt:intro}, Eq.~(\ref{eqn:Intro_TLL_H4}))
\begin{eqnarray} \label{eqn:H_LL}
{H_{\rm TLL} = \frac {u} {2\pi} \int {dx [\frac{1}{\kappa} (\bigtriangledown \phi)^2 +
\kappa (\bigtriangledown \theta)^2}]},
\end{eqnarray}
depends on $\phi$ only through $(\bigtriangledown \phi(x))^2$, 
it is easy to show that by a redefinition of the field 
$\tilde{\phi}=\phi-\frac{\kappa}{u}\int^x dy \eta(y)$ 
one can incorporate the $H_{\rm fs}$ term inside $H_{\rm TLL}$ and get a similar form of Hamiltonian.
Therefore, the forward scattering term is not expected to change the physics of the system.

Nevertheless, it was shown that this redefinition of the field has an effect on the
correlation functions \cite{giamarchi03}. 
This results in a decay of the density-density correlation function,
which is, practically, the quantity we measure, and this decay
is not related to the conductance. It is an exponential decay of the form
$e^{-x/l}$, where $l = \frac {1}{2 D_{\rm f}} (\frac{u}{\kappa})^2$, 
and $D_{\rm f}$ is the forward scattering strength of the disorder (defined in the non-interacting case).

For the decay described by the characteristic length $l$, one can find, using the
Bethe ansatz solution, the factor $u/\kappa$ for each value of $I$.
It is easy to show that $u/\kappa$, and thus $l$, are monotonically increasing functions of $I$, 
as opposed to the FO decay length (see Fig.~\ref{fig:LL_scaling_I} in the inset).

Moreover, one can estimate $l$ quantitatively for the system we consider.
The factor $u/\kappa$ found from Bethe ansatz solution 
ranges from $u/\kappa=2$ for $I=0$ to $u/\kappa \approx 4.5$ for $I=1.5t$. 
Denoting the amplitude of the disorder correlation function by $D$, 
i.e., $\langle V(x)V(x') \rangle = D\delta(x-x')$, one finds that
$D_f$ and $D_b$ (the forward and backward scattering disorder strengths) are 
of the same order of magnitude as $D$. 
For non-interacting spinless electrons in a 1D lattice \cite{shreiber}
$1/D_b \approx 100/W^2$. Substituting $W=0.1$, one gets $l$ of the order of $10^5$,
which is much longer than the observed decay length.

We thus conclude that the backward scattering processes are much more significant
in the model treated, thus $\xi$ is a very good approximation to the localization length, 
and its interaction dependence should be described by Eq.~(\ref{eqn:xi_I}).

Using the prediction of Eq.~(\ref{eqn:xi_I}) with the value
of the disorder we employ along this chapter (order of $10^{-1}$), 
and recalling that without interactions $\xi_0 \approx 100/W^2$,
the localization length should range between $\xi(I=0) \approx 10^4$, which is much larger 
than the lattice sizes we considered, and thus almost doesn't influence the electron density, to $\xi(I=2) \approx 10^2$, 
in which the disorder effect should indeed be much more dominant, in agreement with the 
qualitative results presented in Fig.~\ref{fig:decay_I}.

The quantitative data shown in the inset of Fig.~\ref{fig:LL_scaling_I} fits 
the theoretical predictions of Eq.~(\ref{eqn:xi_I})
for not too weak interactions. For weak interactions ($I \lesssim 0.5$)
no such fit was found, which is however expected, since for this regime
the theoretical localization length is much larger than the wire length.
The fact that the best fit to Eq.~(\ref{eqn:xi_I}) was for $\xi_0 \approx 7000$
($1500$) for $W=0.1$ ($0.2$), and not the expected $\xi_0 \approx 10000$ ($2500$),
can be attributed to the same reason, as well as to the neglected forward scattering 
term which is stronger for weak $I$.
We also note that the exact choice of the wire slices over which the fit is done, can
change slightly the values of $\xi$. This, however, does not change the qualitative results,
showing a monotonic decrease of $\xi$ as a function of the interaction strength.

To summarize, 
the effect of disorder on the FO decay in the Anderson regime can be described by an
extra exponential decay of the FO, which depends on the localization length,
of the form
\begin{eqnarray} \label{eqn:Y_j_2}
\langle \Delta N_j \rangle = A (-1)^j j^{-\kappa} \exp ( -j/\xi(\kappa)),
\end{eqnarray}
where the localization length $\xi(\kappa)$ decreases monotonically
as the interactions increase.

\section{From The CDW Regime Towards Mott Insulator}
\label{sec:FO_CDW}
\subsection{A Clean Lead: Extracting the CDW Correlation Length}

\begin{figure}[htb]\centering
\includegraphics[trim=0mm 0mm 0mm 0mm, clip, width=8cm]{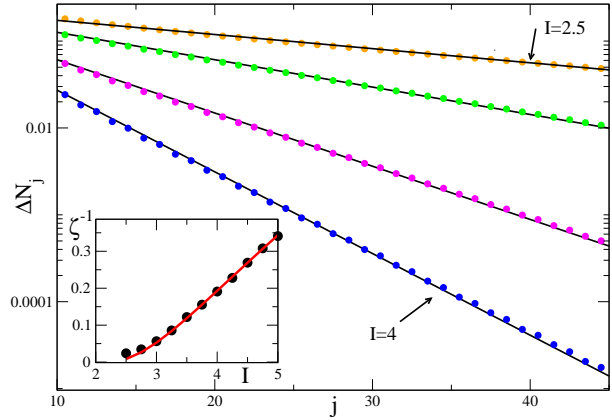}
\caption[Friedel oscillations decay for clean samples (MI)]
{\label{fig:cdw_frid}
The oscillations decay in the CDW regime for various interaction strengths and without disorder
(note the semi-log scale).
As the interaction increases, the correlation length decreases and the decay is faster.
Inset: the inverse correlation length of the CDW state for various interaction strengths (symbols)
fitted to the theory prediction Eq.~(\ref{eqn:zeta}).
}
\end{figure}

Since the CDW is an insulating phase, the decay of the $2k_F$ oscillations without disorder is supposed to 
be exponential and the characteristic length is the CDW correlation length \cite{pang93},
i.e., $\propto \exp(-x/\zeta)$. 
In Fig.~\ref{fig:cdw_frid} such an exponential 
decay of $\Delta N_j$ is shown on a semi-log scale for various interaction strengths.
An exact Bethe ansatz solution \cite{mikeska04} of our model gives the relation between 
the correlation length and the interaction as
\begin{eqnarray} \label{eqn:zeta}
\zeta \sim exp(\pi/ \sqrt{I/(2t) - 1}).
\end{eqnarray}

The correlation lengths extracted from
the DMRG results are presented with a fit to the exact formula in the inset of Fig.~\ref{fig:cdw_frid}.
As can be seen, for $I$ not very close to the TLL-CDW transition point (which occurs at $I=2t$), 
the results fit the theory very well.

\subsection{Decay Length in the Disordered CDW Phase}

\begin{figure}[htb]\centering
\includegraphics[trim=0mm 0mm 0mm 0mm, clip, width=8cm]{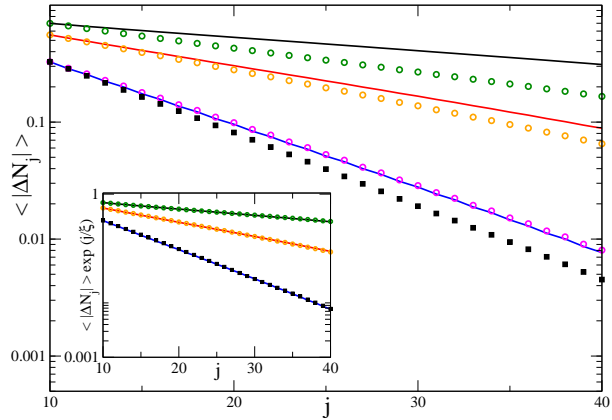}
\caption[Friedel oscillations decay for disordered samples (MI)]
{\label{fig:CDW_decay_I3.5}
The decay of the oscillations of a disordered CDW with $I=2.5,3$ and $3.5$ (top to bottom,  
note the semi-log scale). 
The lines correspond to the clean sample result, and the symbols to the averaged 
disordered data. For $W=0.1$ (circles) the disorder effect 
is clearly seen for $I=2.5$ and $I=3$ but not for $I=3.5$ in which $\xi$ is
much larger than the correlation length $\zeta$. For $W=0.2$ (squares) $\xi$ is small 
enough to affect the decay even for $I=3.5$.
Inset: multiplying $\Delta N_j$ by $e^{x/\xi}$ collapses the disordered data on the clean curves.
}
\end{figure}

For $W \ne 0$, $\Delta N_j$ is averaged over $100$ realizations,
for which we expect a sampling error of the order of one percent. 
Assuming that the disorder adds another exponential term to the oscillations decay,
which is thus proportional to $\exp(-x/\zeta-x/\xi)$,
there are two competing length scales - the decay length due to disorder ($\xi$) vs. the correlation length ($\zeta$). 
For strong interactions and weak disorder $\zeta \ll \xi$ so 
that the disorder effect is hardly seen, but 
increasing the disorder or decreasing the interaction strength 
should result in a combination of the two exponential decays.
The DMRG results, presented in Fig.~\ref{fig:CDW_decay_I3.5}, show the disorder 
effect on the oscillations decay. For $I=2.5$ and $I=3$ one can see faster decay for the disordered samples with $W=0.1$.
For stronger interaction larger disorder is required in order to affect the decay.

Similarly to the AI phase, the extra decay length can be extracted by fitting, 
for each value of $I$,
the $W \ne 0$ curve multiplied by $e^{x/\xi}$ to the $W=0$ one. Such a 
rescaling is presented in the inset of Fig.~\ref{fig:CDW_decay_I3.5}. 

\begin{figure}[htb]\centering
\includegraphics[trim=0mm 0mm 0mm 0mm, clip, width=8cm]{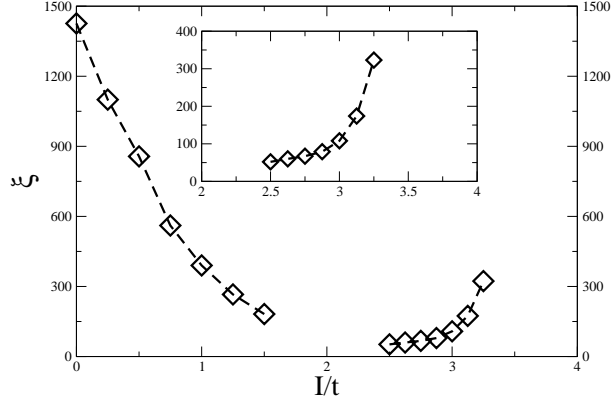}
\caption[Decay length of FO in AI and MI phases]
{\label{fig:xi_I_CDW}
The decay length due to disorder ($\xi$) in the TLL ($I<2t$) and in the CDW ($I>2t$)
phases as a function of the interaction strength. 
Inset: zoom into the CDW regime.
}
\end{figure}

\subsection{The Decay Length Dependence on Interaction: AI vs. MI}

As can be seen in Fig.~\ref{fig:xi_I_CDW}, 
the decay length extracted for the disordered MI regime increases as
a function of the interaction strength (for $2t < I \lesssim 3.5t$), 
an opposite behavior to the AI case ($I<2t$).
Results for stronger values of $I$ are not shown, since
for too strong interactions the correlation length is very small,
and thus the estimate of $\xi$ is less accurate.

These results reveal that as the interaction strength increases in the MI phase, the 
disorder effect decreases. In the AI phase, on the other hand, the disorder effect
is enhanced with increasing interactions.
The difference between these two behaviors results from the difference in the
ground states of the two phases in the clean case. In our model 
there is a competition between the kinetic energy (the hopping term) and the potential
(the interaction). The hopping term prefers the existence of a flat particle
distribution whereas the interaction term prefers a CDW-like form. 
For different values of $I$ the results of that competition are different: for $I<2t$
(the TLL phase) the hopping term wins, and the distribution is flat, while for $I>2t$ 
(the CDW phase) a CDW starts to form.

Inside the clean TLL phase, as $I$ increases, the CDW fluctuations are stronger. Yet, the 
average density profile in the ground state remains flat because of the 
hopping term. But when disorder is introduced, 
the flat density state becomes less favorable than a state with a fluctuating density,
the latter being preferred by both the disorder and the interactions. For a 
constant disorder, as the interactions become stronger, these fluctuations are 
enhanced, so the disorder effect increases.

In the CDW phase, on the other hand, without disorder, the interaction wins over 
the hopping, and the ground state has a CDW form. 
Turning on the disorder might change the particle distribution, e.g. by allowing 
an electron to move into a site with lower on-site energy, but this results in raising 
the interaction energy. As the interaction strength gets stronger, the probability 
of such a process decreases, so that the actual effect of the disorder is getting weaker.

\section{Conclusions and Future Prospects}
\label{sec:FO_con}
In conclusion, we have shown that the FO envelope
is affected by both the interaction strength, and the disorder strength. 
In the TLL phase and the resulting disordered AI phase,
interactions actually enhance the effect of FO since it drops with a 
weaker power law $j^{-\kappa}$, while disorder decreases the FO
since it adds an exponential factor to its decay form. 
We have shown that the length scale for this exponential decay 
is a good approximation to the mobility localization length, since it is only weakly
influenced by forward scattering processes for weak disorder.
Thus we established a convenient way to evaluate the dependence of the
localization length on disorder and interaction using only
the ground-state properties of the system. Qualitatively, the
localization length as a function of interaction for a given weak disorder
always decreases. As long as the localization
length is not much longer than the wire's length, the localization length behavior
is quantitatively described by the renormalization-group results \cite{apel_82}.

However, while the decay length of the $2k_F$ oscillations envelope due to disorder 
is monotonically decreasing in the AI phase, we have shown that it is monotonically 
increasing in the disordered MI phase. The difference between these two regimes is explained by
the difference between the ground states of the clean samples in each case. In 
the AI phase the pure ground state is flat, and both the disorder and the interactions try to introduce fluctuations in it. 
In the MI phase, on the other hand, the pure ground state oscillates with a $2k_F$ 
wave vector, and these oscillations are enhanced by the interactions and reduced by the disorder.
As a result, the disorder effect (for a constant disorder strength) is getting weaker,
in the MI phase, as the interactions are enhanced.

In addition, we have analytically described the dependence of the FO amplitude on
the impurity strength and on the dot-lead coupling, in the non-interacting case.


\begin{center}
*~~~~~*~~~~~*
\end{center}

During the discussion of the disordered MI phase, we have mentioned that the
strength of the disorder has a significant impact on the phase of the wire.
While for a weak enough disorder the wire can still be described by a MI phase, for a strong 
disorder it turns to an AI phase. A few recent studies \cite{orignac99,giamarchi01} 
point out the possibility that another phase exists between the MI and the AI, which 
is the Mott glass (MG) phase.

The MG phase is characterized by partial features of both the AI and the MI phases.
For example, it does not have a long range order, since it includes ordered regions 
which are separated by some kind of domain walls (solitons in the CDW language). In 
addition, the MG energy spectrum is gapless. In both these properties the MG phase is thus 
similar to the AI one. On the other hand, the low-lying excitations are not charged, but 
describe a motion of the domain walls, and as a result the MG phase is incompressible, 
like the MI phase.

The model used in Refs. \cite{orignac99,giamarchi01} consists of spinless electrons in a disordered
potential, as in our case. Nevertheless, in order to assure the existence of a clean MI phase,
as opposed to the TLL phase, it introduces a $2k_F$ commensurate potential as well.
Therefore, when there is no disorder, a MI phase is obtained for the entire range of interactions.
The effect of disorder is now added to that of the periodic potential. The resulting
phase diagram shows that for a non-vanishing disorder strength, a gradual increase of the
commensurate potential causes the wire to turn from an AI (when there is no periodic potential), 
via a MG, towards a MI (strong periodic potential).

It is therefore clear that a finite regime of the MG phase can be obtained using a similar
model to the one we use, by adding a periodic potential to the Hamiltonian. 
An interesting question can then be raised: How will the decay of the FO look like? 
Can the shape of the FO decay help to identify the borders of the MG phase?
The results of this chapter show that one is able to differentiate between the AI and the MI 
phases by looking at the dependence of the decay length on the interactions.
In the case of the MG phase, such an identification is much less clear. However,
the versatile properties of the MG phase might be helpful for that task, 
which we leave for a future study.

\begin{center}
*~~~~~*~~~~~*
\end{center}

Finally we remark to the experimental relevance of this work. The theoretical
treatment of disorder usually involves statistics over an ensemble of
many samples which is usually hard to obtain experimentally.
Furthermore, in the case we deal here, a measurement of FO on a disordered sample 
seems at first sight daunting. However,
the simple method we suggest in order to deal with the disorder, is 
in principle experimentally feasible, and solves these two difficulties.

Once a technical method for measuring the electron density is established,
it should be used twice for each sample, before and after the coupling of the
wire to the dot. In principle, by using a gate it should be possible to eliminate
the coupling between the dot and the wire.
Our results, as can be seen in Figs.~\ref{fig:frid_dis_sample} and \ref{fig:CDW_frid_sample}, 
which present typical results for a particular realization,
point out that the difference between these two measurements 
should show a very clear FO, even for a specific sample.

\cleardoublepage

\chapter{A Disordered QD with Interacting Electrons} \label{cpt:ch3}

In this chapter we explore some properties of a two-dimensional quantum dot, 
without considering its coupling to external leads. We model the quantum dot by a two-dimensional lattice
with either nearest neighbor or Coulombic interactions, which is occupied by spinless particles.
Since the size of the relevant Hilbert space is huge, such a
problem cannot be solved by an exact diagonalization (except for very small systems), 
and we thus use the particle-hole density-matrix renormalization-group (PH-DMRG) 
method, which was detailed in section \ref{sec:num_phd}.

One of the most important quantities involved in any physical system 
is the energy spectrum, and especially the ground-state energy. 
In this chapter we show that the PH-DMRG method can be used in order to approximate the 
ground-state energy of the disordered quantum dot with interacting electrons, and that it leads
to results which are much more accurate than those of the Hartree-Fock approximation. 
Moreover, following a comparison to other approximate methods, we suggest an improvement 
of the PH-DMRG truncation algorithm, which reduces the error rate of the 
traditional\footnote[1]{It's a bit weird to call it traditional since the known number of PH-DMRG applications
can be count currently using one hand. However, by "traditional" we refer to the method as it is
documented in \cite{dukelsky02} and \cite{dukelsky04}.} PH-DMRG by almost $30$ percents.

As an application for the improved PH-DMRG method we calculate the ground-state energies 
for different numbers of electrons, and find the addition spectrum of the system. We compare
the PH-DMRG results to those of the Hartree-Fock approximation in three aspects: the error rate,
the average and the fluctuations of the addition spectrum.

\section{Introduction}
In the last decade there is a growing interest in the low-temperature physics
of disordered many-particle systems, such as electron dephasing due to interactions
\cite{aleiner99} and two-dimensional (2D) 'metal-insulator' transition \cite{abrahams01}. 
Transport properties through quantum dots (QDs) have also been recently investigated and shown 
to exhibit some interesting phenomena in the presence of both interactions
and disorder \cite{alhassid00}.
An analytical treatment of these problems is
difficult, unfortunately, since both the disorder and the interactions cannot be
considered as a small perturbation. A traditional numerical treatment of such problems is
restricted to small systems, since the size of the many-particle Hilbert space
grows exponentially with the system size.

During recent years, a few methods were established in order to 
decrease the size of the Hilbert space to a size which is computationally feasible. 
One way is to define an iterative order in which the system is treated, and use a 
sophisticated truncation method between the iterations to reduce the space size. This is the 
idea behind the ensemble of renormalization group methods, among which the 
density-matrix renormalization-group (DMRG) method is an 
honored member (see section \ref{sec:num_dmrg}).

A different approach uses a predefined constraint which is checked before states are inserted
into the Hamiltonian. The entire system may be treated immediately, yet not all the system states 
are taken into the Hamiltonian. Therefore, the matrix size one needs to diagonalize is smaller 
than the entire Hilbert space dimension, and hopefully small enough to be exactly solved.
Yet, if the constraint is defined on the many-particle space, then although the final matrix size 
might be small, the calculation time required grows exponentially with the system size. Thus 
such a constraint does not solve the main issue.

However, Berkovits in Ref.~\cite{berkovits03-2} has investigated constraints which are checked 
against the single-particle states before they are used to build the multi-particle basis. 
The usage of this approach was demonstrated on a ground-state energy calculation of a 
disordered 2D lattice of $4$ rows and $6$ columns containing $10$ interacting electrons,
for which the full Hilbert space contains $1,961,256$ states. 
The error rate, or the discrepancy, can be defined as $D(x) = \langle |x^\prime-x|/|x| \rangle$,
where $x$ is an exact quantity, $x^\prime$ is an approximation for $x$,
and the average is done over different realizations of the disorder. By calculating the 
value of $D(E_{gs})$, one thus has a good estimate about the accuracy of the approximation method.

In this chapter we quote the results obtained in Ref.~\cite{berkovits03-2} by two methods, 
which we denote as "energy-cut" and "generation-cut". In both methods the first
step is to execute a self-consistent Hartree-Fock (HF) approximation, and obtain a sorted
"list" of wave-functions and energy levels, which partly incorporate the interaction effects. 
In the first method, the energy-cut, one neglects the states with the highest single-particle energies. 
Such states should, intuitively, have the smallest contribution to the many-particle ground-state energy. 
For a system with $N$ sites and $n_e$ electrons, the full Hamiltonian matrix is of size $M=\binom{N}{n_e}$.
Using the energy-cut method, one takes only the lowest $m_R$ single-particle states, and thus diagonalizes a matrix of 
size $M_R = \binom{m_R}{n_e}$. In the results of Ref.~\cite{berkovits03-2}, the energy-cut method was used for
the $4 \times 6$ lattice with $10$ electrons, by taking up to $m_R=18$ (out of the $24$) HF states,
which leads to a maximal matrix size of $M_R = 43,758$. The best discrepancy obtained by this method
was $D(E_{gs}) \approx 2.5$ percent, a significant improvement of the HF results, for which $D(E_{gs}) \approx 4.5$ percent.

The second method, generation-cut, has obtained better results than the energy-cut method. 
This method is based on the localization of the Fock space. Since the interaction term is a two-body 
operator, only many-body states which differ by at most two electron-hole pairs are coupled by 
the Hamiltonian. It was proven \cite{altshuler97} that the average contribution of a state 
containing $k$ electron-hole pairs to the exact ground state is proportional to $\exp(-k/{\xi_F})$, 
where $\xi_F$ is the Fock space localization length. Considering also the number of
states in the $k$-th electron-hole generation, $\binom{N-n_e}{k} \binom{n_e}{k}$, one finds out 
\cite{berkovits03-2} that the weight of generations falls off exponentially as long 
as $\xi_F^{-1} > \ln[(N-n_e)n_e/(k+1)^2]$.

Therefore one can consider in the approximated Hamiltonian only states with a small 
number of particle-hole pairs. For exactly $f$ particle-hole pairs, the number of states is
$\binom{N-n_e}{f} \binom{n_e}{f}$. In the results of Ref.~\cite{berkovits03-2}, the 
generation-cut method was used by taking up to $3$ particle-hole pairs, which leads
to a maximal matrix size of $47,916$. The lowest error obtained by this method was 
$D(E_{gs}) \approx 1.5$ percent. 

Unfortunately, one cannot simply 'cut and paste' these couple of methods for larger systems. 
As the size of the system increases, the number of important HF states (for the energy-cut method),
and the number of important particle-hole generations (for the generation-cut) are
expected to increase. Moreover, even for the same number of chosen HF states and particle-hole
generations, the matrix size is exponential in the system size, and it becomes soon too 
large to be solved exactly. 
Therefore, a treatment of larger systems will experience similar
problems to those of the exact diagonalization technique. 

Thus there is a great need for an alternative method, which can give an accurate 
approximation to the ground-state energy of a 2D QD, yet can be extended to larger systems.
In this chapter we check if the PH-DMRG method is suitable for that task.

\subsection{Chapter's Outline}
The rest of the chapter is organized as follows. In the following section
we briefly describe the Hamiltonian model and the numerical PH-DMRG method,
which was extensively discussed in section \ref{sec:num_phd}.
In section \ref{sec:ph_res} our main results are presented.
We check the accuracy of the PH-DMRG method in approximating the ground-state energy of
an ensemble of disordered lattices with interaction.
We show that the PH-DMRG method 
can get an accuracy which is similar to that of the energy-cut
technique, whereas the generation-cut method obtains better results.
Analyzing some disadvantages of the PH-DMRG algorithm, we then suggest 
an improvement of the PH-DMRG truncation method. This leads to an improvement 
of more than $30$ percents of the PH-DMRG accuracy, making it
comparable to the generation-cut technique as well.

In section \ref{sec:ph_cnn} we compare the PH-DMRG accuracy between different schemes: 
Short-range vs. long-range interactions, and intermediate vs. strong interactions.

As an additional application we show in section \ref{sec:ph_delta2} the results 
for a calculation of the addition spectrum using PH-DMRG, when either nearest neighbor (NN) 
or Coulombic interactions are present.
We show that in both cases the accuracy
of the PH-DMRG method is much better than that of the HF approximation.

We conclude the chapter in section \ref{sec:ph_end}, in which 
we discuss the possibility to use the PH-DMRG for larger systems, and
we point out an optional future research.

\section{Model}
\label{sec:ph_model}
The system we treat, as a QD model, is composed of $n_e$ spinless electrons on a 2D disordered 
lattice of $A$ columns and $B$ rows. The electrons can hop from one lattice site to one of its NNs,
and either NN or Coulombic interactions are considered. Therefore, the Hamiltonian describing the 
physics of the system can be written as $\hat H = \hat H_0 + \hat H_{\rm int}$,
where $\hat H_0$ is given by Eq.~(\ref{eqn:H_0_phdmrg}), and $\hat H_{\rm int}$,
for NN interactions, is given by Eq.~(\ref{eqn:H_int_nn}):
\begin{eqnarray} \label{eqn:H_phdmrg2}
{\hat H}_0 &=& \sum_{m} \epsilon_{m}{\hat a}^{\dagger}_{m}{\hat a}_{m} 
-t \displaystyle \sum_{\langle m,n \rangle}({\hat a}^{\dagger}_{m}{\hat a}_{n} + H.c.), \\ \nonumber
{\hat H}_{int} &=& 
\displaystyle V \sum_{\langle m,n \rangle}{\hat a}^{\dagger}_{m}{\hat a}^\dagger_{n}{\hat a}_{n}{\hat a}_{m}.
\end{eqnarray}
Here, $\hat a^\dagger_m$ and $\hat a_m$ denote creation and annihilation operators
of an electron in lattice site $m$, where $\langle m,n \rangle$ represents NN sites $m$ and $n$. 
$\epsilon_{m}$ is the on-site energy of site $m$, chosen randomly from a uniform distribution
$[-W/2,W/2]$, $t$ is the hopping matrix element between NNs, and $V$ is the NN interaction strength. 
$t$ is conventionally taken as the energy unit.

The ground-state energy of the system is calculated using the PH-DMRG 
method, which was described in details in section \ref{sec:num_phd}. The 
general idea is to divide the energy levels to those above (particle-states) 
and below (hole-states) $E_F$, and then treat these states iteratively. Starting from the
vacuum state, in which all levels below $E_F$ are filled, and all the
others are empty, we add in each iteration one particle-state and one
hole-state, starting from $E_F$ and proceeding in both directions. 
We diagonalize the superblock composed of the states we already
added, maintaining the number of particles constant. Each iteration ends
with a truncation of the Hilbert space of both the particle block and the hole
block, using their corresponding density matrices, after which a new couple of 
states can be added again. The iteration process stops after all of the states
were added\footnote[2]{A comprehensive description is presented in section \ref{sec:num_phd}.}.

As was emphasized in section \ref{sec:num_phd}, the accuracy of the PH-DMRG method depends 
mostly on the number of states, $p$, that are kept between successive iterations. In other words,
in each iteration, $p$ eigenvectors of the density matrix are taken, while all the others are 
neglected. Except the accuracy, $p$ influences dramatically the computational resources 
required during the process. 

In general, as the Hamiltonian is more complicated, and as the number of both particle-states 
and hole-states increases, the size of the superblock Hamiltonian, whose creation and 
diagonalization are the main bottle-neck of the 
algorithm, is enlarged. Therefore, for different Hamiltonians, and even for different number of 
particles in a given model, the upper limit of $p$ can change.
In this study the maximal number of block states that we keep is $p=120$, which is of the
order of the number of states used in typical PH-DMRG applications\footnote[3]{$p_{\rm max} \approx 100$
was used in most implementations. However, for simpler models, which lead systematically
to small superblock Hamiltonians, PH-DMRG was utilized with $p_{\rm max} \approx 400$. It should also
be noted that the presence of disorder in our model requires also averaging over realizations.}.

\section{Ground-State Energy Calculation}
\label{sec:ph_res}
We begin by presenting the PH-DMRG results for the calculation of the ground-state energy in an interacting
disordered dot. The Hamiltonian for such a model is given by Eq.~(\ref{eqn:H_phdmrg2}).
We investigate the case of $10$ interacting electrons in a disordered $4 \times 6$ lattice, with the
same ensemble used in Ref.~\cite{berkovits03-2}. In this section we restrict ourselves to 
the case of NN interactions, with strength $V=3t$, and we set the disorder strength to be $W=5t$.

\begin{figure}[htb]
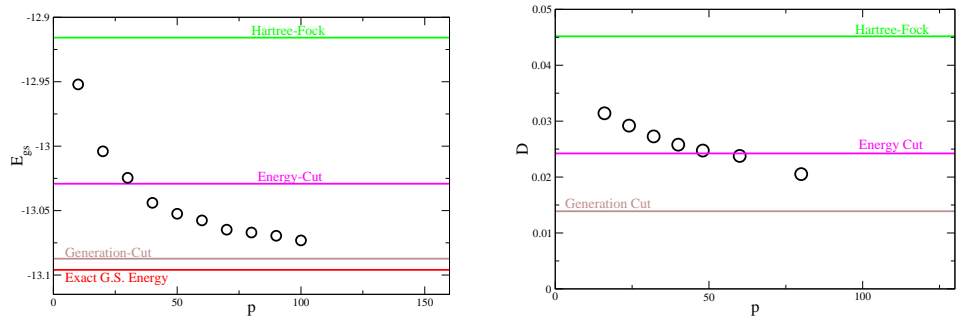
\centering
\begin{minipage}[t]{.45\textwidth}
\includegraphics[trim=0mm 0mm 0mm 0mm, clip, width=.95\textwidth]{Egs_sample1_a}
\end{minipage}
\hfil
\begin{minipage}[t]{.45\textwidth}
\includegraphics[trim=0mm 0mm 0mm 0mm, clip, width=.95\textwidth]{res_ph_a}
\end{minipage}
\caption[Energy calculation and discrepancy using PH-DMRG]
{\label{phfig_1}
PH-DMRG results for the ground-state energy calculation as a function of $p$, 
the number of states kept during the renormalization process.
The results are compared to various approximated results (see text).
(a) The ground-state energy of a specific realization.
The exact result is presented by the red line.
(b) The discrepancy $D(E_{gs})=\langle|E_{gs}'-E_{gs}|/|E_{gs}|\rangle$ averaged
over an ensemble of $100$ realizations. 
}
\end{figure}

Typical results, for a specific realization, are shown in Fig.~\ref{phfig_1}(a),
as a function of $p$, the number of block states kept. 
The results are compared to the HF results (green line), to the results of the energy-cut method, 
in which the lowest $18$ HF states were used (magenta), and to those of the generation-cut method, 
in which up to $3$ particle-hole generations were kept (brown). The exact results are also 
drawn (a red line).
As $p$ increases, the PH-DMRG approximation is better. Typical values of $p$ in which the PH-DMRG 
method is already more accurate than the energy-cut method are $40-50$
(for the sample shown $p \gtrsim 30$ was sufficient).
In rare samples larger values of $p$ were required in order to improve the energy-cut results.
However, a very slow convergence is seen, and the generation-cut method obtains
the best results for up to $p = 80$, for all samples.

Averaging over realizations makes the picture more clear, as one can see in Fig.~\ref{phfig_1}(b).
The average accuracy of the PH-DMRG calculation is improved very slowly by increasing $p$.
The best PH-DMRG results shown (keeping $80$ states) gets lower discrepancy 
than the energy-cut method, yet the generation-cut method obtains much more accurate results.

The comparison between the PH-DMRG results and those of the generation-cut method raise the
following question:
Can one find a way to improve the PH-DMRG process 
(besides the option to increase the block size)? In the following we present such a possible
improvement, motivated by an analysis of the PH-DMRG truncation method.
Recalling the DMRG principles, the main difference between the DMRG and previous numerical 
renormalization group (NRG) methods, is the truncation algorithm. In the NRG method, the truncation 
of states is based on their energies, while in the DMRG it is based on the density-matrix eigenvalues.
The density-matrix eigenvectors with the highest eigenvalues are considered as the most 
important, and the rest of the states are neglected.

Nevertheless, the difference between these criteria, i.e., the NRG's lowest energy and the 
DMRG's highest density-matrix eigenvalues, is not the main reason for the DMRG success.
This success originates mainly from the physical situation to which the density matrix emulates. 
While the NRG truncation method is based only on the sites which were already iteratively added,
in the DMRG algorithm other sites are also included in the superblock.
The superblock is composed of a "system" coupled to an "environment",
which represents all the sites which were not yet included in the iteration process,
and thus the truncation is based on a future prospect, which leads to better results.

On the other hand, the superblock in the PH-DMRG does not consider any "future" states.
The PH-DMRG process couples only the already-used particle-states and hole-states, but
other states, those which were not yet inserted, are not part of the superblock. 
Therefore, the truncation does not take them into account, and this might limit its
success.

In order to give the truncation decision a wider point of view, one may
use an additional condition to help choosing the states to continue with.
Such a condition should thus take into account some further considerations. 
Based on the great success of the generation-cut in Ref.~\cite{berkovits03-2},
we try here to consider the number of particle-hole generations as an extra 
condition taken during the truncation step. 

As we have discussed above, the idea of the Fock space localization indicates that
the weight of successive particle-hole generations decreases exponentially.
Moreover, in a comparison between specific states at a given time during the process, 
those with smaller number of particle-hole generations are apriori more important.



There are a few ways in which one can incorporate this criterion into the 
density-matrix truncation process. For example, when the states are weighted by their
density-matrix eigenvalues, one can add a multiplicative factor for each eigenstate,
based on the number of particle-hole pairs it consists of. Such a method should be 
accompanied by a systematic investigation of the weighting procedure, e.g. by a
calculation of the relative probability for each number of particle-hole pairs
according to the Fock space localization length, and we leave this subject for 
future studies.

Here we use a simpler condition. We define a maximal number $k$ of preferred 
particle-hole generations, and states with $f \le k$ are kept in the first stage. If there are
too many such states, we use the density-matrix eigenvalues, and take the states
with the highest eigenvalues. After this first round, if there is still a room for more 
states, we take also states with $f > k$, according to their density-matrix order,
until the maximal number of states $p$ is reached.

\begin{figure}[htb]
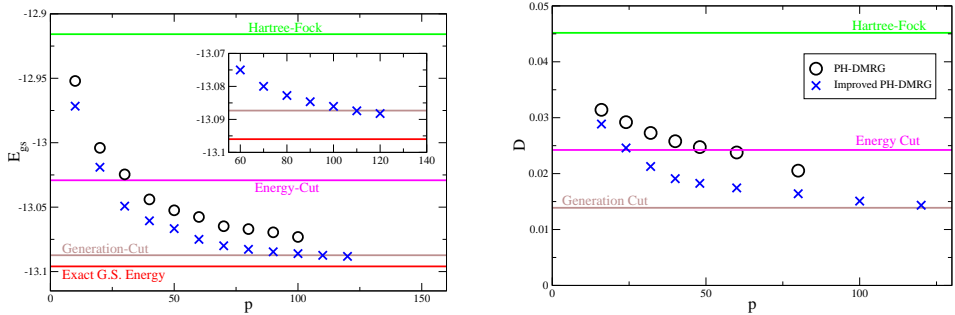
\centering
\begin{minipage}[t]{.45\textwidth}
\includegraphics[trim=0mm 0mm 0mm 0mm, clip, width=.95\textwidth]{Egs_sample1_b}
\end{minipage}
\hfil
\begin{minipage}[t]{.45\textwidth}
\includegraphics[trim=0mm 0mm 0mm 0mm, clip, width=.95\textwidth]{res_ph_b}
\end{minipage}
\caption[Energy calculation and discrepancy using improved PH-DMRG]
{\label{phfig_1b}
The same as Fig.~\ref{phfig_1}, together with the results of the improved PH-DMRG method
(traditional version - black circles, improved version - blue crosses).
Inset: zoom into the region of $p \ge 60$. 
Both panels show that the improved PH-DMRG method obtains similar results as the generation-cut
method for $p \approx 110-120$.}
\end{figure}

For the $4 \times 6$ system with $10$ electrons, we've executed the suggested
truncation method for $k=0,1,2,3,4$, and for each realization we have picked the
lowest energy. However, it turns out that in $95$ percents of the samples the $k=2$ case 
resulted in the most accurate result, so that using a constant $k=2$ would lead to 
similar results\footnote[4]{In the following sections we thus use $k=2$.}.
Fig.~\ref{phfig_1b} presents the same curves of Fig.~\ref{phfig_1}, together with the
results of the improved algorithm. In Fig.~\ref{phfig_1b}(a) the results of $k=2$ are
compared to the traditional results for a specific realization, while in Fig.~\ref{phfig_1b}(b)
the averaged discrepancy is shown.
As can be clearly seen, the results obtained by the improved method, which are better 
by almost $30$ percents than the regular PH-DMRG results, are comparable to the 
results of the generation-cut method of Ref.~\cite{berkovits03-2}.

From these results it is quite clear that the PH-DMRG can, in principle, obtain the 
ground-state energy with an accuracy of the order of $1.5$ percent, when the number of states $p$ 
is of the order of $100$. In the last section we discuss how these results are affected when
larger dots are considered. We will conclude that the enlargement of the system has much more dramatic influence
for the generation-cut method, compared to the PH-DMRG. Therefore, the PH-DMRG is a good candidate
for the task of approximating ground-state energies of disordered QDs.

Nevertheless, we still need to check how sensitive the accuracy of the PH-DMRG method is when 
the interactions have long range, or when they are much stronger than those considered in this section. 
This will be discussed in the following section.

\section{Long-Range Interactions; Strong Interactions}
\label{sec:ph_cnn}

The Hamiltonian used in the previous section, as well as that of section \ref{sec:num_phd},
utilized only short-range interactions. In the framework of the real-space DMRG method 
there is a huge impact when the interaction range increases. The DMRG iterations add
subsequent sites one after another, and if long-range interactions are considered, much more 
data should be stored from previous steps. Practically, therefore, real-space DMRG applications 
traditionally consider only nearest neighbors, or next nearest neighbors interactions.

On the other hand, the interactions range does not affect the PH-DMRG method almost at all.
It does of course change the first stages of the method, i.e., the single-particle diagonalization 
and the self consistent HF stage, because the Hamiltonian is changed. For example, for Coulombic 
interactions the Hamiltonian is (compare to $\hat H_{int}$ in Eq.~(\ref{eqn:H_phdmrg2}))
\begin{eqnarray} \label{eqn:H_int_c}
{\hat H}_{int}^{(C)} = 
\displaystyle \frac{1}{2} \sum_{m \ne n}\frac{V_c}{|r_m-r_n|}{\hat a}^{\dagger}_{m}{\hat a}^{\dagger}_{n}{\hat a}_{n}{\hat a}_{m},
\end{eqnarray}
where $|r_m-r_n|$ is the distance between sites $m$ and $n$, measured in lattice units.
Accordingly, the definition of the antisymmetric interaction element
$V_{\alpha \beta \gamma \delta}$ in Eq.~(\ref{eqn:V_abcd}) should be
modified to
\begin{eqnarray} \label{eqn:V_abcd_C}
V_{\alpha \beta \gamma \delta}^{(C)} &=& \frac{1}{2} \sum_{m \ne n} \frac{V_c}{|r_m-r_n|} \{
~ \phi^*_\alpha(m) \phi^{*}_\beta(n) \left[ \phi_\gamma(n) \phi^{}_\delta(m) - 
\phi_\delta(m) \phi^{}_\gamma(n) \right] \\ \nonumber
&~& ~~~~~~~~~~~~~~~~ - \phi^*_\alpha(n) \phi^{*}_\beta(m) \left[ \phi_\gamma(n) 
\phi^{}_\delta(m) - \phi_\delta(m) \phi^{}_\gamma(n) \right] ~ \} \\ \nonumber
&=&  \frac{1}{2} \sum_{m \ne n} \frac{V_c}{|r_m-r_n|} \left[ \phi^*_\alpha(m) \phi^{*}_\beta(n)  - \phi^*_\alpha(n) \phi^{*}_\beta(m) \right] 
 \left[ \phi_\gamma(n) \phi^{}_\delta(m) - \phi_\delta(m) \phi^{}_\gamma(n) \right].
\end{eqnarray}
However, except for that slight change, the entire iterations process, which is the most taxing stage of the 
PH-DMRG algorithm, remains the same. The bottom line is that since the iteration is done 
over states in momentum-space, it does not matter what the real-space range of interactions is.

Therefore, the range of interactions does not change the feasibility of the PH-DMRG method at all.
However, one should ask what effect does the interactions range have on the PH-DMRG accuracy. 
A related question is how the accuracy is changed when the interactions become stronger,
and does the enhancement of interactions in the short-range and long-range cases have the 
same influence.

\begin{figure}[ht]\centering
\includegraphics[trim=0mm 0mm 0mm 0mm, clip, width=7cm,height=!]{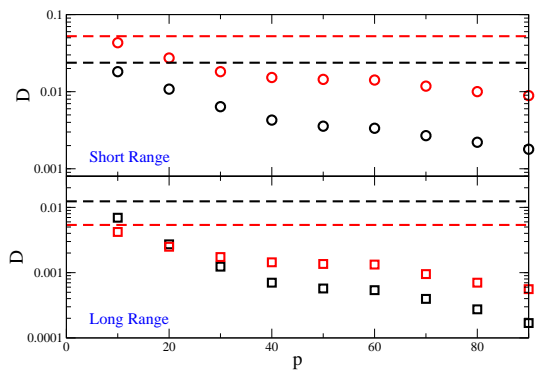}
\caption[PH-DMRG results for short-range and long-range interactions]
{\label{phfig_1c}
The discrepancy $D(E_{gs})$ obtained by the PH-DMRG calculation as a function of the number of states kept,
for an ensemble of $4 \times 6$ disordered lattices occupied by $6$ electrons. The accuracy is compared between 
short-range (upper panel) and long-range (lower panel) interactions, and between 
intermediate interactions strength ($V=V_c=3t$, black symbols) and a strong one ($V=V_c=10t$, red symbols).
The results of the HF are presented in dashed lines.
Note the semi-log scale.
}
\end{figure}

In Fig.~\ref{phfig_1c} we present the results of the discrepancy obtained for short-range (upper panel)
and long-range (lower panel) interactions, as a function of the number of states kept, $p$. 
In both cases we compare the results for intermediate strength of interactions (black) and a strong one (red).
These results were obtained by the PH-DMRG method for $4 \times 6$ lattice with $6$ electrons. Since 
the number of electrons is small (compared to the previous section), there are less hole-states, and 
the PH-DMRG accuracy is better. It should be noted that the same is true for the result of the HF
approximation (shown in dashed lines).

For both short-range and long-range interactions, we find that the accuracy is reduced when the 
interaction strength is enhanced.
Therefore, as the interaction strength increases, the number of states $p$ 
which is required in order to obtain the same accuracy, increases.

A more striking feature is the difference between the short-range and the long-range 
interactions in the discrepancy. For small values of $p$, the better results of the 
long-range case are explained by the fact that the HF approximation is known to work
better for long-range interactions. Nevertheless, the improvement of the long-range results by the 
PH-DMRG method is fascinating, and the obtained averaged discrepancy for $V_c=3t$ with $p=90$ states is 
$D(E_{gs}) \approx 10^{-4}$, more than an order of magnitude better than the accuracy in the 
short-range case. 

We thus close this section with the following conclusions: The PH-DMRG can be 
used as an accurate method for calculating ground-state energies of disordered QDs
with interactions. The physical parameters of the system can affect the accuracy
significantly. The best accuracy is obtained for long-range interactions which are not
too strong, when the number of electrons is not too large.

\section{Addition Spectrum Calculation}
\label{sec:ph_delta2}
As we have shown in the previous sections, the PH-DMRG method can be used
in order to get accurate results for the ground-state energy of interacting
2D systems. As a useful application we present in this section a calculation
of the addition spectrum of a QD, accompanied by a comparison of the PH-DMRG results
to those of exact diagonalization and of the HF method.

The addition spectrum can be defined by (see section \ref{sec:intro_delta2})
\begin{eqnarray} \label{eqn:ph_Delta2_exact}
\Delta_2 = E_{gs}(N_e) - 2E_{gs}(N_e-1) + E_{gs}(N_e-2).
\end{eqnarray}
Therefore, for a calculation of the addition spectrum one needs the ground-state
energies of $3$ successive electron numbers for each realization. The results
shown in the current section are for the ensemble of $4 \times 6$ samples used in the previous sections, 
occupied by $4,5$ and $6$ electrons, with either NN or Coulomb interactions. 
In general, the results were
better in the Coulombic case, because of a higher accuracy for each energy calculation. 

\begin{figure}[ht]
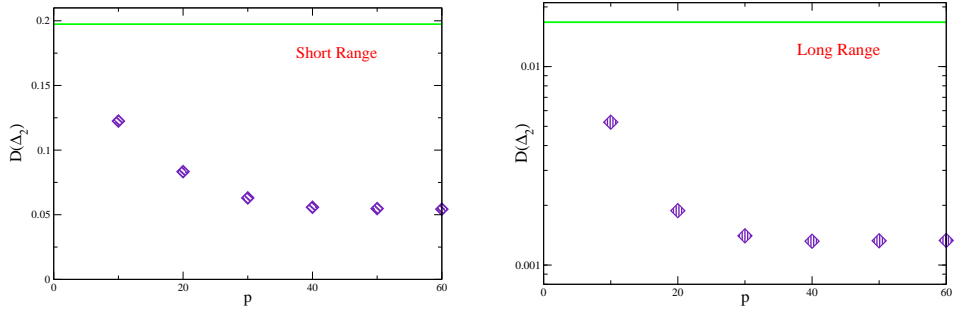
\centering
\begin{minipage}[t]{.45\textwidth}
\includegraphics[trim=0mm 0mm 0mm 0mm, clip, width=.95\textwidth]{d2err_nn}
\end{minipage}
\hfil
\begin{minipage}[t]{.45\textwidth}
\includegraphics[trim=0mm 0mm 0mm 0mm, clip, width=.95\textwidth]{d2_disc_n456_C}
\end{minipage}
\caption[Discrepancy of $\Delta_2$ calculation using PH-DMRG]
{\label{phfig_2}
The averaged discrepancy $D(\Delta_2)$ obtained by the improved 
PH-DMRG calculation, for a $4 \times 6$ lattice occupied by $4,5$ and $6$ electrons, 
with NN (left panel) or Coulombic (right panel) interactions. The PH-DMRG results are shown
as a function of $p$ (symbols), together with the HF results (lines).
Note the semi-log scale in the right panel.
}
\end{figure}

In Fig.~\ref{phfig_2} the results for the averaged discrepancy,
$D(\Delta_2)$, are shown, as a function 
of the number of block-states kept in the PH-DMRG calculation, for both NN (left panel) and
Coulombic (right) interactions. The HF approximation obtains, in the Coulombic case, $D(\Delta_2) \approx 1.7$ percents.
As can be seen, the PH-DMRG method, even with a very small number of states ($p \ge 30$), reduces
significantly the error rate to a level of $\sim 0.13$ percents, an improvement of more 
than an order of magnitude.

\begin{figure}[ht]
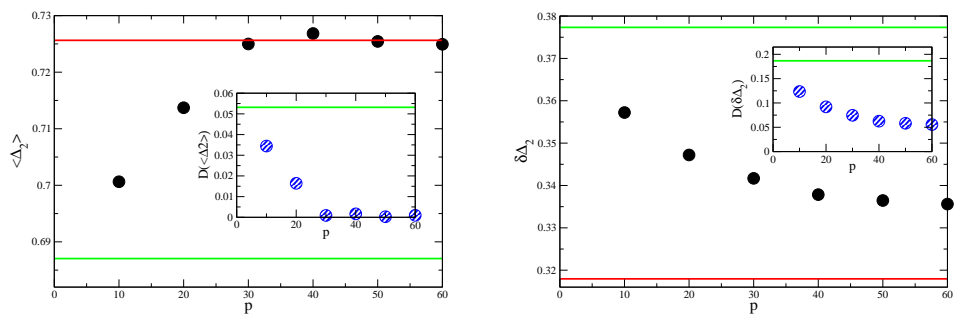
\centering
\begin{minipage}[t]{.45\textwidth}
\includegraphics[trim=0mm 0mm 0mm 0mm, clip, width=.95\textwidth]{d2_avg_n456}
\end{minipage}
\hfil
\begin{minipage}[t]{.45\textwidth}
\includegraphics[trim=0mm 0mm 0mm 0mm, clip, width=.95\textwidth]{dd2_n456}
\end{minipage}
\caption[Calculation of $\langle \Delta_2 \rangle$ and $\delta \Delta_2$ with short-range interactions]
{\label{phfig_3}
The results of the PH-DMRG calculation for (a) $\langle \Delta_2 \rangle$ and
(b) $\delta \Delta_2$ of the $4 \times 6$ system occupied by $4,5$ and $6$ 
electrons with NN interactions, as functions of $p$, the number of states kept. 
The green lines correspond to the HF results, and the red lines to the exact ones.
Insets: $|\langle \Delta_2' \rangle - \langle \Delta_2 \rangle| / |\langle \Delta_2 \rangle| $
and $|\delta \Delta_2' - \delta \Delta_2| / |\delta \Delta_2|$ as functions of $p$.
}
\end{figure}

\begin{figure}[ht]
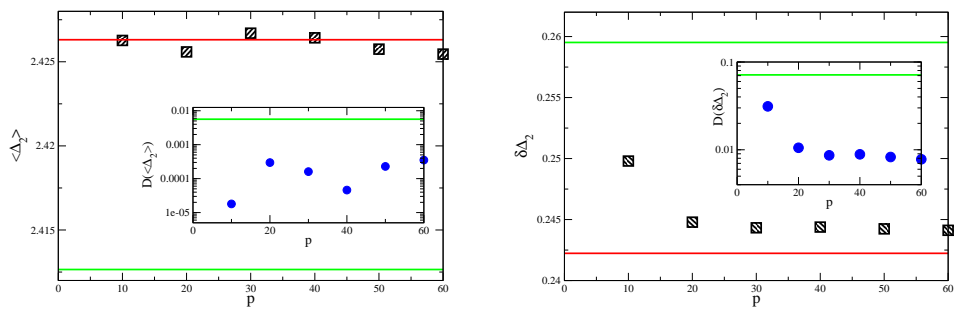
\centering
\begin{minipage}[t]{.45\textwidth}
\includegraphics[trim=0mm 0mm 0mm 0mm, clip, width=.95\textwidth]{d2avg_n456_C}
\end{minipage}
\hfil
\begin{minipage}[t]{.45\textwidth}
\includegraphics[trim=0mm 0mm 0mm 0mm, clip, width=.95\textwidth]{dd2_n456_C}
\end{minipage}
\caption[Calculation of $\langle \Delta_2 \rangle$ and $\delta \Delta_2$ with long-range interactions]
{\label{phfig_4}
The same as Fig~\ref{phfig_3}, but with Coulombic interactions.
Note the semi-log scale in the insets.
}
\end{figure}

However, in the NN case the results are quite poor. The starting point of the PH-DMRG algorithm, i.e.,
the HF results, give an average error of almost $20$ (!) percents. The PH-DMRG improves it by a factor of $4$, 
to the order of $5$ percent, which is still a very high error rate.
It is thus interesting to check whether the calculation of 
$\langle \Delta_2 \rangle$ and $\delta \Delta_2$ can give more accurate results.
Typical results are shown in Fig.~\ref{phfig_3} ($\langle \Delta_2 \rangle$ (left panel)
and $\delta \Delta_2$ (right panel), for NN interactions),
and in Fig.~\ref{phfig_4} (for Coulomb interactions). In all cases the corresponding PH-DMRG results
are compared to the exact solution and to the HF results. In the insets 
we present $|\langle \Delta_2' \rangle - \langle \Delta_2 \rangle| / |\langle \Delta_2 \rangle| $
and $|\delta \Delta_2' - \delta \Delta_2| / |\delta \Delta_2|$.

As can be seen, in the NN case the PH-DMRG results for $\langle \Delta_2 \rangle$
are very accurate for $p \ge 30$. On the other hand, $p=60$ is still not sufficient
in order to get accurate results of $\delta \Delta_2$, which shows very slow convergence.
In both cases, however, the PH-DMRG results are significantly more close to the exact results
than the results obtained by the HF method.

In the Coulombic case, the results for $\langle \Delta_2 \rangle$ are similar to those of
the NN case,
and very small value of $p$ is sufficient to get very accurate results. Notice that the results
continue to fluctuate around the exact result. This results from the fact that the approximation
for $\Delta_2$ is done using $3$ different ground-state approximations, which have different
convergence rates. However, for all values of $p$, the error is less than $0.1$ percent.
For $\delta \Delta_2$, on the other hand, the convergence with increasing $p$ is clearly seen. 
As can be seen, the Coulombic case leads to much more accurate results than those of
the NN interactions, giving an error rate of only $0.8$ percents.

To conclude this section, we define the improvement-factor of the PH-DMRG 
by the ratio between the HF discrepancy, and that of the PH-DMRG. 
Table \ref{tbl:phdmrg_res_d2} summarizes the error rates obtained by the PH-DMRG (using $p=60$)
and the respective improvement factors (in parentheses) for the different cases we've treated.

\begin{table}[ht] 
\centering 
\begin{tabular}{|c||c|c|}
\hline
 & & \\ [0.2ex]
&{\bf NN}&{\bf Coulomb}\\ [1ex]
\hline \hline  
 & & \\ [0.2ex]
$D(\Delta_2)$& 0.054314 (3.6) & 0.001330 (12.6) \\ [1ex]
$\langle \Delta_2 \rangle$& 0.000965 (55.1) & 0.000352 (16.0) \\ [1ex]
$\delta \Delta_2$& 0.055554 (3.4) & 0.007794 (9.2) \\ [1ex]
\hline
\end{tabular}
\caption{\label{tbl:phdmrg_res_d2} PH-DMRG error rates for $\Delta_2$ calculation}
\end{table}

It is easy to see that for Coulomb interactions the PH-DMRG improves all results 
related to the addition spectrum by an order of magnitude, and leads to error rates of
less than $1$ percent for all cases. However, for the NN interactions, a small error rate
and a significant improvement factor are seen only for $\langle \Delta_2 \rangle$,
while modest factors are obtained for $D(\Delta_2)$ and $\delta \Delta_2$,
with error rates larger than $5$ percents.

\section{Conclusions and Future Prospects}
\label{sec:ph_end}
In conclusion, we have seen in this chapter that the PH-DMRG method
can be used for a calculation of the ground-state energy in disordered
systems with interactions. We have analyzed some disadvantages of the
method and suggested an improvement of the traditional implementation.
We have also compared the accuracy of the ground-state calculation between long-range
and short-range interactions, and between intermediate and strong interactions.
We have found that the PH-DMRG works better when the HF approximation is better.
The best accuracy is obtained for long-range interactions which are weak or intermediate, 
when the number of electrons is not near half filling.
 
In each of the cases we have checked, the PH-DMRG leads to a significant improvement
of the ground-state energy approximation from that of the HF method. For example,
while the HF results, for a $4 \times 6$ lattice occupied by $10$ electrons with NN interactions,
show a discrepancy of $4.5$ percents, the PH-DMRG results, which were obtained by keeping up to $p=100$ 
block-states between successive iterations, decrease it to $\sim 1.5$ percent from the 
exact solution. We have compared this method to two other methods which were reported 
in Ref.~\cite{berkovits03-2}, and which can be used to approximate the ground-state energy 
with a truncated Hilbert space. The obtained PH-DMRG error rate is better than the
method we've denoted as energy-cut, and is similar to that of the generation-cut technique.

Nevertheless, for a full comparison one must also consider the feasibility of these methods 
when a treatment of larger systems is required. To understand the difference 
between the methods, let's start with the small lattice size we've treated, $4 \times 6$. 
The energy-cut, based on $18$ energy levels, and the generation-cut, with up to $3$ particle-hole
generations, require the diagonalization of matrices of sizes $43,758$ and $47,916$, respectively. 
The largest superblock diagonalization in the PH-DMRG process, for $p=120$, 
was for a matrix size of $13,494$.

When a larger lattice is treated, and in order to get the same accuracy as for the small system,
the energy-cut and generation-cut methods must include more levels and generations. Moreover,
even was it sufficient to take the same number of levels and generations as in the small case, 
the size of matrices would have grown exponentially with the lattice size. Therefore, the
lattice enlargement makes these methods infeasible even for modest lattice sizes. In the PH-DMRG method, 
on the other hand, the size of the matrix may remain constant, since it depends on the block size, 
and not on the system size.

Yet, for larger systems, the number of single-particle states is larger,
and the discrepancy of the PH-DMRG is expected to increase, unless more block
states are constantly kept. Therefore $p$, and thus the matrix size being diagonalized, are 
expected to increase in the PH-DMRG method as well. 

\begin{figure}[ht]\centering
\includegraphics[trim=0mm 0mm 0mm 0mm, clip, width=7cm,height=!]{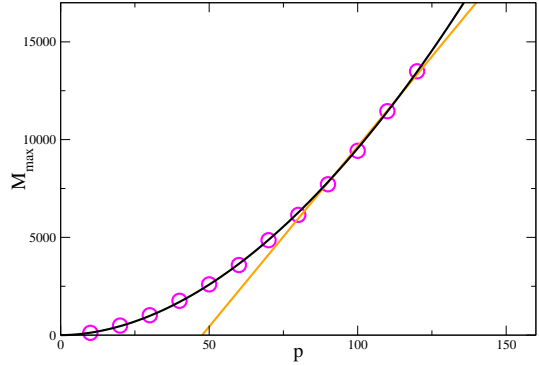}
\caption[Maximal superblock size]
{\label{phfig_6}
The maximal size of the superblock Hamiltonian needed to be diagonalized, for the
$4 \times 6$ lattice with $10$ electrons. The orange line is a linear fit of the
$p>80$ points, while the black line is a power law fit in the entire range.
}
\end{figure}

The largest matrix size needed to be diagonalized in the PH-DMRG process is shown
in Fig.~\ref{phfig_6} as a function of $p$. The dependence of $M_{max}$ on $p$, for large values of $p$,
was empirically found to be linear \cite{dimitrova02}. From our results one can see that 
although a linear fit of the large-$p$ points (orange line) is possible, yet 
a power law including the entire $p$ regime (black line) seems more appropriate,
resulting in $M_{max} \sim p^{1.89}$.
In any case, it is clear that the matrix size is less than quadratic in $p$.
Furthermore, since the largest matrix size used in our current PH-DMRG application is still much 
below the technology limit, its increase should not be a problem.
It is thus clear that in principle the PH-DMRG is capable of treating larger systems.
Indeed, initial studies we have already performed show that the PH-DMRG 
is feasible for systems of the order of $10 \times 10$.

\begin{center}
*~~~~~*~~~~~*
\end{center}

Beyond the ground-state energies, we have used the improved PH-DMRG method in 
order to calculate the addition spectrum, and we have checked its accuracy.
The best accuracy we have obtained is for $\langle \Delta_2 \rangle$, in 
which a rapid convergence is exhibited, and a very small number of block-states
is sufficient, for both short-range and long-range interactions. 
The accuracy of the $\delta \Delta_2$ calculation was pretty poor in the NN case,
resulting in more than $5\%
$ error and very slow convergence for $p=60$.
Even though, the PH-DMRG 
results of $\delta \Delta_2$ are much more accurate than the corresponding HF results.
Moreover, for the case of long-range interactions, the $\delta \Delta_2$ calculation
was much better, getting an accuracy of $\sim 0.8$ percent.
The $\Delta_2$ discrepancy, averaged over the different realizations, has obtained 
in the long-range case
$\sim 0.13$ percents of error, much better than the $\sim 1.7$ percents of the HF results.

A comprehensive study of the addition spectrum was performed a few years ago, 
using the self-consistent HF method, on 2D lattices of the order of $10 \times 10$ \cite{walker99}. 
It was found that for strong interactions the fluctuations of $\Delta_2$ do not scale with 
the mean level spacing $\Delta$ for both short-range and long-range interactions, as opposed
to a single-parameter scaling argument. In the case of short-range interactions, 
a saturation of $\delta \Delta_2$ was obtained for very high values of interaction  
and it was explained as an appearance of charge density modulations.
For Coulombic interactions the increase of $\delta \Delta_2$ was shown to be faster than 
linear in $\Delta$.

Based on our results, however, one might question the ability of the HF method 
to provide an accurate approximation of the addition spectrum. Our results point 
out that the usage of the HF method may lead to quite high error rates,
especially for short-range interactions. It can thus be
interesting to investigate the addition spectrum for such systems using the PH-DMRG, 
which is appropriate for such system sizes, and which can improve the accuracy significantly. 
A PH-DMRG research of the addition spectrum is thus left for a future study.

\cleardoublepage
\chapter{Two-Electron Magnetization in Quantum Dots} \label{cpt:ch4}

In this chapter we study the magnetization of an interacting quantum dot occupied by spin $1/2$ electrons,
in the presence of spin-orbit coupling. Modeling the dot by a two-dimensional lattice, we
utilize exact calculations for lattices with a small number of electrons. 
For an $N$-sites lattice, which is occupied by $n_e$ electrons, the Hilbert space 
size is $M=\binom{2N}{n_e}$, so that if the size of the system is not too large, and it is occupied by 
relatively small number of electrons, 
one can still use an exact diagonalization method and get the exact lowest wave-functions,
even when interactions between the electrons are considered. 

We calculate the expectation value of the spin operators in the ground-state, and by a comparison of 
the system's energy with and without a magnetic field, we also calculate the g-factor.
For the case in which there are two electrons occupying an interacting quantum dot, we find a level crossing
between the two lowest many-body eigenfunctions as a function of the spin-orbit scattering rate, 
resulting in a finite magnetization of the ground-state. 
This is a clear evidence for the importance of the interplay between spin-orbit scattering and interactions,
and can have a significant influence on g-factor 
measurements.

\section{Introduction}
The effects of spin-orbit (SO) coupling on the energy spectrum of quantum dots (QDs) have attracted notable attention in the 
recent years \cite{halperin86,beenakker97,ralph95,davidovic99}. 
Much work has been concentrated, both experimentally and theoretically, on the magnetization 
of mesoscopic samples. Specifically many studies focus on g-factor measurements,
and try to understand them theoretically. The g-factor is
defined through the splitting of the Kramers' doublets \cite{kramers30,Merzbacher} in the presence of a weak magnetic field, 
$\epsilon_{i}^{(H)} - \epsilon_{i}^{(0)} = \pm \frac{1}{2} g \mu_B H$, where 
$\epsilon_i^{(H)}$ ($\epsilon_{i}^{(0)}$) is the $i$-th single-particle energy level with (without) a 
magnetic field, $\mu_B$ is the Bohr magneton and $H$ is the magnetic field.

For free electrons the g-factor is constant, $g=2$, and this value is more or less correct also for
bulk measurements in various metals \cite{halperin86}. However, in experiments done on mesoscopic samples,
values which are significantly less than the free value of the g-factor were obtained. Moreover, large
fluctuations in the measured values were seen. These findings attracted theoretical attention, and resulted
in two analytical studies which have obtained, within the framework of the random matrix theory (RMT), 
a description of the g-factor probability distribution in the presence of SO coupling and disorder 
but without interactions \cite{brouwer00,matveev00}. 
In a more recent study, the statistical properties of these distribution functions were related to several 
physical observables \cite{mucciolo06}. 
According to these studies, the SO coupling influences the probability distribution of the g-factors of the 
discrete energy levels. The distribution function was shown to be universal, where the 
width is expressed in terms of various physical parameters.
The presence of strong SO coupling and disorder causes the g-factor to fluctuate from sample to sample, 
even when both have the same characteristic strength of SO coupling. Moreover, the g-factor is expected to fluctuate also 
between different levels of a specific sample, according to the RMT distribution function. 

The definition of the g-factor through the splitting of the Kramers' doublets thus
relates the g-factor to the change in the energy of a {\bf specific} energy level when
a weak magnetic field is applied. Therefore,
\begin{eqnarray} \label{eqn:g_gdef}
g_{i,\sigma} = \frac{2 \left[ \epsilon_{i\sigma}^{(0)}-\epsilon_{i\sigma}^{(H)} \right]}{\mu_B H},
\end{eqnarray}
where $g_{i,\sigma}$ is the g-factor of the $i$-th level, with spin $\sigma$. The 
spin index $\sigma \in \{+,-\}$ is used to relate the states $|i,+\rangle$ and $|i,-\rangle$ via a
time reversal operation.
In the absence of a magnetic field each level is two-fold degenerate (Kramers' degeneracy \cite{kramers30,Merzbacher}), 
and this degeneracy is lifted by the magnetic field, which causes the energy of one of the levels
to increase and the energy of the other to decrease. Therefore, $g_{i,\sigma}$ as defined by this formula can have either
sign, depending on the direction of the energy change. The ground-state energy always decreases
when a magnetic field is applied, thus the g-factor of the ground-state obtained by Eq.~(\ref{eqn:g_gdef}) 
is positive. Usually, the value of $g$ does not depend on the spin index, at least
to zeroth order in $H$, so that one can denote the g-factor of the $i$-th level as $\pm g_i$,
with the convention that $g_i \ge 0$.

According to Eq.~(\ref{eqn:g_gdef}), a measurement of the g-factor should compare the specific 
energy level before and after the magnetic field is applied. 
However, practical measurements are usually related to the total energy of the system, and not
to that of a specific level. Nevertheless, if there is no interaction between particles,
the change of the total ground-state energy 
due to magnetic field may be equivalent to that of the highest filled level. This is the case when the
number of electrons $n_e$ is odd, i.e. $n_e=2p+1$, with an integer $p$. 
The lowest $n_e-1$ levels are composed of $p$ Kramers' pairs, 
where in each pair one level increases and the other decreases in the presence of a magnetic field, so that their
total contribution vanishes. Therefore, the motion of the highest level, $p+1$, which is singly occupied, 
will determine the g-factor, so that one can write
\begin{eqnarray} \label{eqn:g_gdef2}
g(n_e) = \frac{2 \left[ E_{gs}^{(0)}(n_e)-E_{gs}^{(H)}(n_e) \right]}{\mu_B H},
\end{eqnarray}
where $g(n_e)$ denotes the g-factor of the ground-state with $n_e$ electrons, and $E_{gs}^{(H)}(n_e)$
represents its energy in the presence of a magnetic field $H$.

In addition, when the number of electrons $n_e$ is even, the total ground-state energy is not expected to
change when a magnetic field is applied, since all the filled levels divide into pairs, in which 
the movement of one level is compensated by the other\footnote[1]{Actually, in the presence of SO coupling and
magnetic field, there are some second order corrections to the energy, which lead to
first order corrections to the g-factor. This issue will be addressed in section \ref{sec:g_I0}. 
However, to first order in $H$ the ground-state energy of an even-occupied non-interacting system 
is constant, and $g=0$.}. 
Therefore, for an even number of electrons, 
a calculation of the g-factor using Eq.~(\ref{eqn:g_gdef2}) gives $g=0$. 
One can thus use Eq.~(\ref{eqn:g_gdef2}) as a practical definition of the g-factor.

However, returning to the experimental point of view, one should notice that 
when the measurement of the energy is indirect, an interpolation of the results 
is also required. For example, using tunneling spectroscopy one can measure the position in which
a conductance peak of a QD occurs. Such an event relates the energies of the QD with $n_e-1$ and $n_e$ 
electrons, with the gate voltage $V_g$, by the relation (see section \ref{sec:intro_CB})
$eV_g = E_{gs}(n_e)-E_{gs}(n_e-1)$.
When a magnetic field is applied, the peak will move as a function of $H$. Therefore, 
by denoting the measured g-factor by $\tilde g$, one can analyze the
peak motion in order to determine the g-factor, by calculating
\begin{eqnarray} \label{eqn:g_gdef_tot}
\tilde g &=& \frac{2 \left[ eV_g(0)-eV_g(H) \right]}{\mu_B H} \\ \nonumber
&=& \frac{2 \left[ E_{gs}^{(0)}(n_e)-E_{gs}^{(H)}(n_e) \right]}{\mu_B H}
- \frac{2 \left[ E_{gs}^{(0)}(n_e-1)-E_{gs}^{(H)}(n_e-1) \right]}{\mu_B H} \\ \nonumber
&=& g(n_e) - g(n_e-1).
\end{eqnarray}
Since either $n_e$ or $n_e-1$ is even, its corresponding g-factor vanishes, and thus $\tilde g$
is equivalent to the other. Namely, $\tilde g = g(n_e)$ or $\tilde g = -g(n_e-1)$. 
Actually, since each peak is divided, in the presence of a magnetic field, into two peaks moving in opposite directions,
an extraction of $\tilde g$ from successive peaks results in a set of the single-particle g-factors, 
i.e. $g_1, -g_1, g_2, -g_2, \dots$.

Indeed, tunneling spectroscopy measurements have obtained many reasonable results for g-factors of 
nano-particles. For example, several experimental studies of metallic three dimensional nano-particles have shown the reduction of the 
measured g-factor as a function of the spin-orbit scattering rate, in accordance with the RMT predictions. 
For Aluminum nano-particles, in which the SO coupling is negligible, the measured g-factor values are approximately 
those of free electrons ($g=2$) \cite{ralph95}, while for Gold nano-particles, in which the SO coupling is strong, the 
measured g-factors were in the range of $0.28-0.45$ \cite{davidovic99}. Furthermore, by extracting several g-factors from 
each sample, Petta and Ralph have succeeded to present an impressive confirmation of the theoretical 
RMT distribution function \cite{petta01}. Nevertheless, it should be mentioned that while the average 
g-factor is expected to be reduced by the SO coupling for three dimensional samples, the RMT predicts an enhancement 
of the g-factor in two dimensional (2D) samples \cite{matveev00} as a function of SO coupling. Yet, no experiment which
measured this increase has been performed to date. 

However, as we have mentioned, Eq.~(\ref{eqn:g_gdef2}) is entirely equivalent to Eq.~(\ref{eqn:g_gdef}) 
only when the system is non-interacting.
Once interactions between the electrons are important, 
it should be emphasized that Eq.~(\ref{eqn:g_gdef2}) is a definition of a {\bf many-particle g-factor}, 
which depends on the total magnetization of the ground-state wave-function. For example, one can obtain  
values of g-factors which are larger than $2$, even in the absence of orbital effects (e.g., when
the magnetic field is in-plane), a phenomena that cannot happen for a single-particle g-factor.

Nevertheless, the RMT-based theoretical studies cited above were performed in the context of one-body levels, 
neglecting any effect of the electron-electron interactions. Indeed, by adding to the RMT 
Hamiltonian an interaction term, using the constant interaction model, an increase of the 
g-factor fluctuations was reported \cite{gorokhov03,gorokhov04}. It was shown that the interactions result in 
a possibility of getting non-trivial spin values in the ground-state, and accordingly in an optional 
enhancement of the g-factor, to values greater than $2$. 

Although the theoretical studies of Refs.~\cite{gorokhov03,gorokhov04} were done for an odd-electron occupation, 
their results suggest the possibility of a non-trivial spin polarization for the even-electron case as well. 
Similar phenomenon was found in disordered dots without spin-orbit coupling and
with infinitely large Hubbard interaction, where 
occupation of an even number of electrons caused non-vanishing spin values \cite{berkov99}.
If, for any reason, this is the case, and the g-factor of an even-electron ground-state differs from zero,
then {\it the quantity measured in tunneling spectroscopy may not equal the single-level g-factor
nor the many-particle g-factor}. In such a case it should be related to the difference between two 
many-particle g-factors, as shown in Eq.~(\ref{eqn:g_gdef_tot})


In this chapter we investigate the ground-state magnetization properties, such as the spin polarization 
and the g-factor of QDs with an even number of particles. Without interactions, such states have $g=0$,
as well as $\langle S_z \rangle = 0$, as predicted. Nevertheless, we show that the interplay between 
spin-orbit scattering and electron-electron interactions may result in a level crossing (LC) between the two lowest
many-body levels. When these states are close in energy, the magnetic field splits them into 
two polarized states with a finite magnetization. Therefore, there exist a possibility to have non-vanishing 
$\langle S_z \rangle$ and g-factor in the two-particle ground-state.


\subsection{Chapter's Outline}
The rest of the chapter is organized as follows. In the next section we describe
the model Hamiltonian we use in order to incorporate, beside the magnetic field, both 
SO coupling and interactions between electrons. In section \ref{sec:g_I0} we present 
results for a non-interacting system, for both single- and double-occupation,
which are shown to reproduce some known ground-state properties. 
We find that there are specific values of the SO coupling strength, in which the Kramers' 
doublet remains degenerate even when a magnetic field is applied. In such points
both states of the doublet have $\langle S_z \rangle = 0$.

The results for the case in which interactions are also considered
are presented in section \ref{sec:g_I}. Our results point out that a finite
magnetization can be obtained for systems with an even-particle occupancy. In section
\ref{sec:g_phys} we discuss the experimental relevance of this finding, and we show that
it might affect g-factor measurements. In the last section we
conclude and address some future possibilities to continue the research.

\section{Model}
In order to model the QD we choose a tight-binding description of a finite 2D
lattice with $A$ columns and $B$ rows (the number of sites is denoted by $N=AB$), with open
boundary conditions, which is occupied by $n_e$ spin $1/2$ electrons. 
In some sections of the chapter, only point interactions 
are considered, i.e., the Hubbard term which couples up and down spins occupying the same 
lattice site. In other parts we add to the Hubbard term either nearest neighbor (NN) or 
Coulomb interactions. In addition, spin-flips during hopping processes are possible, 
with a finite probability, as a result of a coupling between the spin degree of freedom
and the orbital motion. 
Separating the interactions from the free part, one can write the Hamiltonian as 
$\hat H_{\rm QD} = \hat H_0 + \hat H_{\rm int}$, 
where the free part can be divided, if disorder effects are neglected, to a hopping term and a Zeeman term, i.e.
$\hat H_0 = \hat H_{\rm hop}+ \hat H_B$.
Each of these terms was discussed in details in section \ref{sec:intro_Hqd}.
The hopping part of the Hamiltonian is thus
\begin{eqnarray} \label{eqn:g_Hdot_0}
\hat H_{\rm hop} =
- \sum_{m,n,\sigma,\sigma^\prime} ( V_{x} \hat a^\dagger_{m,n,\sigma} \hat a_{m,n+1,\sigma^\prime} +
 V_{y} \hat a^\dagger_{m,n,\sigma} \hat a_{m+1,n,\sigma^\prime} + H.c.),
\end{eqnarray}
where $\hat a^\dagger_{m,n,\sigma}$ ($\hat a_{m,n,\sigma}$)
is a creation (annihilation) operator of an electron with spin $\sigma$ in the
lattice site placed in row $m$ and column $n$. The matrices $V_x$ and $V_y$ are 
defined (in the absence of a magnetic field) by \cite{ando89}
\begin{equation} \label{eqn:Intro_Hso_V2}
V_x = \left( \begin{array}{cc}
	V_1 & V_2 \\
	-V_2 & V_1 \\
	\end{array} \right)~~;~~~
	V_y = \left( \begin{array}{cc}
	V_1 & -iV_2 \\
	-iV_2 & V_1 \\
	\end{array} \right)~,
\end{equation}
where $V_1$ ($V_2$) is the hopping matrix element, for events which conserve (flip) the spin.

When a perpendicular magnetic field is applied, it adds a phase to the hopping matrix element. 
We take the field direction to be perpendicular to our 2D sample, i.e. along the $\hat z$ axis,
and we choose a gauge in which
the vector potential is $A=Hy \hat x$. With that gauge,
one has to modify the hopping elements is the $\hat x$ direction (see section \ref{sec:intro_Hqd}), 
so that $V_x \rightarrow V_x e^{i \theta m}$, with $m$ being the row number
and $\theta = \frac {2 \pi H s^2}{\phi_0}$. Here 
$s$ is the lattice constant and $\phi_0 = hc/e$ is the magnetic flux quantum.
In both cases, i.e. with and without a magnetic field, the overall hopping amplitude, 
$t = \sqrt{V_1^2 + V_2^2}$, is taken as the energy scale of the problem. 
In other words, all energy terms are expressed in units of $t$.

The strength of the SO coupling can be expressed by the ratio between the 
the spin-flip amplitude and the total hopping element (excluding phases).
Using a dimensionless parameter 
$\lambda = \frac {V_2}{\sqrt{V_1^2+V_2^2}}= V_2/t$, we examine the entire range of $\lambda$, between 
very weak ($\lambda \rightarrow 0$) and very strong ($\lambda \lesssim 1$) spin-orbit 
coupling.
A similar approach is usually utilized within the RMT framework, by writing 
$H = (1-\alpha) H_{GOE} + \alpha H_{GSE}$. Here GOE (GSE) denotes the Gaussian orthogonal 
(symplectic) ensemble, which corresponds to the case in which the SO coupling is very weak (strong). 
Changing $\alpha$ from $0$ to $1$ modifies the Hamiltonian between these two 
limits\footnote[2]{Note that the definitions of $\lambda$ and $\alpha$ are different. A strong SO coupling 
which is related to the symplectic ensemble of RMT ($\alpha \rightarrow 1$), corresponds, 
in our model, to $\lambda \approx 0.5-0.7$. The physical relevance of higher values of $\lambda$ is questionable.}.

With the choice of a perpendicular magnetic field, the Zeeman term in the Hamiltonian is diagonal, 
and can be written as
\begin{eqnarray} \label{eqn:g_Hdot_B}
\hat H_B = \displaystyle \mu_B H \sum_{m,n,\sigma} \sigma \hat a^\dagger_{m,n,\sigma} \hat a_{m,n,\sigma},
\end{eqnarray}
where $\sigma = \pm 1$. 

The Zeeman energy can be related to the hopping phase $\theta$ and to the hopping amplitude $t$ 
by the following consideration. The phase $\theta$
is a dimensionless parameter, measuring the magnetic flux 
throughout a lattice unit cell, in units of the quantum flux $\phi_0$.
One can express the absolute value of the Zeeman energy as
$\mu_B H = \mu_B \phi_0 \frac {\theta} {2 \pi s^2}$. Substituting the physical constants 
$\mu_B \phi_0 = \frac {\pi \hbar^2}{m_0}$, $m_0$ being the electron mass, 
and using the relation $t=\frac{\hbar^2}{2m_{\rm eff}s^2}$,
where $m_{\rm eff}$ is the effective mass,
one gets $\mu_B H = \frac {\theta \hbar^2} {2 m_0 s^2} = \theta t \frac{m_{\rm eff}}{m_0}$.
For the metallic nano-particles used in several experiments \cite{ralph95,davidovic99,petta01},
the ratio between the effective mass and the electron mass is close to unity.
We will assume $m_{\rm eff} = m_0$;
Deviations from this value will not affect our main results. Finally, 
since all energies are measured in units of $t$, the strength of the magnetic field 
$\mu_B H / t$ determines exactly the hopping phase.

The Hubbard term results in
\begin{eqnarray} \label{eqn:g_Hdot_int}
\hat H_{\rm int}^{\rm (Hubbard)} = U_H \displaystyle \sum_{m,n}
{\hat a}^{\dagger}_{m,n,\uparrow}{\hat a}_{m,n,\uparrow}
{\hat a}^\dagger_{m,n,\downarrow}{\hat a}_{m,n,\downarrow},
\end{eqnarray}
where $U_H$ is the Hubbard interaction strength.
When an increase of the range of interactions is considered, by either NN or Coulomb interactions,
one of the following terms is added to the Hubbard term:
\begin{eqnarray} \label{eqn:g_Hint_C_NN}
{\hat H}_{\rm int}^{\rm (NN)} &=& 
\displaystyle \sum_{<m_1,n_1;m_2,n_2>;\sigma_1,\sigma_2} U_{\rm NN}~
{\hat a}^{\dagger}_{m_1,n_1,\sigma_1}{\hat a}^{\dagger}_{m_2,n_2,\sigma_2}{\hat a}_{m_2,n_2,\sigma_2}{\hat a}_{m_1,n_1,\sigma_1} \\ \nonumber
{\hat H}_{\rm int}^{\rm (C)} &=& 
\displaystyle \sum_{m_1,n_1 \ne m_2,n_2;\sigma_1,\sigma_2}\frac{U_C}{|r_{m_1,n_1}-r_{m_2,n_2}|}
{\hat a}^{\dagger}_{m_1,n_1,\sigma_1}{\hat a}^{\dagger}_{m_2,n_2,\sigma_2}{\hat a}_{m_2,n_2,\sigma_2}{\hat a}_{m_1,n_1,\sigma_1},
\end{eqnarray}
where $<m_1,n_1;m_2,n_2>$ denotes that the sites $m_1,n_1$ and $m_2,n_2$ are NNs, and $U_{\rm NN}$ is the
interaction strength between NN. For the Coulomb interactions, the distance between sites
$m_1,n_1$ and $m_2,n_2$, expressed in lattice constant units, is denoted by $|r_{m_1,n_1}-r_{m_2,n_2}|$,
and $U_C$ is the Coulomb interaction strength between sites which are one lattice constant apart.

In order to calculate the spin polarization of the QD we apply a weak magnetic field 
along the $\hat z$ axis and calculate $\langle \hat S_z \rangle$ for the lowest levels. 
For g-factor calculations, we compare the ground-state energies with and without the magnetic 
field for each sample, and use Eq.~(\ref{eqn:g_gdef2}).
We use $\mu_B H/t \sim 10^{-4} - 10^{-3}$, for dots in which
the mean level spacing is of the order of $0.1t$.
For an experimental system in which the mean level spacing is $0.1-1 ~ meV$, it
is equivalent to a magnetic field of $10-1000 ~ G$, in correspondence with practical measurements \cite{petta01}. 

The Hamiltonian $\hat H_{\rm QD}$ is exactly diagonalized using Lanczos
procedure, for lattices of up to $11 \times 10$ sites, occupied by $1$ or $2$ electrons. 
A discussion of the lattice sizes will be given in section \ref{sec:g_phys}.
It will be shown that some of the significant results, to be detailed 
in the following sections, are not an artifact 
of small sizes. From checking the size dependence of these results one can see
that they will not disappear, even for larger QD sizes. Moreover, they 
are expected to be even more pronounced. 


\section{Non-Interacting Electrons}
\label{sec:g_I0}
We start by presenting the non-interacting results for the spin polarization,
by taking $U_H=U_{\rm NN}=U_C=0$ . The expectation value of the operator $\hat S_z$ in the low lying states, 
i.e. $\langle \hat S_z \rangle$, is calculated when a weak magnetic field ($\mu H \ll t$) 
is applied along the $\hat z$ axis.

\begin{figure}[htbp]
\centering
\includegraphics[trim=0mm 0mm 0mm 0mm, clip, width=3in,height=!]{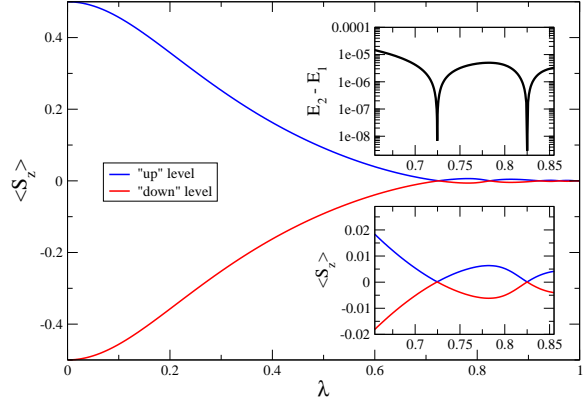}
\caption[Single-level spin polarization of non-interacting electrons]
{\label{fig:sz_n1} 
The spin projection $\langle \hat S_z \rangle$ of the lowest two single-particle 
levels are shown as a function of the spin-orbit coupling strength $\lambda$,
for a non-interacting system of $8 \times 7$ sites.
Insets: zoom into the strong spin-orbit regime shows signs of level crossings.
The upper inset shows the energy difference between the first two levels when a 
magnetic field is applied (notice the logarithmic scale), 
and the lower one shows $\langle \hat S_z \rangle$ of each.
}
\end{figure}

The results for $\langle S_z \rangle$ as a function of the SO parameter $\lambda$, of singly-occupied 
states in a $8 \times 7$ lattice, are shown in Fig.~\ref{fig:sz_n1}. 
When it is needed, we denote by $\langle S_z^{(m)} \rangle$ the expectation value of the
operator $\hat S_z$ in the $m$-th eigenfunction ($m=1$ being the ground-state).
In places where expressions with different numbers of particles are related, we use
$\langle S_z^{(m)}(n_e=n) \rangle$ to denote that the expectation value is evaluated 
in the $n$-particle Hilbert space.

We start with the single-particle levels, taking $n_e=1$.
Without the magnetic field, all single-particle states (and in 
particular the ground-state) are doubly-degenerate (the Kramers' degeneracy) \cite{kramers30,Merzbacher}. 
When a magnetic field is applied, it splits this degeneracy, and one gets to zeroth order
in the magnetic field, $\langle S_z^{(1)}\rangle = - \langle S_z^{(2)}\rangle$.
For $\lambda \rightarrow 0$, $\left| \langle S_z\rangle \right| \rightarrow \frac{1}{2}$,
and increasing the SO coupling leads to a decrease of 
$\left| \langle S_z\rangle \right|$, as can be expected.
For different levels one gets similar, although not identical curves, with the 
same qualitative limits for weak and strong spin-orbit coupling. Results for the g-factor
calculation are similar, i.e. for $\lambda \rightarrow 0$ we get (for the ground-state) $g = 2$, and
as the SO strength increases $g$ drops monotonically towards $g=0$.

In the insets of Fig.~\ref{fig:sz_n1} we zoom into the regime of strong SO coupling, 
showing the difference in the energies of the two lowest states (upper inset), and each of their 
$\langle S_z \rangle$ (lower inset). Both plots point towards
LCs between the first two single-particle levels. At the crossing points both 
states have $\langle \hat S_z \rangle = 0$, so that the magnetic field does not break  
Kramers' degeneracy, an unusual result. It should be noted, however, that
the crossing points are obtained for values of $\lambda > \frac{\sqrt{2}}{2}$, which correspond to the
situation in which a spin-flip hopping process is more probable than a spin-conserving one, and
the existence of such a regime in practice is improbable.

\begin{figure}[htbp]
\centering
\includegraphics[trim=0mm 0mm 0mm 0mm, clip, width=3in,height=!]{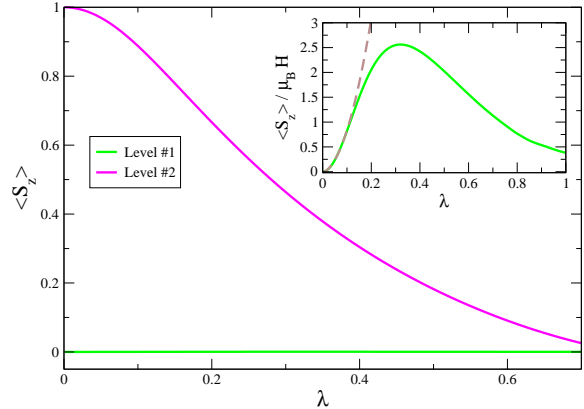}
\caption[Spin polarization of two non-interacting electrons]
{\label{fig:sz_n2_I0} 
The spin projection $\langle \hat S_z \rangle$ of the lowest two doubly-occupied levels 
are shown as a function of the spin-orbit coupling strength $\lambda$,
for a non-interacting system of $8 \times 7$ sites.
Inset: the spin polarization of the two-particle ground-state, which has 
a first-order dependence on the magnetic field $H$. For weak SO coupling, it also has 
a quadratic dependence on $\lambda$ (the dashed brown line represents a quadratic fit for small $\lambda$).
}
\end{figure}

For $n=2$, without interactions, the many-body state is a Slater determinant
of the single-particle states, and since the operator $S_z$ is additive, one can write 
$\langle S_z^{(1)}(n_e=2)\rangle = \langle S_z^{(1)}(n_e=1)\rangle + \langle S_z^{(2)}(n_e=1)\rangle$, and
$\langle S_z^{(2)}(n_e=2)\rangle = \langle S_z^{(1)}(n_e=1)\rangle + \langle S_z^{(3)}(n_e=1)\rangle$.
To zeroth order in the magnetic field, as noted above, the two contributions to
$\langle S_z^{(1)}(n_e=2)\rangle$ cancel each other, and this term vanishes. 
The results are shown in Fig.~\ref{fig:sz_n2_I0}. The results for the g-factor
calculation are similar, i.e. for the ground-state we get $g \approx 0$, while
for the first excited state $g=4$ for $\lambda=0$, and it decreases monotonically towards $g=0$
when $\lambda$ increases.

Since for the g-factor measurements the limit $H \rightarrow 0$ is taken, the approximation
$\langle S_z^{(1)}(n_e=2)\rangle \approx 0$ is usually sufficient. 
However, as can be seen in the inset of Fig.~\ref{fig:sz_n2_I0}, 
there is also a higher order term. For weak spin-orbit coupling and weak magnetic field, 
using first-order perturbation theory in both $\lambda$ and $H$, 
it was shown that there is an additional contribution to $S_z$, which is proportional to 
$\lambda^2 H$ \cite{sone76}. This explains the quadratic increase of $\langle S_z^{(1)}(n_e=2) \rangle / \mu_B H$
as a function of $\lambda$ which is shown in Fig.~\ref{fig:sz_n2_I0}(inset) for weak SO coupling,
in which the perturbation theory is valid. For moderate values of SO coupling,
our results suggest that this trend is reversed, and $\langle S_z^{(1)}(n_e=2) \rangle / \mu_B H$ 
starts to decrease\footnote[3]{Whereas the exact point of the maximum in $\langle S_z^{(1)}(2) \rangle / \mu_B H$ 
is different for different lattice sizes, the qualitative shape for all the sizes checked was the same.}.


Returning to the points of LC shown in the insets of Fig.~\ref{fig:sz_n1},
it is important to notice that such crossings occur between states which are
the time reversal of each other, or, in other words, 
states which belong to the same Kramers' pair. 
No such crossings occur between states which originate from 
different pairs, although the single-particle level-spacing is significantly reduced by the SO coupling. 
Yet, the energy difference between subsequent Kramers' pairs
is much larger than the energy contribution of the weak magnetic field we apply. 
Correspondingly, for the double-occupation case, the lowest two states (those which are presented in Fig.~\ref{fig:sz_n2_I0})
should not exhibit any crossing.
This can change once interactions are considered, as will be shown in the next section.


\section{Interplay Between Spin-Orbit Coupling and Interactions}
\label{sec:g_I}
\begin{figure}[htbp]
\centering
\includegraphics[trim=0mm 0mm 0mm 0mm, clip, width=3in,height=!]{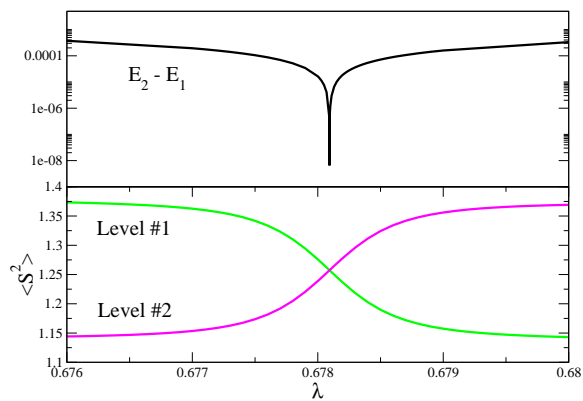}
\caption[Level crossing of the doubly-occupied states with interaction]
{\label{fig:dE_S2_n2} 
Typical results of the level crossing of the two lowest doubly-occupied states. The results 
shown were obtained for a system of $8 \times 7$ sites, with $U_H=10$. 
Upper panel: the energy difference $E_2 - E_1$ (without a magnetic field) is shown as a 
function of the spin-orbit coupling strength $\lambda$ (notice the semi-log scale). 
The dip shows the crossing point.
Lower panel: the switch of $\langle \hat S^2 \rangle$ (in the presence of a
magnetic field) between these two states, which occurs at the same place.
Note the tiny scale of $\lambda$.
}
\end{figure}

We now move to study the effect of interactions, by turning on the Hubbard interaction term,
for which an interaction energy $U_H>0$ is paid for a couple of electrons occupying the same lattice site. 
Calculating the ground-state energies of the two lowest doubly-occupied states, one finds that
there is a LC between these states, a feature which does not exist for the non-interacting case.
In the non-interacting case different levels may approach each other when the
SO coupling increases, yet the minimal distance between them is much larger than the magnetic energy.
The presence of interactions enhances this tendency, 
towards the situation in which a LC is possible.
This crossing happens at a certain value of the SO coupling, i.e. at $\lambda = \lambda_c$,
and in its vicinity, the expectation value of $\hat S^2$ switches
between these states, as can be seen in Fig.~\ref{fig:dE_S2_n2}.

\begin{figure}[htbp]
\centering
\includegraphics[trim=0mm 0mm 0mm 0mm, clip, width=3in,height=!]{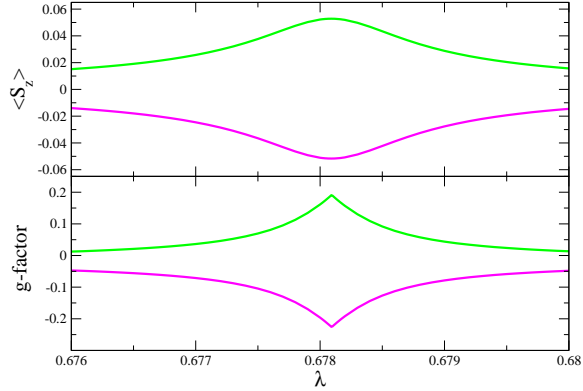}
\caption[Spin polarization and g-factors in the level crossing regime]
{\label{fig:Sz_n2_I} 
Typical results of the spin polarization $\langle \hat S_z \rangle$ (upper panel) 
and the g-factor, calculated using Eq.~(\ref{eqn:g_gdef2}) (lower panel),
of the two lowest doubly-occupied states, in the regime of the level crossing between them. 
The results shown were obtained for a system of $8 \times 7$ sites, with $U_H=10t$
and $\mu_B H=10^{-4} t$. Note the tiny scale of $\lambda$.
}
\end{figure}

Looking at the energy curves and the switching of $\langle \hat S^2 \rangle$, one 
would naively expect that the magnetization properties, e.g. $\langle \hat S_z \rangle$ 
and the g-factor, will switch as well at $\lambda_c$.
However, as the energies of these two states become close enough to each other, the
energy associated with the magnetic field becomes more and more important, resulting in a 
polarization of the spins of both states. This leads to an enhancement of
$\langle \hat S_z \rangle$ and the g-factor values in the crossing region.
As can be seen in Fig.~\ref{fig:Sz_n2_I}, both $\langle \hat S_z \rangle$ and the g-factor 
can reach significant values.

\begin{figure}[htbp]
\centering
\begin{minipage}[t]{.45\textwidth}
\includegraphics[trim=0mm 0mm 0mm 0mm, clip, width=.95\textwidth]{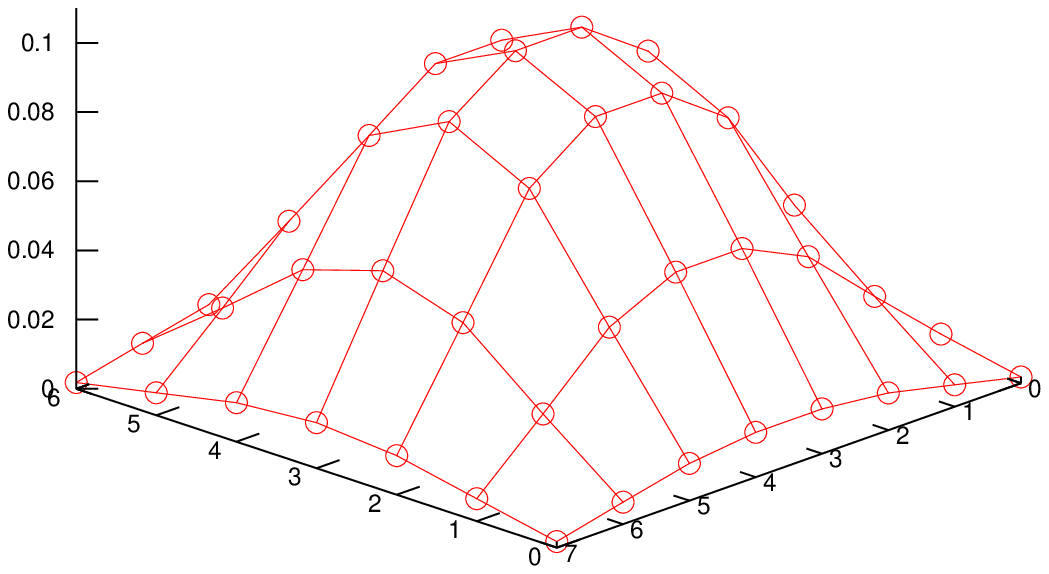}
\end{minipage}
\hfil
\begin{minipage}[t]{.45\textwidth}
\includegraphics[trim=0mm 0mm 0mm 0mm, clip, width=.95\textwidth]{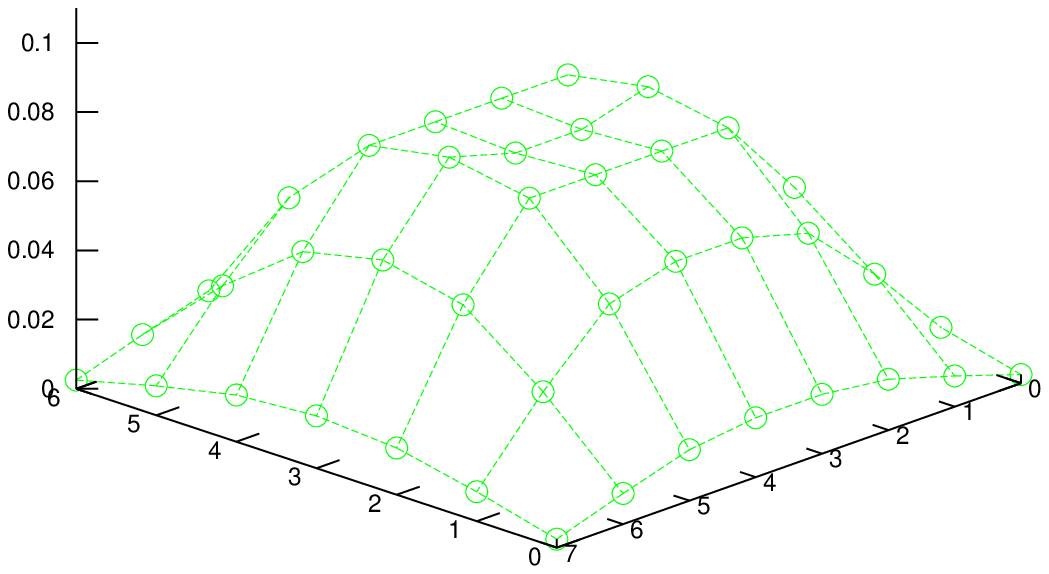}
\end{minipage}
\caption[Electron distribution for an arbitrary SO coupling]
{\label{fig:pop_0.550} 
The electron distribution of the first two doubly occupied states 
(left - ground state, right - first excited state) for a 
$8 \times 7$ lattice with Hubbard interactions of strength $U_H=10t$ and 
SO coupling $\lambda=0.550$.
}
\end{figure}

\begin{figure}[htbp]
\centering
\begin{minipage}[t]{.45\textwidth}
\includegraphics[trim=0mm 0mm 0mm 0mm, clip, width=.95\textwidth]{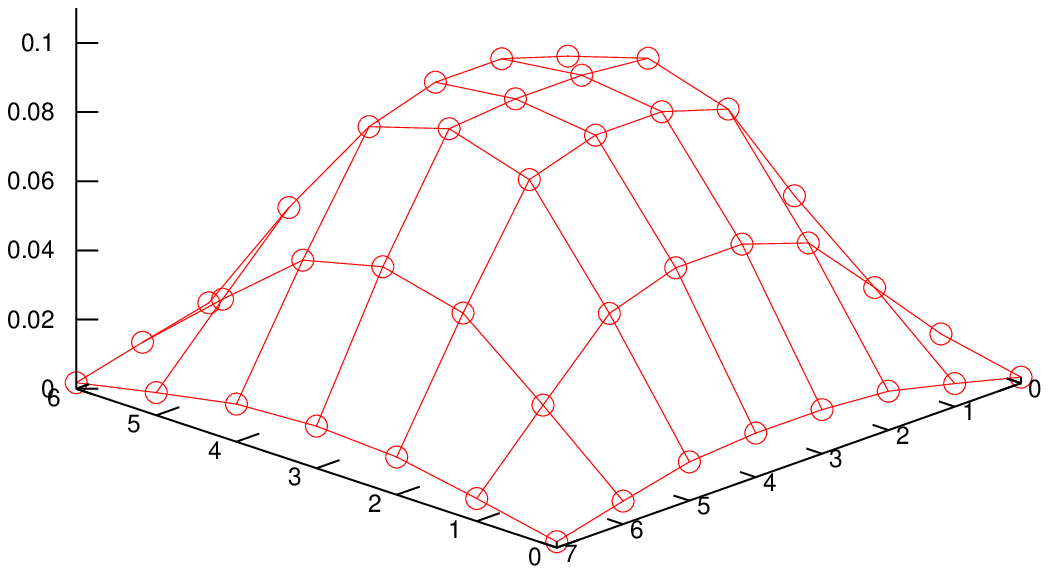}
\end{minipage}
\hfil
\begin{minipage}[t]{.45\textwidth}
\includegraphics[trim=0mm 0mm 0mm 0mm, clip, width=.95\textwidth]{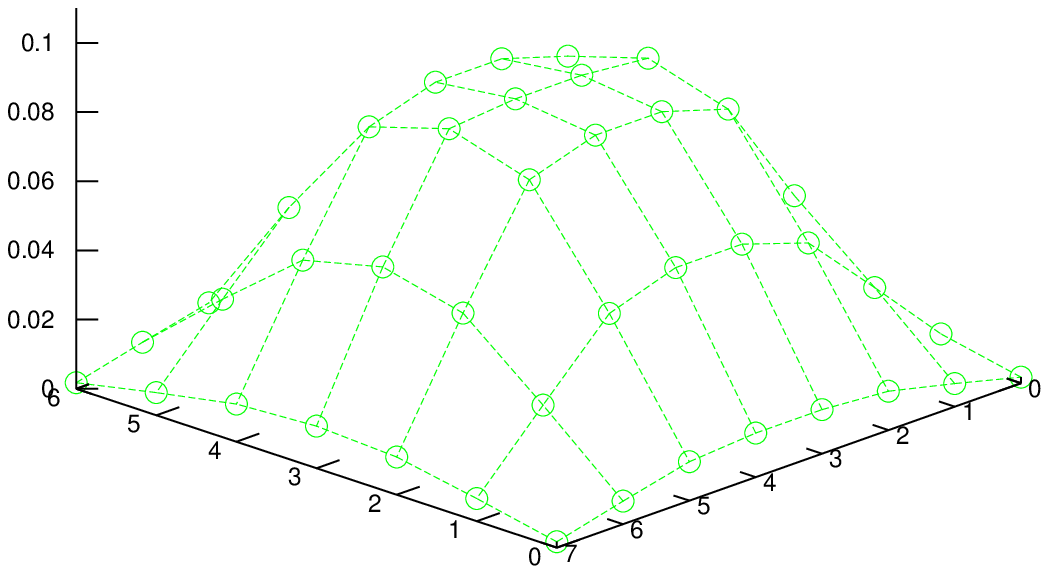}
\end{minipage}
\caption[Electron distribution in the level crossing vicinity]
{\label{fig:pop_0.678} 
The same as Fig.~\ref{fig:pop_0.550}, with
$\lambda=\lambda_c=0.678$, resulting in identical charge distributions.
}
\end{figure}

\begin{figure}[htbp]
\centering
\begin{minipage}[t]{.45\textwidth}
\includegraphics[trim=0mm 0mm 0mm 0mm, clip, width=.95\textwidth]{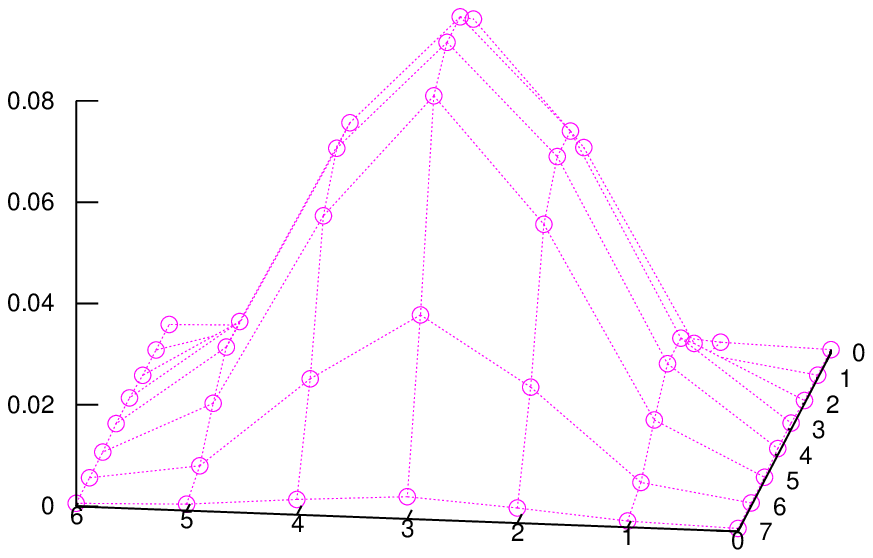}
\end{minipage}
\hfil
\begin{minipage}[t]{.45\textwidth}
\includegraphics[trim=0mm 0mm 0mm 0mm, clip, width=.95\textwidth]{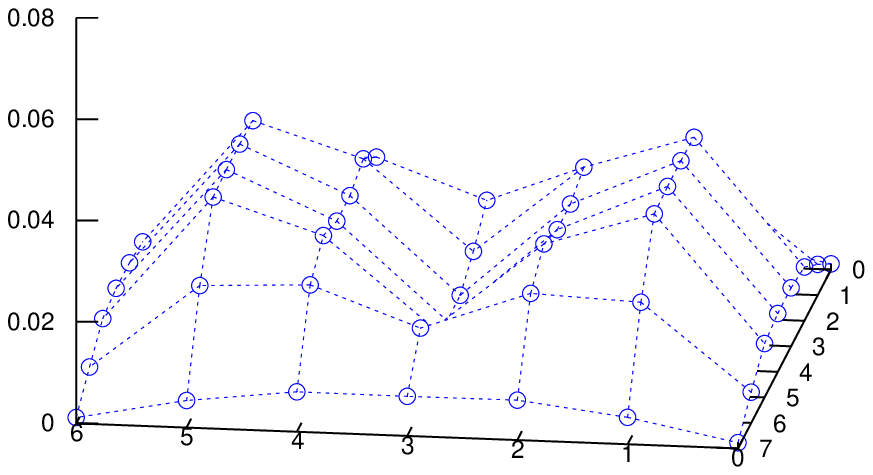}
\end{minipage}
\caption[Electron distribution separated between up and down spins]
{\label{fig:pop_0.678_updown} 
The electron distribution of the ground state of Fig.~\ref{fig:pop_0.678} separated between
spin-up (left) and spin-down (right) electrons.
In the first excited state these distributions are exchanged.
}
\end{figure}

Looking at the particle distribution throughout the lattice can shed some light over
the mechanism of the LC. For most values of $\lambda$ the
first two doubly-occupied states are different in their density distribution,
as can be seen in Fig.~\ref{fig:pop_0.550}. However, in the vicinity of the LC, 
these states are identical in their spatial components (Fig.~\ref{fig:pop_0.678}). 
Yet, their spin degree of freedom gives rise to a polarization of both states, 
in opposite spin directions (Fig.~\ref{fig:pop_0.678_updown}).
One may conclude that a specific combination of the SO and the interaction strengths can 
lead to a separation of the spatial and the spin degrees of freedom. In such a case the lowest 
two states are identical in their spatial coordinates, whereas the spin degree of freedom
is responsible for their polarization. 
In other words, states which experience a LC are 
time-reversal of each other, similarly to the non-interacting case.

\begin{figure}[htbp]
\centering
\includegraphics[trim=0mm 0mm 0mm 0mm, clip, width=3in,height=!]{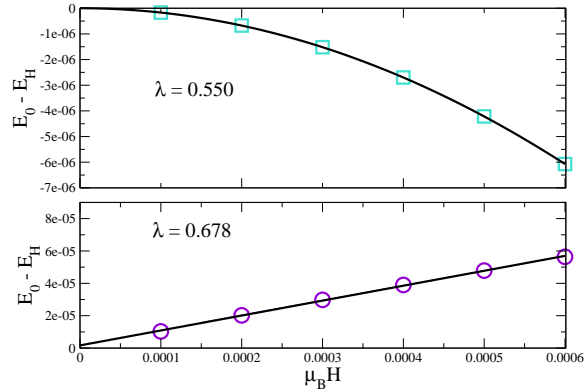}
\caption[Energy dependence on magnetic field in the level crossing vicinity]
{\label{fig:E_H_n2_I} 
The dependence of the energy on the magnetic field is compared between the regime of a
level crossing (lower panel), to another arbitrary point (upper panel). The results shown by symbols were obtained 
for a system of $8 \times 7$ sites, with $U_H=10$, and the solid lines represent quadratic
(upper panel) and linear (lower panel) fits. 
}
\end{figure}

We thus see that in the vicinity of a LC a finite value of the g-factor can be 
obtained, in contrast to the ordinary assumption of $g=0$ for a doubly-occupied system.
In general, the vanishing of the g-factor results from the quadratic dependence of 
the ground-state energy on the magnetic field. 
Such a dependence is shown in the upper panel of Fig.~\ref{fig:E_H_n2_I},
for an arbitrary value of the SO coupling. 
However, near a LC point, as shown above, the ground-state has a finite spin 
polarization. As a result, the dependence of the energy on the magnetic field is linear
(lower panel of Fig.~\ref{fig:E_H_n2_I}), and the g-factor is finite
(lower panel of Fig.~\ref{fig:Sz_n2_I}).
The clear linear dependence in the exact point $\lambda_c$
is actually limited to the region in which $\mu_B H$ is greater than the energy difference 
of the two many-particle states. The same restriction holds for the peaks in 
$\langle \hat S_z \rangle$ and $g$, which are thus getting wider as the magnetic field 
is enhanced.

From these results one can conclude that whereas the g-factor vanishes by definition
for most values of $\lambda$, it has a finite value near $\lambda_c$. Calculating
the g-factor using Eq.~(\ref{eqn:g_gdef2}) leads to the typical results shown in the
lower panel of Fig.~\ref{fig:Sz_n2_I}. As will be discussed in the next section,
such g-factor values can be significantly large, and thus they cannot be neglected.

\section{Experimental Relevance}
\label{sec:g_phys}
The enhancement of the g-factor discussed in the previous section, was obtained for various
system sizes. However, since the exact diagonalization technique used is limited by size, the 
scalability question, i.e. the question whether such a finite g-factor can be experimentally 
measured, is important. The peak, in both $\langle \hat S_z \rangle$ and the g-factor, 
can be characterized mainly by two properties, namely the peak height and its width.
In addition, the peak occurrence is characterized by the value of $\lambda_c$, 
and by the corresponding interaction strength and lattice size for which it occurs. 

To answer the scalability question one should check if for realistic sample sizes with a 
reasonable strength of interaction the crossing point $\lambda_c$ is small enough  
to be obtained by realistic doping with magnetic impurities. In addition, one has to find
whether the width and the height of the predicted g-factor peak at this point are experimentally
measurable.

\begin{figure}[htbp]
\centering
\includegraphics[trim=0mm 0mm 0mm 0mm, clip, width=4in,height=!]{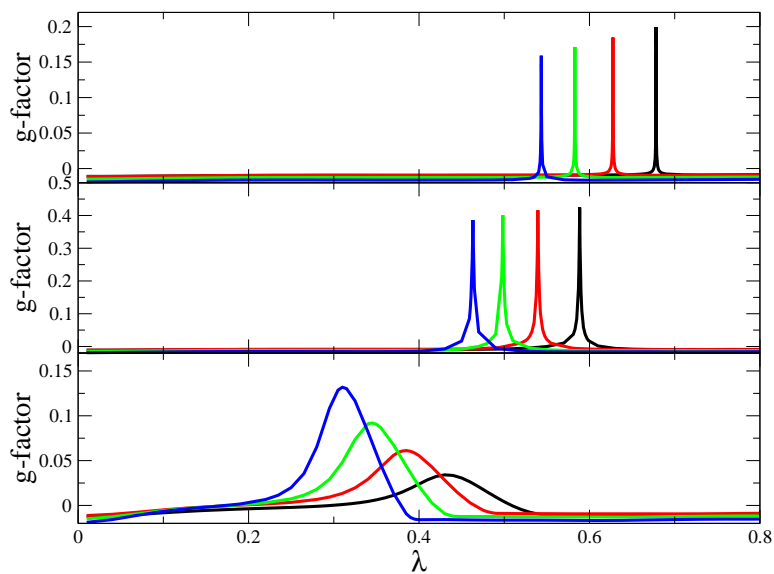}
\caption[The g-factor peak for various system sizes and interaction schemes]
{\label{fig:g_so_UCNN} 
The g-factor of the doubly-occupied ground-state as a function of the SO coupling strength $\lambda$,
for lattices sizes of $8 \times 7$ (black), $9 \times 8$ (red), $10 \times 9$ (green) and
$11 \times 10$ (blue). Different panels correspond to different type of interactions: 
Hubbard interactions with $U_H=10t$ (upper panel), Hubbard and NN interactions with $U_H=10t$ and
$U_{NN}=5t$ (middle panel), Hubbard and Coulomb interactions with $U_H=10t$ and
$U_{C}=5t$ (lower panel). 
}
\end{figure}

Since a substantial enlargement of the system is not numerically possible, in the following
we present results for different system sizes, from which the trend can be clearly understood.
For example, in the upper panel of Fig.~\ref{fig:g_so_UCNN}, the g-factor peak is shown for system sizes
ranging from $8 \times 7$ to $11 \times 10$. As can be seen, the value of $\lambda_c$
decreases with increasing system size, suggesting that for a sufficiently large system size
the crossing occurs for moderate value of the SO coupling.
On the other hand, since the peak height decreases slowly with increasing system size,
the question if it will not be negligible for realistic system sizes remains open.

Nevertheless, the effect of other types of interactions can change this picture. Using 
NN interactions (in addition to the Hubbard term) results in much higher peaks, so that
there is a larger chance to find a finite g-factor for larger systems. This is shown in the 
middle panel of Fig.~\ref{fig:g_so_UCNN}. In the lower panel we show the results obtained when Coulomb 
interactions were added to the Hubbard term. Although the peak heights are smaller, they become 
much wider, and most important, the peak height increases with system size. 

Further investigation of the Coulomb case shows that the maxima in the g-factor curves
are not accompanied by LCs between the first two levels. Yet,
they occur when the energies of these levels are close enough. Furthermore,
these maxima, and their different shape, result from the second order term (i.e., the $H^2$ term) 
in the energy. The corresponding g-factors are thus of first order in the magnetic field
(and not constant as in the Hubbard and NN cases). Nevertheless, significant values are obtained, 
even for a magnetic field as weak as $\mu H/t \approx 10^{-4}$, corresponding to $10-100$ Gauss.

Moreover, since the maxima are the result of avoided crossings, and the energy difference between the
levels is expected to reduce when larger samples are treated,
the values of the g-factor in the maxima region should further increase. 
This is in contrast to the other cases (Hubbard, NN)
where the peak results from a true LC and decreases with increasing system size,
probably due to a decrease in the inter-level Hamiltonian matrix element
caused by the applied magnetic field.

In order to get insight into the other differences between the three cases, namely the
LC and the sharp g-factor peak in the Hubbard and in the NN cases, which are absent in the Coulombic case,
the role of the spin component is further explored. By considering $U_{\uparrow \uparrow}=U_{\downarrow \downarrow} = U_1$
and $U_{\uparrow \downarrow}=U_{\downarrow \uparrow} = U_2$ (for either $U_{NN}$ or $U_C$),
one can check whether these phenomena are observed for different parameter regimes,
and specifically for the limits of parallel spin interactions ($U_2 \rightarrow 0$ with a finite $U_1$), and 
anti-parallel spin interactions ($U_1 \rightarrow 0$ and $U_2$ is finite).

For the NN case, we find that the g-factor peak is enhanced when the interactions are
only between anti-parallel spins, while it disappears for the case in which only
interactions between parallel spins are considered.
For the Coulombic case, when only anti-parallel interactions are used, 
a sharp g-factor peak does appear. However, when parallel interactions are considered,
either with or without the anti-parallel ones, the sharp peak disappears.
In all cases an appearance of the sharp g-factor peak is accompanied by a LC.
The different cases are summarized in Table \ref{tbl:sum_ucnn}.

\begin{table}[ht] 
\centering 
\begin{tabular}{|c||c|c|c|}
\hline
 & & & \\ [0.2ex]
&{\bf Spin independent} & {\bf Parallel} & {\bf Anti-parallel} \\
&{\bf ($U_1=U_2$)}&{\bf ($U_2 \rightarrow 0$)}&{\bf ($U_1 \rightarrow 0$)}\\ [1ex]
\hline \hline  
 & & & \\ [0.2ex]
NN & LC & - & LC \\ [1ex]
C & -  & - & LC \\ [1ex]
\hline
\end{tabular}
\caption{\label{tbl:sum_ucnn} Level crossing occurrence for different interaction types} 
\end{table}

These results can be understood in the following way. As we have previously shown,
the crossing levels are the time reversal of each other, and the spin degree of freedom 
is responsible for their spin polarization. Therefore the question whether a LC can 
occur is crucially related to the possibility to polarize the lowest states. 
The polarization of these states might become improbable when interactions between
parallel spins are considered.

Therefore, when there is only anti-parallel interactions, for both interaction 
types (NN, Coulomb) a polarization of the ground state is possible, and thus a LC does occur. 
In the opposite case, when the interactions are only between parallel spins, 
the probability of spin polarization is reduced because of the interaction. 
For the case of Coulomb interactions, 
the polarization of the lowest states is totally blocked, and a LC cannot occur.
In the NN case, however, since the interaction is only short ranged, single-particle states which are 
spatially separated can be combined, in a rough approximation, to a polarized many-particle state. 
However, a two-particle state composed of anti-parallel spins is energetically preferable,
for any strength of $\lambda$, so that a LC of the lowest states does not occur.

In the case of spin-independent interactions the difference between Coulomb and NN interactions
stems from the interaction range. The presence of both parallel and anti-parallel interactions types
causes the re-appearance of a polarized ground-state in the NN case, combined 
of spatially separated single-particle states.
On the other hand, the long range of the Coulomb interactions prevents the ground-state polarization,
and thus a LC between the lowest two levels does not occur. It should be noted,
however, that a strong avoided crossing does appear (lower panel of Fig.~\ref{fig:g_so_UCNN}).
In addition, a polarization of an excited state, as well as LC between excited states, are still possible.


\section{Conclusions and Future Prospects}
In this chapter we have shown that the combination of interactions and spin-orbit 
scattering can cause unexpected magnetization of states with an even number 
of electrons. We have also shown that for realistic sizes of QDs, such a result
can be experimentally observed, and might be relevant for understanding some
measurements.

As we have noted, one of the popular methods for measuring the g-factor is by tunneling
spectroscopy. The number of electrons in the dot in such an experiment is changed by one during
each tunneling event, involving a transition between an even electron number and an odd one.
According to our results, such a measurement might present the result for the difference
in g-factor between the two states, Eq.~(\ref{eqn:g_gdef_tot}). 
If the even-electron state has a non-vanishing g-factor, like in the vicinity of the 
LCs we have presented, the measured quantity $\tilde g$ may not equal the 
g-factor of the odd electron state, to which it is usually attributed.

\begin{figure}[htbp]
\centering
\includegraphics[trim=0mm 0mm 0mm 0mm, clip, width=4in,height=!]{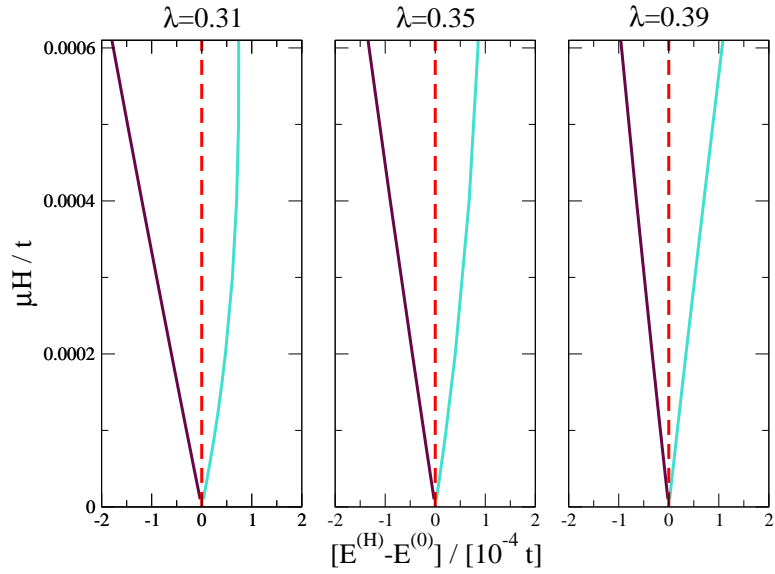}
\caption[Motion of Coulomb peaks with magnetic field]
{\label{fig:E0_conc} 
The motion of the first Coulomb peaks ($n_e=1,2$) as a function of the magnetic field
is shown for a lattice of $11 \times 10$ with $U_H=10t$ and $U_{C}=5t$ (for which
$\lambda_c\approx 0.31$).
}
\end{figure}

In such cases, a trace of the LC may be seen experimentally. In the
regular case (as opposed to the LC scenario), the two levels which belong to the same 
Kramers' doublet have the same g-factor, and the motion of the two energies as a function of 
a magnetic field is symmetric. However, in the region of a LC, the two levels get 
contributions from different even-particle states.
Explicitly, the $p$ Kramers' pair is divided by the magnetic field to the measured values 
$g(2p-1)-g(2p-2)$ and $g(2p)-g(2p-1)$. In general, this motion, as a function of
the magnetic field, is not symmetric. An example is presented in Fig.~\ref{fig:E0_conc}.
As one can see, the most clear non-symmetric motion is obtained for $\lambda \approx \lambda_c$
(left panel), but such a motion can be seen for a region in its vicinity as well (middle panel). 
Far enough from this region (right panel) the symmetric motion reappears.

\begin{center}
*~~~~~*~~~~~*
\end{center}

As a last remark we note that the model we have used in this chapter has neglected any kind of disorder 
inside the QD. This is in contrast with the usual experimental configuration, in which the
fabrication of clean samples is difficult. It is thus interesting to check
the influence of disorder on the two-particle g-factor peak described above, for example to investigate 
how its place and shape vary between different realizations. 
We leave this question to future research.


\newcommand{\newchapp}[1]    
	{                   
	\chapter{#1}
	  \markboth{\hfill \thechapter. \MakeUppercase{{{#1}}}} {}
	  \markright{\thechapter. \MakeUppercase{{{#1}}} \hfill }
	}

\cleardoublepage
\newpage 

\newchapp{Summary}
\label{conc}

In this thesis we have examined two types of mesoscopic systems which are
commonly used nowadays in innovative physical studies. The first type we have
investigated is a quantum dot (QD) which is coupled to a one-dimensional (1D) 
wire, and the second type is an isolated two-dimensional (2D) QD.

When a QD is coupled to one end of a semi-infinite 1D lead, it may have an 
important influence on the wire's characteristics. Using the numerical
density matrix renormalization group (DMRG) method we have shown 
in chapter \ref{cpt:ch1} that the
thermodynamic properties of the entire system are sensitive to the dot 
properties: its energy level and the strength of its coupling to the wire.
The wire itself can be in one of three phases, the ferromagnetic (FM),
the Tomonaga-Luttinger liquid (TLL) and the charge density wave (CDW) phases.
When the wire is in the TLL phase, a change of the dot's level causes
a continuous change in the level occupation, in the total occupation
and in the free energy. We have shown that this change can be explained
within the random phase approximation.

On the other hand, when the wire is described by one of the other couple 
of phases, the FM or the CDW, these thermodynamic quantities behave in a
different way. There is an abrupt jump in the dot occupation when its 
orbital crosses the chemical potential of the lead. This jump is
accompanied by an inversion of the occupation of each site in the lead,
and in an abrupt change of the first derivative of the free energy. 
We have proven that this is a result of a simple level crossing in 
the FM case, whereas it is a sign of a first order quantum phase 
transition in the case of a CDW.

Another influence of the dot on the coupled system is the Friedel oscillations (FO)
in the wire, which are investigated in chapter \ref{cpt:ch2}. In metallic
phases, such as the TLL one, these oscillations decay as a power law, with
an exponent depending on the interaction strength. If the wire is an
insulator, as the CDW phase is, the FO decay exponentially.

Once disorder is introduced in the wire, its phase changes to an 
Anderson insulator (AI). However, for a very weak disorder applied onto a CDW, 
a finite wire may still be described as a Mott insulator (MI). The effect
of the disorder on the FO decay can be described as another exponential 
decay factor, with a characteristic decay length. We have shown that 
for a fixed weak disorder, an enhancement of the interactions 
in the AI and the MI phases leads to different results: the decay length 
decreases in the AI phase, while it increases in the MI case.
This difference was explained according to the interplay between
interactions and disorder.

In the AI regime, we have proven that the decay length can be associated 
with Anderson localization length. Our results, presenting a decrease of
the localization length with increasing interactions, confirm previous
predictions.

The other type of mesoscopic systems we investigate is a 2D QD.
Considering such a disordered QD, with interacting spinless electrons,
we suggest in chapter \ref{cpt:ch3} to use the numerical particle-hole
DMRG (PH-DMRG) method in order to approximate the system's ground state. 
We have shown that an improvement of the PH-DMRG truncation method leads 
to results which are much more accurate than those obtained by Hartree-Fock 
approximations. Furthermore, a significant improvement of the accuracy was 
exhibited when the revised PH-DMRG method was used to calculate the addition 
spectrum of the QD. The suggested method thus opens a door to accurate 
calculations of ground-state properties in two dimensions.

In chapter \ref{cpt:ch4} we have investigated the lowest states of the 
2D QD occupied by spin $1/2$ electrons with interactions and in the presence of
spin-orbit coupling. We have shown that at certain values of the spin-orbit 
coupling one can obtain a level crossing between the lowest two many-body levels 
of a doubly occupied dot. At the crossing point these two states have identical
charge distributions, whereas they are different in their spin degree of freedom.
Therefore, the level crossing is accompanied by a finite magnetization of the 
ground state, and a finite g-factor is obtained, in contrast to the
usual $g=0$ case for an even number of electrons. An investigation of
the size dependence of this phenomenon suggests that it might have
significant impact on g-factor measurements.












\cleardoublepage
\addcontentsline{toc}{chapter}{References}
\bibliographystyle{}


\begin{thebibliography}{}
\bibitem{schollwock05-2}
A short review of the PH-DMRG method and some of its implementations can be found in page 36 
of the DMRG review of U. Schollw\"ock (Ref.~\cite{schollwock05}).
\end{thebibliography}

\end{document}